\documentclass[epj]{svjour}
\usepackage[maxfloats=256]{morefloats}
\usepackage{url}
\usepackage{amsmath}
\usepackage{graphicx}
\usepackage{dcolumn}
\usepackage{physics}
\usepackage{bm}
\usepackage{hyperref}
\usepackage{epsfig,colordvi}
\usepackage{hyperref}
\usepackage{graphicx}
\usepackage{dcolumn}
\usepackage{xspace}
\usepackage{times}
\usepackage{bm}
\usepackage{soul}
\usepackage{ulem}
\usepackage{xcolor}
\usepackage{color}

\usepackage{float}
\usepackage{booktabs} 

\RequirePackage[numbers,sort&compress]{natbib}

 \def\be{\begin{equation}}
 \def\ee{\end{equation}}
 \def\bea{\begin{eqnarray}}
 \def\eea{\end{eqnarray}}
 \def\bean{\begin{eqnarray*}}
 \def\eean{\end{eqnarray*}}

\begin{document}

\title{Study on the equation-of-state with light clusters and hypernuclei}

\author{Elena Bratkovskaya\inst{1,2,3} and Jörg Aichelin\inst{4,5} }
%
%
\institute{ 
  GSI Helmholtzzentrum f\"{u}r Schwerionenforschung GmbH, Planckstrasse 1, 64291 Darmstadt, Germany \and  
  Institute for Theoretical Physics, Johann Wolfgang Goethe Universit\"{a}t, Frankfurt am Main, Germany  \and
  Helmholtz Research Academy Hessen for FAIR (HFHF), GSI Helmholtz Center for Heavy Ion Physics. Campus Frankfurt, 60438 Frankfurt, Germany  \and 
  SUBATECH, Nantes University, IMT Atlantique, IN2P3/CNRS, 4 rue Alfred Kastler, 44307 Nantes cedex 3, France \and
  Frankfurt Institute for Advanced Studies, Ruth Moufang Str. 1, 60438 Frankfurt, Germany             }
\date{Received: date / Revised version: date}
%
\abstract{
Heavy-ion collision experiments offer a unique opportunity to explore the early stages of the Universe by creating matter under extreme conditions of high temperature and baryon density. The properties of such matter are governed by the equation-of-state (EoS), which remains a central focus of investigation from both experimental and theoretical perspectives. Flow harmonics are among the most sensitive observables for probing the EoS, as they strongly reflect the underlying interactions and degrees of freedom of the system.
In this article, we review the current status of our understanding of the EoS based on microscopic transport models, emphasizing comparisons with experimental data in the few-GeV energy range.
\PACS{{ } heavy-ion collisions, transport models, cluster production, collective flow}
} 

\titlerunning{Study on equation-of-state with light clusters and hypernuclei }
\authorrunning{ E. Bratkovskaya  and J. Aichelin}
\maketitle
\section{Introduction}
\label{intro}

The quest for the equation-of-state (EoS) of strongly interacting matter is one of the major objectives of present-day nuclear physics. Its knowledge is a cornerstone for the understanding of heavy-ion reactions, the mass-radius relation of neutron stars, and neutron star mergers. Experimentally, its study started with the first heavy-ion experiments at the Bevalac accelerator in Berkeley \cite{Gustafsson:1984ka}, where nuclear densities well above normal nuclear matter density, $\rho_0$, could be obtained in the laboratory. Later, complementary information has been obtained from the observed mass-radius relations of neutron stars \cite{Koehn:2024set,Ozel:2010bz,Lastowiecki:2011hh,Bogdanov:2021yip,Bogdanov:2019ixe,Bogdanov:2019qjb,Raaijmakers:2019qny} and, more recently, from gravitational waves emitted during such mergers \cite{Bauswein:2012ya,Most:2022wgo,LIGOScientific:2018cki,LIGOScientific:2020aai}.

On the theoretical side, the EoS of nuclear matter can be calculated by the Brueckner G-matrix approach \cite{Botermans:1990qi,Fuchs:1998zz} or by chiral effective theories \cite{Holt:2016pjb}. However, both rely on expansion schemes that limit their predictive power to densities below or close to $\rho_0$. Lattice QCD calculations face the sign problem for finite chemical potential $\mu$, making them unreliable at the high densities reached in heavy-ion collisions. Therefore, the theoretical interpretation of these experiments is a primary way to study the EoS systematically \cite{Gale:1987zz,Aichelin:1987ti,Ko:1987gp,Aichelin:1991xy,Weber:1993et,Fuchs:1998zz,Sahu:1998vz,Danielewicz:1998vz,Kuhrts:2000zs,Sahu:1999mq,Cassing:2000bj,Sahu:2002ku,Nara:2020ztb}.

In this situation, comparing experimental heavy-ion data with results from transport approaches, where the EoS is an input, which can be varied, is the only systematic method for its exploration. By varying beam energy, system size, and centrality, the EoS can be probed at densities up to $\sim 3\rho_0$ and temperatures up to $T=120$ MeV, covering a region relevant for neutron star mergers. At beam energies around 1 GeV/nucleon, where densities up to three times $\rho_0$ can be reached and meson production is infrequent, the nucleon dynamics is very sensitive to the potential interaction, directly related to the EoS \cite{Gale:1987zz,Aichelin:1987ti,Ko:1987gp,Aichelin:1991xy,Weber:1993et,Fuchs:1998zz,Sahu:1998vz,Danielewicz:1998vz,Kuhrts:2000zs,Sahu:1999mq,Cassing:2000bj,Sahu:2002ku,Nara:2020ztb}. These studies revealed that the directed ($v_1$) and elliptic ($v_2$) flow, along with subthreshold kaon production, are among the most promising experimental signals for determining the EoS.

Theoretical predictions within microscopic transport appro\-aches are challenging because flow coefficients are sensitive not only to the potentials (reflecting the EoS) but also to the properties of hadrons in the medium, their collisions, and the reaction centrality. In earlier transport calculations, the EoS was considered static. From the first Plastic Ball data \cite{Gustafsson:1984ka}, it was concluded that the nuclear EoS is rather hard \cite{Molitoris:1986pp}. Later, it was realized that the strong momentum dependence of the nucleon-nucleon potential must be included \cite{Cooper:2009zza}. 

This has been investigated within Quantum Molecular Dynamics (QMD) models \cite{Aichelin:1987ti,LeFevre:2015paj,Hillmann:2019wlt}, where the implementation is performed in a semi-classical framework using non-relativistic two-body potentials, and within BUU-type models \cite{Welke:1988zz,Gale:1987zz,Gale:1989dm,Zhang:1994hpa,Sahu:1998vz,Tarasovicova:2024isp,Mohs:2024gyc}, where the interaction is treated at the mean-field (MF) level. In the latter case, the potential (with scalar and vector parts) can be connected to the  self-energies of particles in the medium from  Dirac–Brueckner many-body theory, which enables a fully covariant formulation of the transport theory \cite{Cassing:1992gf}.
Two common conclusions emerged \cite{Aichelin:1987ti,Welke:1988zz,Gale:1987zz,Gale:1989dm,Sahu:1999mq,Cassing:2000bj,Sahu:2002ku,Zhang:1994hpa,Tarasovicova:2024isp,Mohs:2024gyc}: 
i) the nucleon flow depends strongly on momentum dependence—a static hard and a soft momentum dependent interaction give similar directed flow, while a static soft EoS gives lower flow values; 
ii) the experimentally measured flow is approximately described by a soft momentum dependent interaction. For reviews, see \cite{Bleicher:2022kcu,TMEP:2022xjg,Sorensen:2023zkk}.

The strategy to determine the static part of the nuclear EoS involves employing parametrized nucleon-nucleon or mean-field potentials in transport approaches, corresponding to different EoS at zero temperature. The compressibility modulus $K$ (often the third parameter after fixing binding energy at saturation density) traditionally characterizes the EoS curvature at $\rho_0$:
\begin{equation}
\kappa = 9\rho \left. \frac{d P}{d\rho}\right\vert_{\rho=\rho_0}=\left. 9 \rho^2 \frac{{\rm \partial}^2(E/A(\rho))}{({\rm \partial}\rho)^2} \right\vert_{\rho=\rho_0}.
\end{equation}
Studies of monopole vibrations \cite{Mekjian:2011wut} suggest $\kappa \approx 200$ MeV (soft EoS, ``S''), while early Plastic Ball data \cite{Gustafsson:1984ka} could be explained by $\kappa \approx 380$ MeV (hard EoS, ``H''). 
Including the momentum dependence of the nucleon-nucleon potential one can keep the same form but has to modify the parameters.

Recently, transport approaches have advanced, and new data sets for Au+Au collisions at similar energies from the HADES \cite{HADES:2020lob,HADES:2022osk} and FOPI \cite{FOPI:2011aa} collaborations, as well as STAR \cite{STAR:2023uxk,STAR:2021yiu,STAR:2024znc,STAR:2021ozh,STAR:2021orx,STAR:2022fnj} at $\sqrt{s_{\rm{NN}}}=3$ GeV, have become available. These experiments measure the directed $v_1$ and elliptic $v_2$ flow  of not only nucleons but also light clusters (deuteron, triton, ${}^{3}\rm{He}$, ${}^{4}\rm{He}$) and hypernuclei, allowing EoS studies to be extended to clusters. This is possible despite the debated origin of cluster production at midrapidity, found from a few hundred MeV per nucleon up to LHC energies \cite{Reisdorf:2010aa,NA49:2004mrq,NA49:2016qvu,STAR:2019sjh,ALICE:2015wav}. The slope of transverse momentum spectra ($\sim 100$ MeV) and the smooth excitation function of multiplicities suggest an energy-independent formation mechanism \cite{Adam:2019wnb,HADES:2020ver,STAR:2019sjh,STAR:2022hbp}.

Several mechanisms have been proposed for light cluster production at midrapidity:
\begin{enumerate}
    \item[i)] Statistical hadronization at a given temperature and chemical potential \cite{Andronic:2010qu}, which cannot predict flow without additional assumptions.
    \item[ii)] Coalescence \cite{Butler:1963pp,Zhu:2015voa,Zhao:2020irc,Sun:2021dlz,Sombun:2018yqh,Kireyeu:2022qmv}, where nucleons form clusters if their relative distance in momentum and coordinate space is below thresholds $\Delta r_0$ and $\Delta p_0$.
    \item[iii)] Dynamical production by the same potential interaction governing the baryon evolution in transport approaches \cite{Aichelin:1991xy,Aichelin:2019tnk,Glassel:2021rod,Coci:2023daq}.
    \item[iv)] Formation via three-body collisions (e.g., $NNN \to dN$, $NN\pi \to d\pi$) and destruction by inverse reactions \cite{Danielewicz:1991dh,Danielewicz:1992mi,Oliinychenko:2020znl,Staudenmaier:2021lrg,Wang:2023gta,Coci:2023daq,Sun:2021dlz,Ege:2024vls}.
   
\end{enumerate}
While observables sensitive to these mechanisms exist, current data do not allow experimental distinction based on rapidity and $p_T$ distributions alone \cite{Kireyeu:2024woo}.

In this paper we review the current status of the EoS study based on microscopic transport models in the few-GeV energy range.  We will focus on two main problems:
\begin{itemize}
    \item study of the EoS of strongly interacting hadronic matter by analyzing the directed flow $v_1$ and elliptic flow $v_2$ of protons and light clusters;
    \item exploration of the influence of momentum dependent potentials on the flow observables;
    \item investigation to which extend $v_1$ and $v_2$ can distinguish between different cluster production models.
\end{itemize}
As a main theoretical tool we employ Parton-Hadron-Quantum-Molecular Dynamics (PHQMD) \cite{Aichelin:2019tnk,Glassel:2021rod,Kireyeu:2022qmv,Coci:2023daq,Kireyeu:2024woo,Kireyeu:2024hjo,Zhou:2025zgn}, a microscopic n-body transport approach. 
PHQMD allows for a direct comparison of dynamical (potential + kinetic) and coalescence production mechanisms applied to the same events, enabling a study of their consequences. We will base our review on two recent PHQMD studies at lower beam energies (1.2--1.5 AGeV) \cite {Kireyeu:2024hjo} to $\sqrt{s_{\rm{NN}}}=3$ GeV \cite{Zhou:2025zgn}, exploring the influence of momentum dependent potentials.
There we confront our calculations with HADES \cite{HADES:2020lob,HADES:2022osk}, FOPI \cite{FOPI:2011aa}, and STAR data \cite{STAR:2023uxk,STAR:2021yiu,STAR:2024znc,STAR:2021ozh,STAR:2021orx,STAR:2022fnj}.

We will relate the PHQMD observations to recent results from UrQMD \cite{Hillmann:2019wlt,Steinheimer:2024eha} and SMASH \cite{Tarasovicova:2024isp,Mohs:2024gyc}, as well as discuss the results for collective flow observables of other models for cluster production  as BUU \cite{Danielewicz:1999zn,Danielewicz:2002pu}, IQMD \cite{Hartnack:1997ez}, dcQMD \cite{Cozma:2024cwc}.

\phantom{a}\\
This paper is organized as follows: In Section \ref{PHQMD}, we introduce the basic concepts of the PHQMD model and describe the implementation of static and momentum-dependent potentials. Section \ref{Models} reviews the constraints on the equation-of-state derived from flow observables, comparing results from PHQMD, pBUU, RBUU, IQMD, dcQMD, UrQMD, and SMASH. Finally, our main conclusions are summarized in Section \ref{conclusions}.

\section{Model description: PHQMD}
\label{PHQMD}

The Parton-Hadron-Quantum-Molecular Dynamics (PHQMD) \cite{Aichelin:2019tnk,Glassel:2021rod,Kireyeu:2022qmv,Coci:2023daq,Kireyeu:2024woo} is a microscopic n-body transport approach that combines the baryon propagation from the Quantum Molecular Dynamics (QMD) model \cite{Aichelin:1991xy,Aichelin:1987ti,Aichelin:1988me,Hartnack:1997ez} with the dynamical properties and interactions of hadronic and partonic degrees-of-freedom from the Parton-Hadron-String-Dynamics (PHSD) approach \cite{Cassing:2008sv,Cassing:2008nn,Cassing:2009vt,Bratkovskaya:2011wp,Linnyk:2015rco,Moreau:2019vhw}. Here we recall the basic concepts for the implementation of the potential in PHQMD.

\subsection{QMD Propagation}

The QMD equations-of-motion (EoM) for an N-body system are derived using the Dirac-Frenkel-McLachlan variational principle \cite{raab:2000,BROECKHOVE1988547}, originally developed in chemical physics and later applied to nuclear physics for QMD-like models \cite{Feldmeier:1989st,Aichelin:1991xy,Ono:1992uy,Hartnack:1997ez}. The time evolution of the N-body wave function $\psi$ is obtained from the variation
\begin{equation}
\delta \int_{t_1}^{t_2} dt \langle \psi(t) | i\frac{d}{dt} - H | \psi(t) \rangle = 0,
\label{varS}
\end{equation}
where $H$ is the N-body Hamiltonian. Equation (\ref{varS}) is solved by approximating the N-body wave function as the direct product of single-particle trial wave functions (without antisymmetrization), $\psi = \prod_i \psi_i$.

Assuming Gaussian wave functions with time-independent width $L$, one obtains equations-of-motion for the centroids \\ 
$({\bf r_{i0}, p_{i0}})$ of the Gaussian single-particle Wigner densities, which resemble the EoM of classical particles with phase space coordinates ${\bf r_{i0}, p_{i0}}$ \cite{Aichelin:1991xy}:
\begin{equation}
\dot{r}_{i0} = \frac{\partial \langle H \rangle}{\partial p_{i0}}, \qquad
\dot{p}_{i0} = -\frac{\partial \langle H \rangle}{\partial r_{i0}}.
\label{prop}
\end{equation}
The key difference from classical EoM is that the expectation value of the quantum Hamiltonian is used rather than a classical Hamiltonian.

In PHQMD, the single-particle Wigner density of the Gaussian wave function for a nucleon is given by
\begin{eqnarray}
&& f({\bf r_i, p_i, r_{i0}, p_{i0}}, t) = \label{Wignerdens} \\
&& = \frac{1}{8\pi^3 \hbar^3}
e^{-\frac{2}{L} ({\bf r_i} - {\bf r_{i0}}(t))^2}
e^{-\frac{L}{8\hbar^2} ({\bf p_i} - {\bf p_{i0}}(t))^2}, \nonumber
\end{eqnarray}
where the Gaussian width is taken as $L = 4.33$ fm$^2$. The corresponding single-particle spatial density is obtained by integrating over momentum and summing over all nucleons:
\begin{eqnarray}
\rho_{sp}({\bf r}, t) &=& \sum_i \int d{\bf p_i} f({\bf r, p_i, r_{i0}, p_{i0}}, t) \nonumber \\
&=& \sum_i \left( \frac{2}{\pi L} \right)^{3/2} e^{-\frac{2}{L} ({\bf r} - {\bf r_{i0}}(t))^2}.
\label{rhosp}
\end{eqnarray}

For a system of N nucleons, the Hamiltonian is the sum of individual nucleon Hamiltonians, composed of kinetic and two-body potential energy terms:
\begin{equation}
H = \sum_i H_i = \sum_i \left( T_i + \sum_{j \neq i} V_{ij} \right).
\label{HTV}
\end{equation}

The total potential energy between nucleons in PHQMD has three components: a local static Skyrme-type interaction, a local momentum dependent interaction, and a Coulomb interaction:
\begin{eqnarray}
V_{ij} &=& V({\bf r}_i, {\bf r}_j, {\bf r}_{i0}, {\bf r}_{j0}, {\bf p}_{i0}, {\bf p}_{j0}, t) \nonumber \\
&=& V_{\rm Skyrme\,loc} + V_{\rm mom} + V_{\rm Coul} \nonumber \\
&=& \frac{1}{2} t_1 \delta({\bf r}_i - {\bf r}_j) + \frac{1}{\gamma+1} t_2 \delta({\bf r}_i - {\bf r}_j) \rho_{\rm int}^{\gamma-1}({\bf r}_{i0}, {\bf r}_{j0}, t) \nonumber \\
&& + \frac{1}{2} V({\bf r}_{i}, {\bf r}_{j}, {\bf p}_{i0}, {\bf p}_{j0}) + \frac{1}{2} \frac{Z_i Z_j e^2}{|{\bf r}_i - {\bf r}_j|}.
\label{eq:ep}
\end{eqnarray}

The expectation value of the potential energy $V_{ij}$ between nucleons i and j is given by
\begin{eqnarray}
&& \langle V_{ij}({\bf r}_{i0}, {\bf p}_{i0}, {\bf r}_{j0}, {\bf p}_{j0}, t) \rangle = \nonumber \\
&& = \int d^3r_i d^3r_j d^3p_i d^3p_j \,
V_{ij}({\bf r}_i, {\bf r}_j, {\bf p}_{i0}, {\bf p}_{j0}) \nonumber \\
&& \times f({\bf r}_i, {\bf p}_i, {\bf r}_{i0}, {\bf p}_{i0}, t) f({\bf r}_j, {\bf p}_j, {\bf r}_{j0}, {\bf p}_{j0}, t).
\label{Vpot}
\end{eqnarray}
The interaction density is defined as
\begin{eqnarray}
&& \rho_{\rm int}({\bf r}_{i0}, t) = \sum_{j \neq i} \int d^3r_i d^3r_j d^3p_i d^3p_j \, \delta({\bf r}_i - {\bf r}_j) \nonumber \\
&& \times f({\bf r}_i, {\bf p}_i, {\bf r}_{i0}, {\bf p}_{i0}, t) f({\bf r_j, p_j, r_{j0}, p_{j0}}, t).
\label{rhoint}
\end{eqnarray}

To extend PHQMD to relativistic energies, Lorentz contraction of the initial nuclei is incorporated through a modified single-particle Wigner density that accounts for contraction in the beam direction:
\begin{eqnarray}
&& \tilde f (\mathbf{r}_i, \mathbf{p}_i, \mathbf{r}_{i0}, \mathbf{p}_{i0}, t) = \label{fGam} \\
&& = \frac{1}{\pi^3}
e^{-\frac{2}{L} [ \mathbf{r}_{i}^T(t) - \mathbf{r}_{i0}^T (t) ]^2}
e^{-\frac{2\gamma_{cm}^2}{L} [ \mathbf{r}_{i}^L(t) - \mathbf{r}_{i0}^L (t) ]^2} \nonumber \\
&& \times e^{-\frac{L}{2} [ \mathbf{p}_{i}^T(t) - \mathbf{p}_{i0}^T (t) ]^2}
e^{-\frac{L}{2\gamma_{cm}^2} [ \mathbf{p}_{i}^L(t) - \mathbf{p}_{i0}^L (t) ]^2}, \nonumber
\end{eqnarray}

where $\gamma_{cm} = 1/\sqrt{1 - v_{cm}^2}$ and $v_{cm}$ is the velocity in the computational frame. The interaction density accordingly modifies to
\begin{eqnarray}
&& \tilde \rho_{\rm int} (\mathbf{r}_{i0}, t) \to C \sum_j \left( \frac{1}{\pi L} \right)^{3/2} \gamma_{cm} \nonumber \\
&& \times e^{-\frac{1}{L} [ \mathbf{r}_{i0}^T(t) - \mathbf{r}_{j0}^T (t) ]^2}
e^{-\frac{\gamma_{cm}^2}{L} [ \mathbf{r}_{i0}^L(t) - \mathbf{r}_{j0}^L (t) ]^2}. \label{densGam}
\end{eqnarray}
Here $C$ is a correction factor needed to compensate for the
lower density in the QMD type approaches compared to the mean-field approaches,
which is adjusted numerically to achieve equality of the two densities - interaction density in QMD and  mean-field density (cf. Ref. \cite{Aichelin:2019tnk}).
For the energies considered here, these relativistic corrections are not significant.

\subsection{Modeling the EoS within PHQMD}

The potential experienced by nucleons traveling through nuclear matter is momentum dependent, as established from elastic pA scattering data \cite{Clark:2006rj,Cooper:1993nx}. These data are typically analyzed by comparing with solutions of the Dirac equation incorporating scalar ($U_s$) and vector ($U_0$) potentials. 

To derive a nucleon-nucleon potential suitable for non-relati\-vistic QMD calculations, we first compute the Schrödinger equivalent potential (optical potential) $U_{opt}$ as described in Ref. \cite{Jaminon:1980vk}:
\begin{equation}
U_{opt}(r, \epsilon) = U_s(r) + U_0(r) + \frac{1}{2m_N} (U_s^2(r) - U_0^2(r)) + \frac{U_0(r)}{m} \epsilon,
\label{uschr}
\end{equation}
where $\epsilon$ is the kinetic energy of the incoming proton in the target rest frame. The imaginary part of the optical potential is neglected since collisions are treated explicitly.

This potential allows rewriting the upper components of the Dirac equation
\begin{equation}
\left( -i\boldsymbol{\alpha} \boldsymbol{\nabla} + \beta m + \beta U_s(r) + U_0(r) \right) \phi = (\epsilon + m) \phi
\end{equation}
in the form of a Schrödinger equation:
\begin{equation}
\left( \frac{-\nabla^2}{2m} + U_{opt}(r, \epsilon) \right) \psi = \frac{2m\epsilon + \epsilon^2}{2m} \psi.
\end{equation}

\begin{figure}[h!]   
\centering
    \includegraphics[scale=0.5]{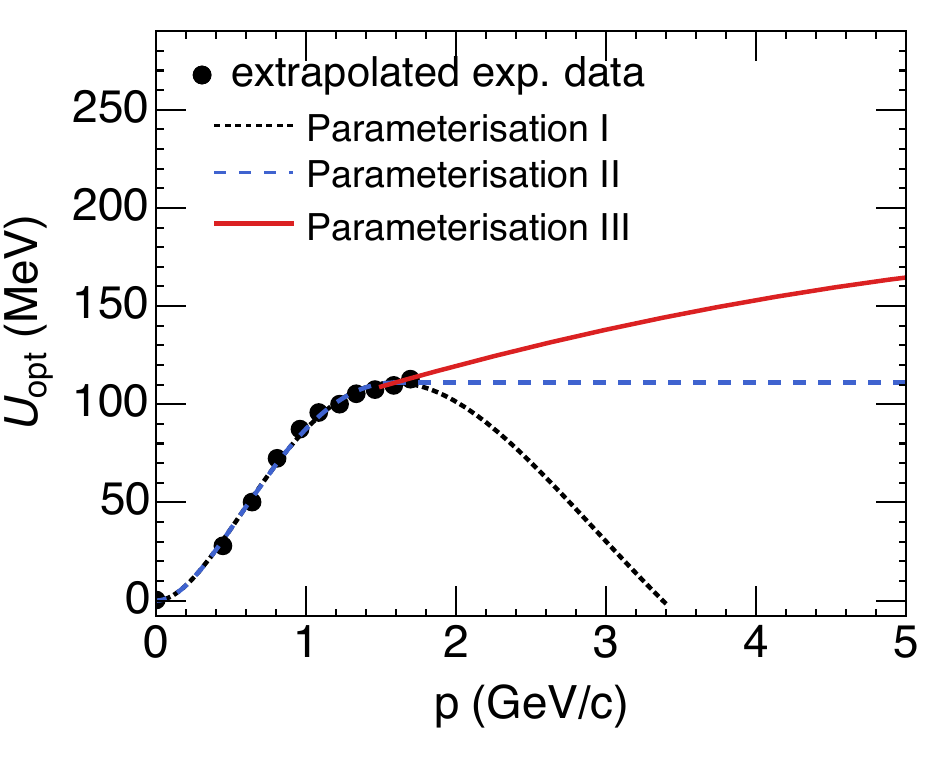}        
   \caption{Schrödinger equivalent optical potential $U_{opt}$ versus total momentum $p$ of the proton extracted from pA collisions \cite{Hama:1990vr, Clark:2006rj,Cooper:1993nx}.
    The figure is adopted from Ref. \cite{Zhou:2025zgn}.}
    \label{fig:uopt}
\end{figure}

Analysis of pA data at different beam energies shows that $U_s$ and $U_0$ depend on $\epsilon$, resulting in a complex $\epsilon$ dependence of $U_{opt}$. Our fit to these experimental data, normalized to $U_{opt}(p=0) = 0$, is shown in Fig. \ref{fig:uopt} alongside $U_{opt}$ extracted from pA scattering data via Dirac equation analysis \cite{Hama:1990vr}. Beyond proton kinetic energies of 1.04 GeV, no data are available, introducing uncertainties for heavy-ion collisions with $\sqrt{s_{\text{NN}}} > 2.32$ GeV.

To study whether different parametrizations affect predictions, we perform calculations with three parameterizations of the momentum dependent potential  -- Fig. \ref{fig:uopt}:
\begin{itemize}
\item[I:] $V({\bf p, p}_1) = (a (\Delta p)^2 + b (\Delta p)^4) \exp[-c \Delta p]$ with $\Delta p = \sqrt{({\bf p} - {\bf p}_1)^2}$
\item[II:] Same as I for $\Delta p \leq 2$ GeV/$c$, but constant for $\Delta p > 2$ GeV/$c$
\item[III:] For $\Delta p < 1.7$ GeV/$c$ same as I, for $\Delta p \geq 1.7$ GeV/$c$: $V(\Delta p) = d + e \Delta p + f (\Delta p)^2$
\end{itemize}
We note in advance that at SIS energies one probes the  $U_{opt} (p)$ for $p<1$ GeV$/c$ while the influence of extrapolation of $U_{opt} (p)$ to higher momentum $p$ is showing up with increasing bombarding energies. 
Parameters are given in Table~\ref{table_eos}.

\begin{figure}[t]
    \centering
    \resizebox{0.45\textwidth}{!}{
        \includegraphics{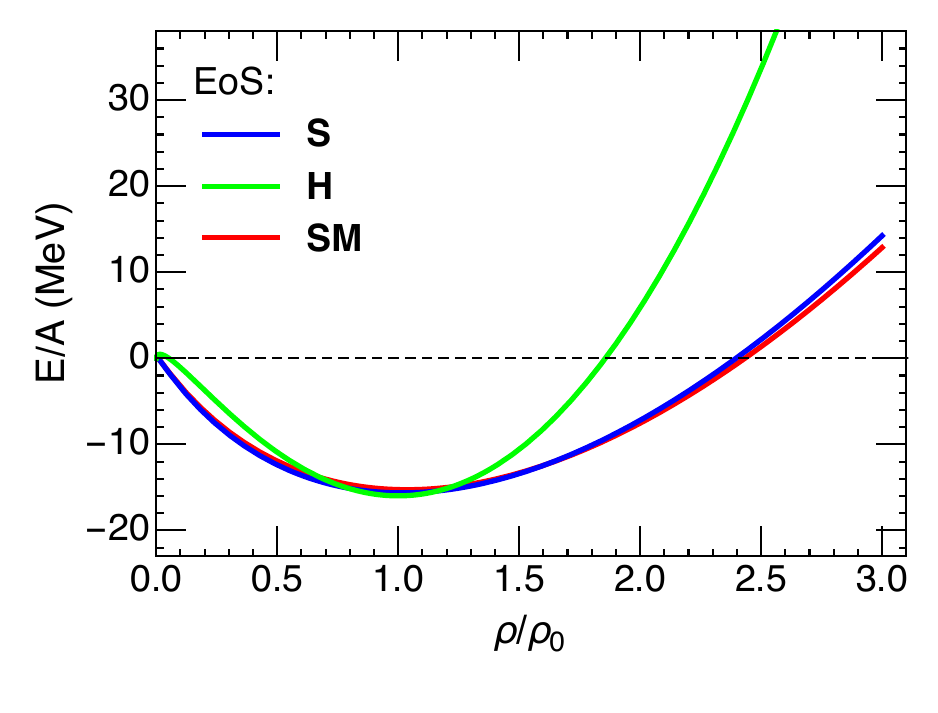}     }
    \caption{Equation-of-state for $T=0$ for the hard (green line), soft (blue line) and the soft momentum dependent potential (red line). 
     The figure is adopted from Ref. \cite{Zhou:2025zgn}.} 
    \label{fig:eos}
\end{figure}

\begin{table}[h]
    \centering
      \begin{tabular}{c c c c c}
      \toprule
      EoS & $\alpha$ [GeV] & $\beta$ [GeV] & $\gamma$ & K [MeV]\\
      \midrule
      S  &  -0.3835 &  0.3295 & 1.15 & 200 \\
      H  &  -0.1253 &  0.071 &  2.0 & 380 \\
      SM & -0.478 & 0.4137 &  1.1 & 200 \\
      \cmidrule{2-4}
       ~ & $a$[GeV$^{-1}$]  & $b$ [GeV$^{-3}$] & $c$  [GeV$^{-1}$]   & ~\\
       ~ & 236.326 & -20.730 &  0.901 & ~ \\
        \cmidrule{2-4}
       ~ & $d$[GeV]  & $e$ & $f$  [GeV$^{-1}$]   & ~\\
       ~ & 72.237 & 27.085 &  -1.722 & ~ \\
      \bottomrule
      \end{tabular}
    \caption{Parameters of the potential used in PHQMD assuming that the momenta
     are given in GeV.} 
    \label{table_eos}
    \end{table}

The Schrödinger equivalent potential is obtained by averaging the two-body potential over the Fermi distribution:
\begin{equation}
U_{opt}(p) = \frac{\int^{p_F} V({\bf p}, {\bf p}_1) d^3p_1}{\frac{4}{3}\pi p_F^3}.
\label{UoptInt}
\end{equation}

In the Dirac analysis, vector and scalar mean-field potentials depend approximately linearly on baryon density. This linear dependence is reproduced by assuming in Eq. (\ref{eq:ep}):
\begin{equation}
V({\bf r}_1, {\bf r}_2, {\bf p}_{10}, {\bf p}_{20}) = V({\bf p}_{10}, {\bf p}_{20}) \delta({\bf r}_1 - {\bf r}_2),
\end{equation}
since the $\delta$ function creates a linear density dependence when averaged over wave functions. The total energy of the system is
\begin{eqnarray}
E &=& \langle \psi(t) | (T + V) | \psi(t) \rangle \nonumber \\
&=& \sum_i \left[ \langle i | \frac{p^2}{2m} | i \rangle + \sum_{i \neq j} \langle ij | V_{ij} | ij \rangle \right] \nonumber \\
&=& \int H(r) d^3r, \label{eq:energy}
\end{eqnarray}
where $|\psi(t)\rangle = \prod_i |\psi_i\rangle$ is the N-body wave function taken as the direct product of single-particle wave functions.

Momentum dependent potentials were introduced in QMD-type approaches in Ref. \cite{Aichelin:1987ti} and further explored in Refs. \cite{Aichelin:1991xy,LeFevre:2016vpp,Hillmann:2019wlt,Nara:2020ztb}, and in BUU-type approaches in Ref. \cite{Gale:1987zz}, with various applications \cite{Ko:1987gp,Weber:1993et,Sahu:1998vz,Danielewicz:1998vz,Sahu:1998vz,Sahu:1999mq,Sahu:2002ku,Cassing:2000bj,Buss:2011mx,Mohs:2020awg} - see discussions and examples in Section 3.

\subsection{Relation of the Potential to the EoS of Nuclear Matter}

In infinite nuclear matter, where momentum and position are uncorrelated, the equation-of-state of cold nuclear matter can be calculated from the potential. The static part of the contribution of the QMD potential to the EOS is given by \cite{Aichelin:2019tnk}:
\begin{equation}
V_{\rm Skyrme\,stat} = \alpha \frac{\rho}{\rho_0} + \beta \left( \frac{\rho}{\rho_0} \right)^\gamma.
\end{equation}
To this, the momentum dependent part for cold nuclear matter is added:
\begin{equation}
V_{\rm mom}(p_F) = \frac{\int^{p_F} \int^{p_F} d^3p_1 d^3p_2 \, V({\bf p}_2 - {\bf p}_1)}{(\frac{4}{3}\pi p_F^3)^2} \frac{\rho}{\rho_0}.
\end{equation}
Since the Fermi momentum $p_F$ is a function of density, the total strong interaction potential becomes:
\begin{equation}
V_{\rm Skyrme}(\rho) = V_{\rm Skyrme\,stat}(\rho) + V_{\rm mom}(\rho).
\end{equation}
The energy per nucleon is obtained by introducing \\ 
$U = \int V(\rho) d\rho$:
\begin{equation}
\frac{E}{A}(\rho) = \frac{3}{5} E_{\rm Fermi}(\rho) + \frac{U}{\rho}.
\end{equation}

The potential contains three parameters $\alpha, \beta, \gamma$ determined by requiring $E/A(\rho_0) = -16$ MeV at normal nuclear density. The compressibility modulus $K$ remains a free parameter. The hard ($K = 380$ MeV), soft ($K = 200$ MeV), and soft momentum dependent ($K = 200$ MeV) equations of state are illustrated in Fig. \ref{fig:eos}, with parameters given in Table \ref{table_eos}. 

We stress that for cold nuclear matter, the soft and soft momentum dependent EoS have identical $E/A(\rho)$ by construction. Similar would hold for hard and hard momentum dependent EoS. As the beam energy increases, the momentum dependence of the potential becomes more important.

\subsection{Cluster Production in PHQMD}

Clusters can be identified in PHQMD using three different algorithms:

\begin{enumerate}

\item \textbf{Potential mechanism:} The attractive potential between ba\-ryons with small relative momentum can form bound nucleon groups. Using the MST clusterization algorithm: nucleons $i$ and $j$ are considered bound if they satisfy
\begin{equation}
| \mathbf{r}_i^* - \mathbf{r}_j^* | < r_{\rm clus},
\label{eq:MSTcond}
\end{equation}
where positions are boosted to the pair center-of-mass frame and $r_{\rm clus} = 4$ fm corresponds approximately to the range of the attractive NN potential. Additionally, clusters must have negative binding energy $E_B < 0$. MST serves as a cluster recognition tool rather than a formation mechanism, since QMD propagates baryons, not pre-formed clusters.

\item \textbf{Kinetic mechanism:} Deuterons are created through catalytic hadronic reactions $\pi NN \leftrightarrow \pi d$ and $NNN \leftrightarrow N d$ in different isospin channels. Quantum nature is considered via excluded volume and projection onto the deuteron wave function in momentum space, reducing production particularly at target/projectile rapidities. Details are in Ref. \cite{Coci:2023daq}.

\item \textbf{Coalescence mechanism:} A proton and neutron form a deuteron if their phase-space distance at freeze-out satisfies $|r_1 - r_2| \leq 3.575$ fm and $|p_1 - p_2| \leq 285$ MeV/$c$. Details are in Ref. \cite{Kireyeu:2022qmv}. This method is used in the PHQMD for model studies and comparison purpose only.
\end{enumerate}

In the calculations presented here, we employ a combination of methods (2) and (3) as detailed in Ref. \cite{Coci:2023daq}.


\section{Model results}
\label{Models}

 In this section, we compare model calculations for light clusters with experimental observables characterizing momentum-space anisotropy in heavy-ion collisions. 
 This anisotropy manifests in the azimuthal distribution of final-state particles, which can be decomposed into a Fourier series:
\begin{equation} 
\label{eq:azim}
\frac{dN}{d\phi} \propto 1 + 2v_1 \cos(\phi - \Psi_R) + 2v_2 \cos(2(\phi - \Psi_R)) + \cdots
\end{equation}
Here, $\phi$ represents the azimuthal angle of a particle measured relative to the event plane (or reaction plane) $\Psi_R$. The flow coefficients $v_n$ ($n = 1, 2, \ldots$) are defined as event-averaged moments over all particles within a given centrality class \cite{Ollitrault:1997di,Poskanzer:1998yz}:
\begin{equation} 
\label{eq:vn}
v_n = \langle \cos(n(\phi - \Psi_R)) \rangle.
\end{equation}
The directed flow $v_1$ and elliptic flow $v_2$ provide sensitive probes of the early-stage dynamics and the equation-of-state of nuclear matter.

We note that we calculate the flow coefficients using the theoretically defined reaction plane ($\Psi_{R}=0$) and compute the directed and elliptic flow as  
\begin{equation}
 \begin{aligned}
  v_1 = \left<\frac{p_x}{p_T}\right> ,  \  \  v_2 = \left<\frac{p_x^2 - p_y^2}{p_T^2} \right>,
 \end{aligned}
\end{equation}
where $p_T$ is the transverse momentum  $p_T=(p_x^2 + p_y^2)^{1/2}$ of the hadron with 4-momentum $p=(E,p_x,p_y,p_z)$.

\subsection{The PHQMD results }

Here we present the highlight of the PHQMD results from Refs. \cite{Kireyeu:2024hjo,Zhou:2025zgn} for the  observables at collision energy of $\sqrt{s_{\text{NN}}}=$1.2 - 3 GeV in order to constrain the momentum dependent interactions and cluster production mechanisms implemented in PHQMD.  

We stress that the EoS significantly influences the nuclear dynamics and affects the rapidity distributions and $p_T$ spectra of baryons, clusters as well as produced mesons. This effect is  especially visible at low energies. For a soft static EoS stopping is less compared to a hard EoS and we find a larger abundances of light clusters at midrapidity. On the other hand, the soft momentum dependent potential acts in the same direction as the hard static EoS for the rapidity distributions. Similar holds for the $p_T$ spectra of hadrons - see examples in Section III of Ref. \cite {Kireyeu:2024hjo} for $\sqrt{s_{\text{NN}}}=1.5$ GeV and in Section V in Ref. \cite{Zhou:2025zgn} for $\sqrt{s_{\text{NN}}}=3$ GeV. 
The flow harmonics provide more distinguished information since they are very sensitive to the pressure building in the system and show a different behavior, which is easy to identify by comparison of the theoretical models with the experimental data.

\subsubsection{PHQMD results for $v_1, v_2$ at SIS energies}

We start by presenting the PHQMD results for the directed flow $v_1$ and elliptic flow $v_2$ of protons and light clusters at SIS energies, obtained with three different equations-of-state: soft static (S), hard static (H), and soft momentum dependent (SM). These results are adopted from Ref. \cite{Kireyeu:2024hjo}, where the PHQMD calculations were confronted with experimental data from the HADES \cite{HADES:2020lob,HADES:2022osk} and FOPI \cite{FOPI:2011aa} collaborations. 
Here we show explicitly the comparison to HADES data stressing that the comparison to FOPI data leads to qualitative similar conclusions.
Also we note that we used parametrization I (cf. Fig. \ref{fig:uopt}) for the $U_{opt}(p)$.

\begin{figure*}[h!]
\centering
\includegraphics[scale=0.29]{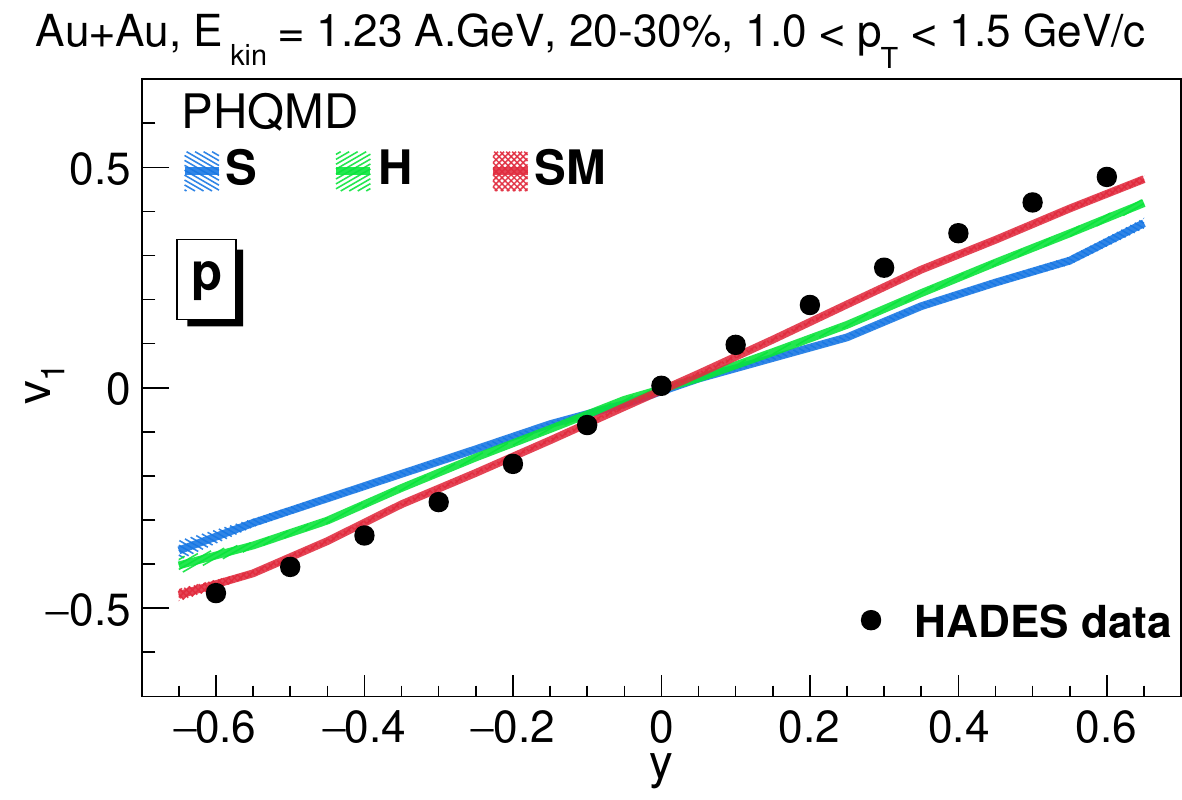}
\includegraphics[scale=0.29]{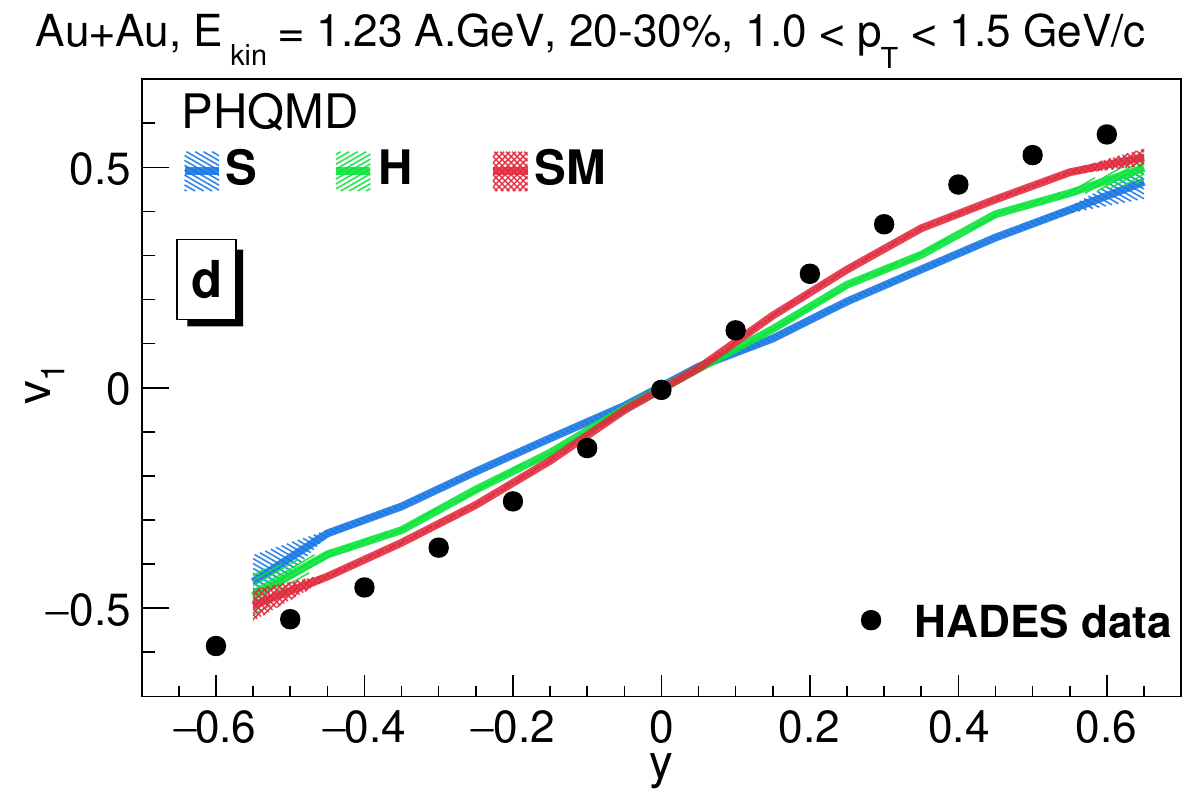}
\includegraphics[scale=0.29]{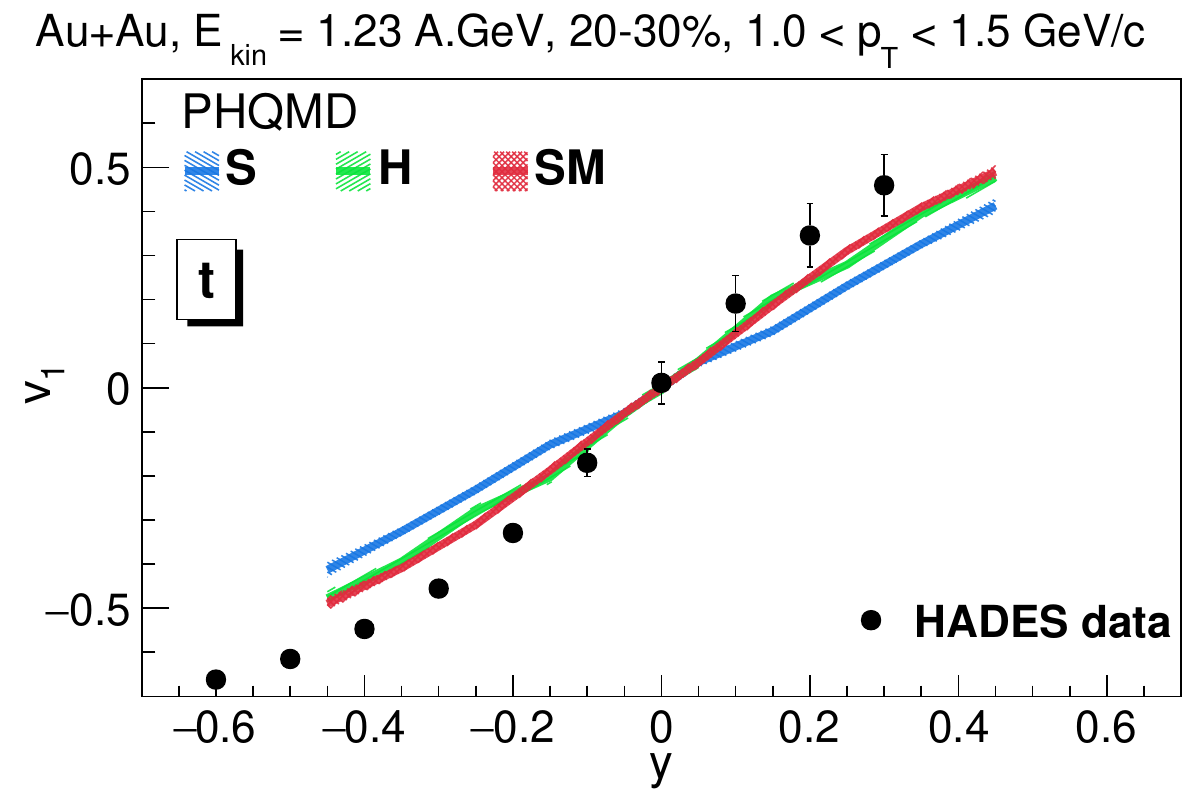} 
    \caption{$v_1$ of protons (left), deuterons (middle) and tritons (right) as a function of rapidity for 20-30\% central Au+Au collisions at $E_{kin}=1.23$ A GeV for  $1.0 < p_T < 1.5$ GeV/c.
   The blue lines "S" correspond to the PHQMD calculations with the "soft" EoS, the green lines "H" show the "hard" EoS, the red lines the "SM" represent the momentum dependent "soft" EoS. 
    The HADES experimental data are taken from Ref. \cite{HADES:2020lob}.
     The figure is adopted from Ref.\cite {Kireyeu:2024hjo}.   }
    \label{fig:hadesv1y}    
\end{figure*}
\begin{figure*}[h!]
\includegraphics[scale=0.29]{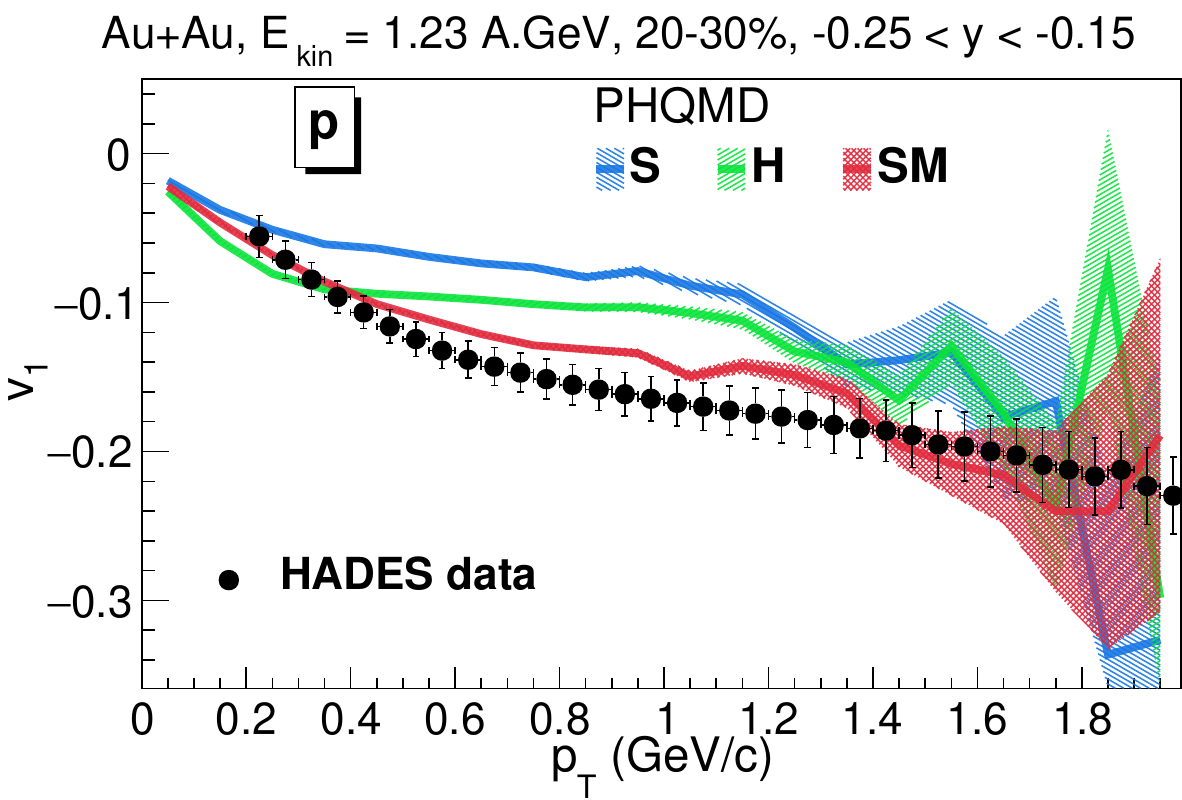}
\includegraphics[scale=0.29]{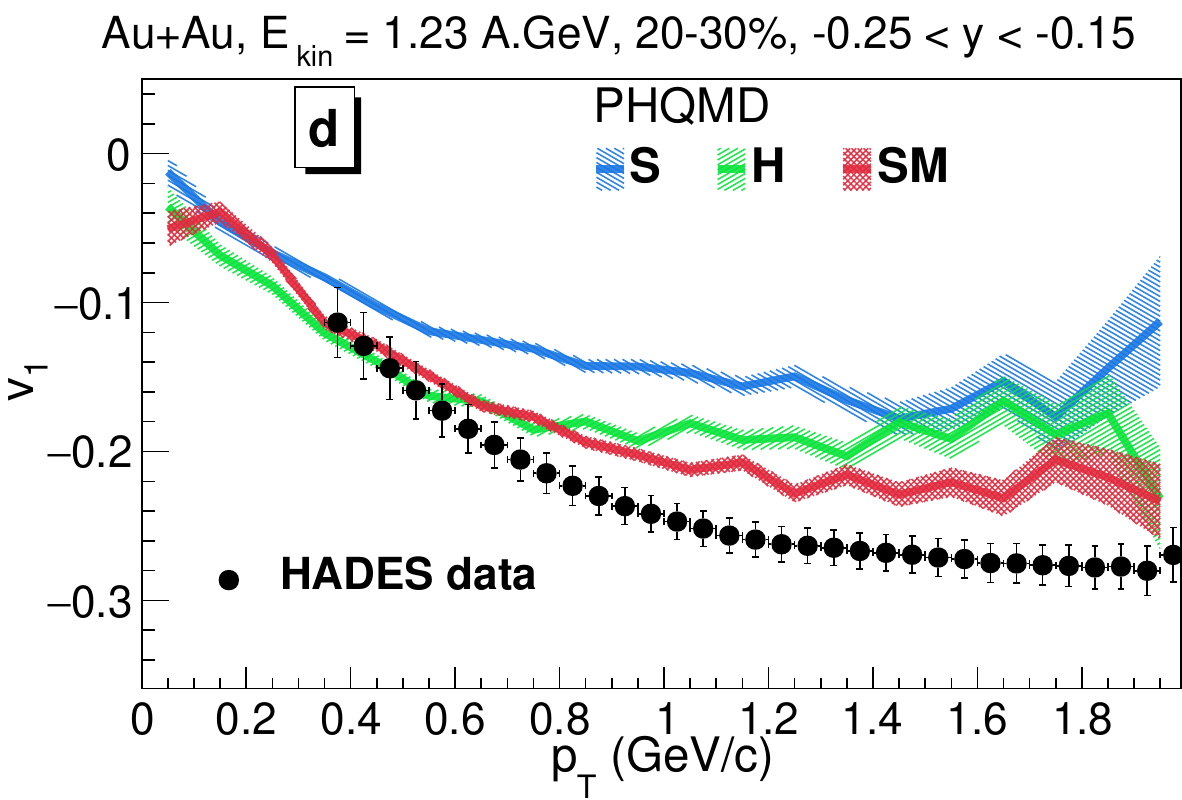}
\includegraphics[scale=0.29]{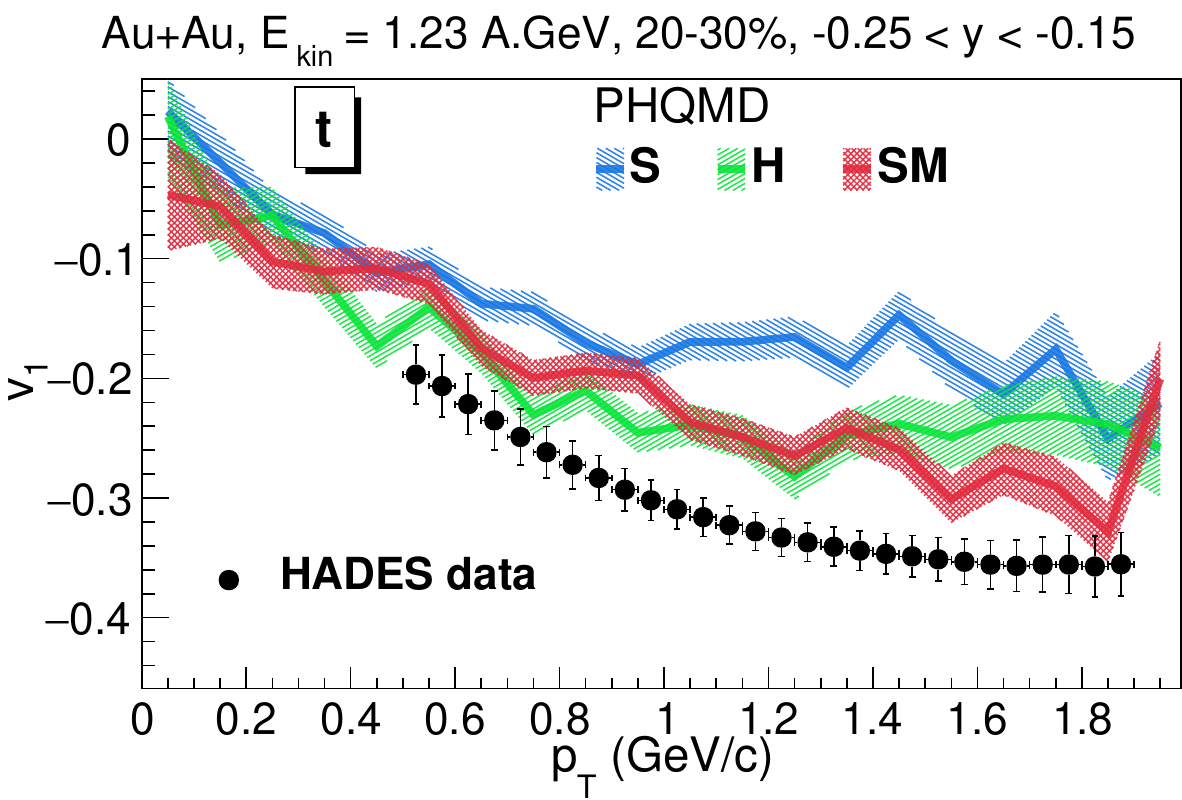}
    \caption{$v_1$ of protons (left), deuterons (middle) and tritons (right) as a function of $p_T$ for 20-30\% central Au+Au collisions at $E_{kin}=1.2$ A GeV  in the rapidity bin $-0.25<y<-0.15$. 
    The colour code is the same as in Fig. \ref{fig:hadesv1y}. 
    The experimental data are taken from Ref. \cite{HADES:2022osk}.
         The figure is adopted from Ref.\cite {Kireyeu:2024hjo}.}
    \label{pdtv1pt}   
\end{figure*}
\begin{figure*}[h!]
    \centering
\includegraphics[scale=0.29]{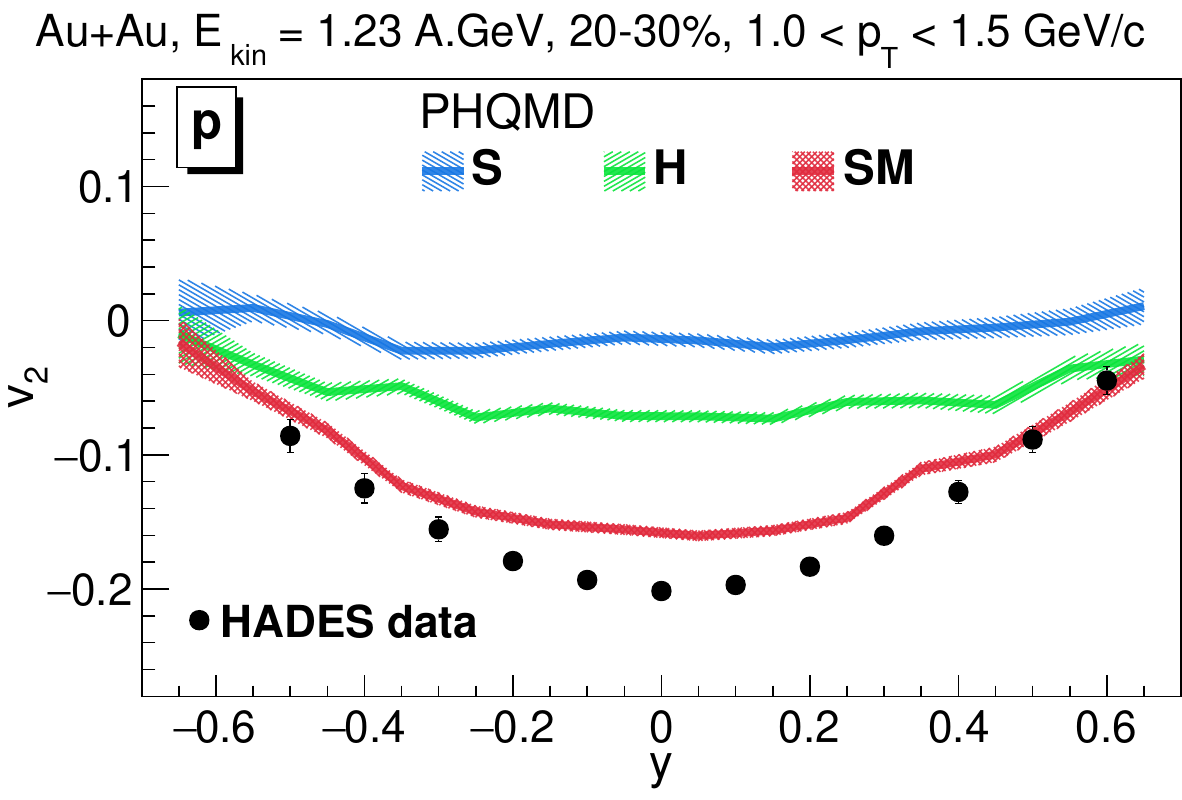}
\includegraphics[scale=0.29]{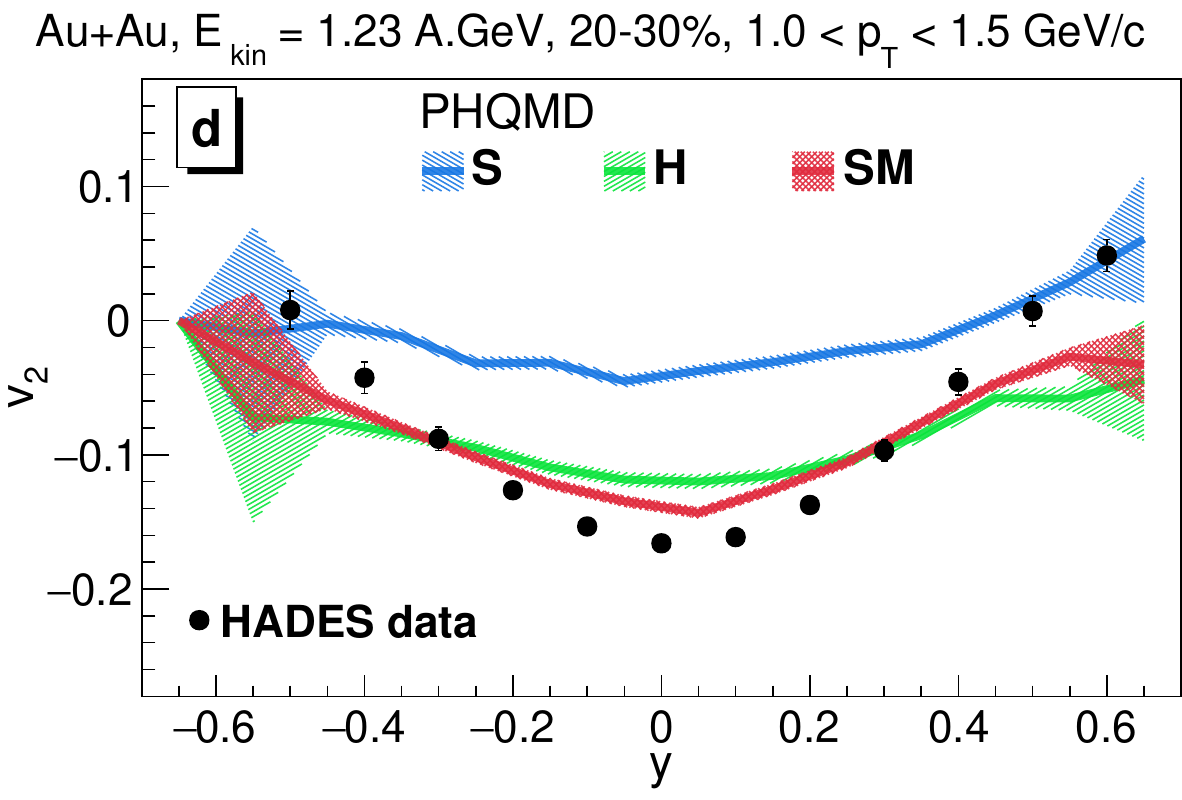}
\includegraphics[scale=0.29]{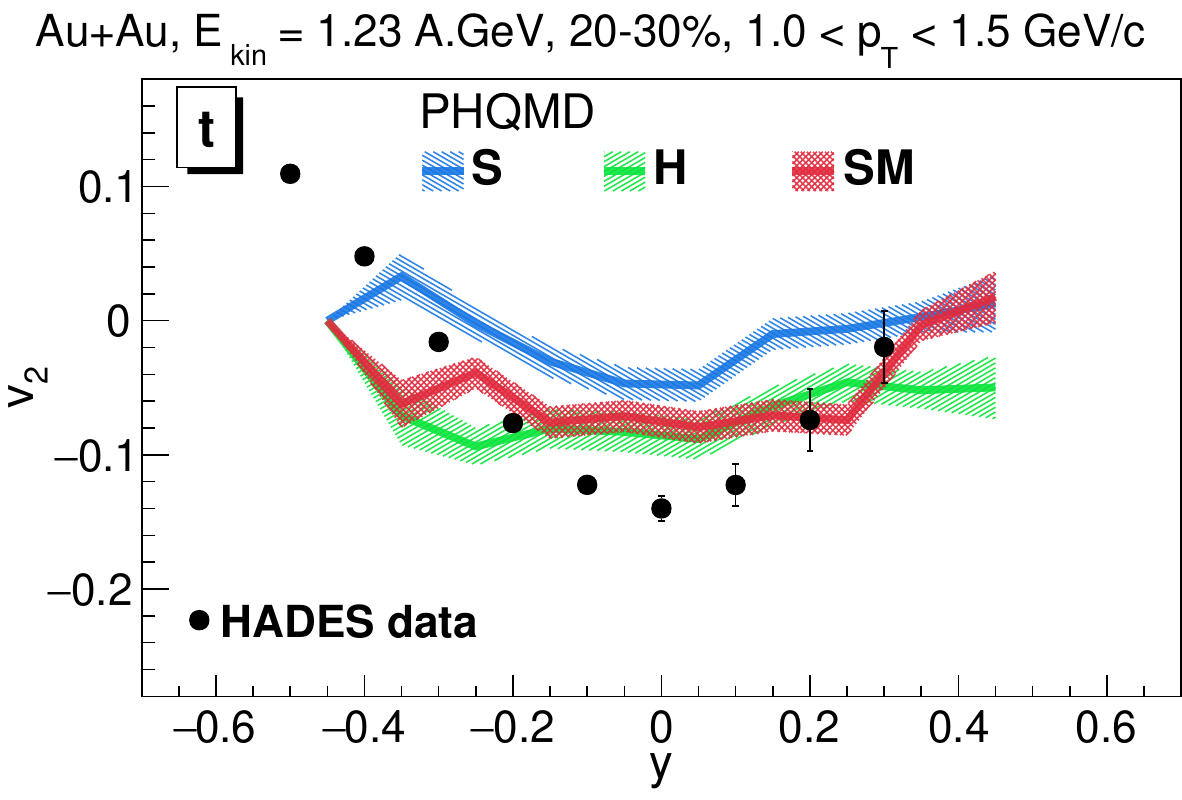}
   \caption{$v_2$ of protons (left), deuterons (middle) and tritons (right) as a function of rapidity for 20-30\% central Au+Au collisions at $E_{kin}=1.23$ A GeV for  $1.0 < p_T < 1.5$ GeV/c. The colour code is the same as in Fig. \ref{fig:hadesv1y}.
   The HADES experimental data are taken from Ref. \cite{HADES:2020lob}.
   The figure is adopted from Ref.\cite {Kireyeu:2024hjo}.    }
    \label{fig:HADESv2yclast}      
\end{figure*}

\begin{figure*}[h!]
\includegraphics[scale=0.29]{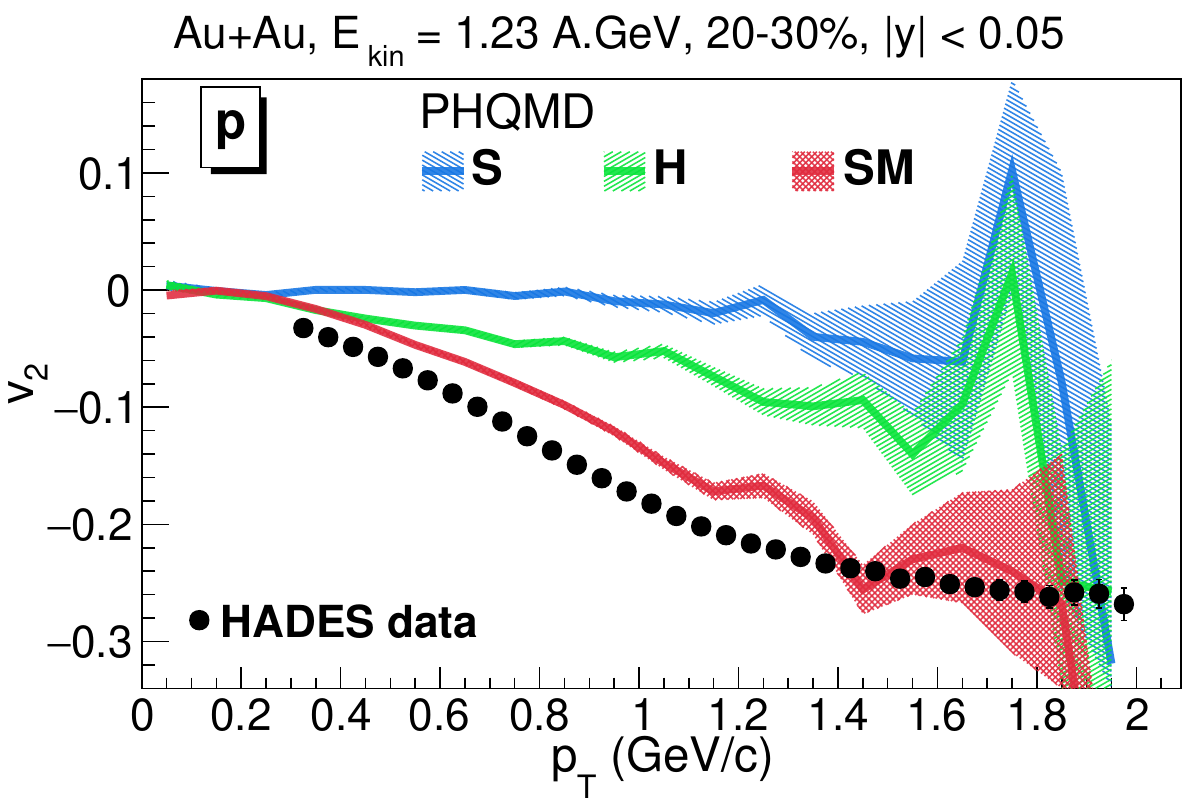}
\includegraphics[scale=0.29]{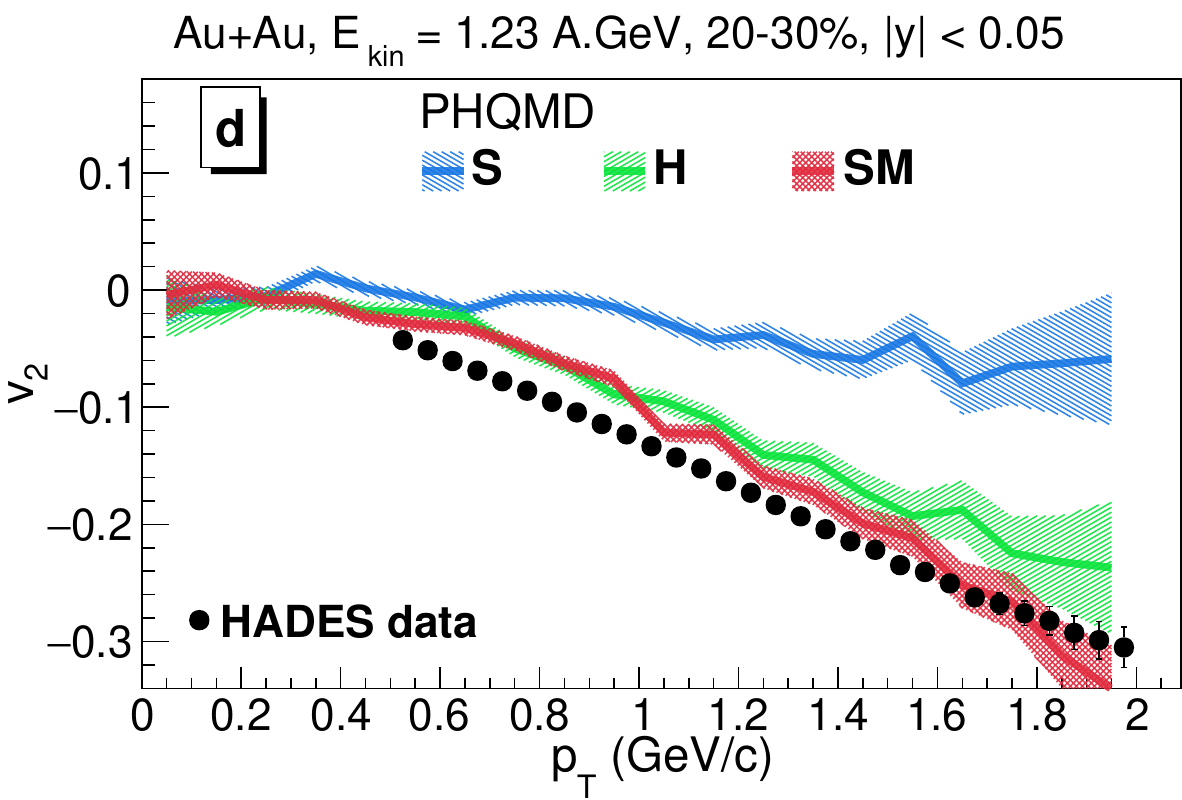}
\includegraphics[scale=0.29]{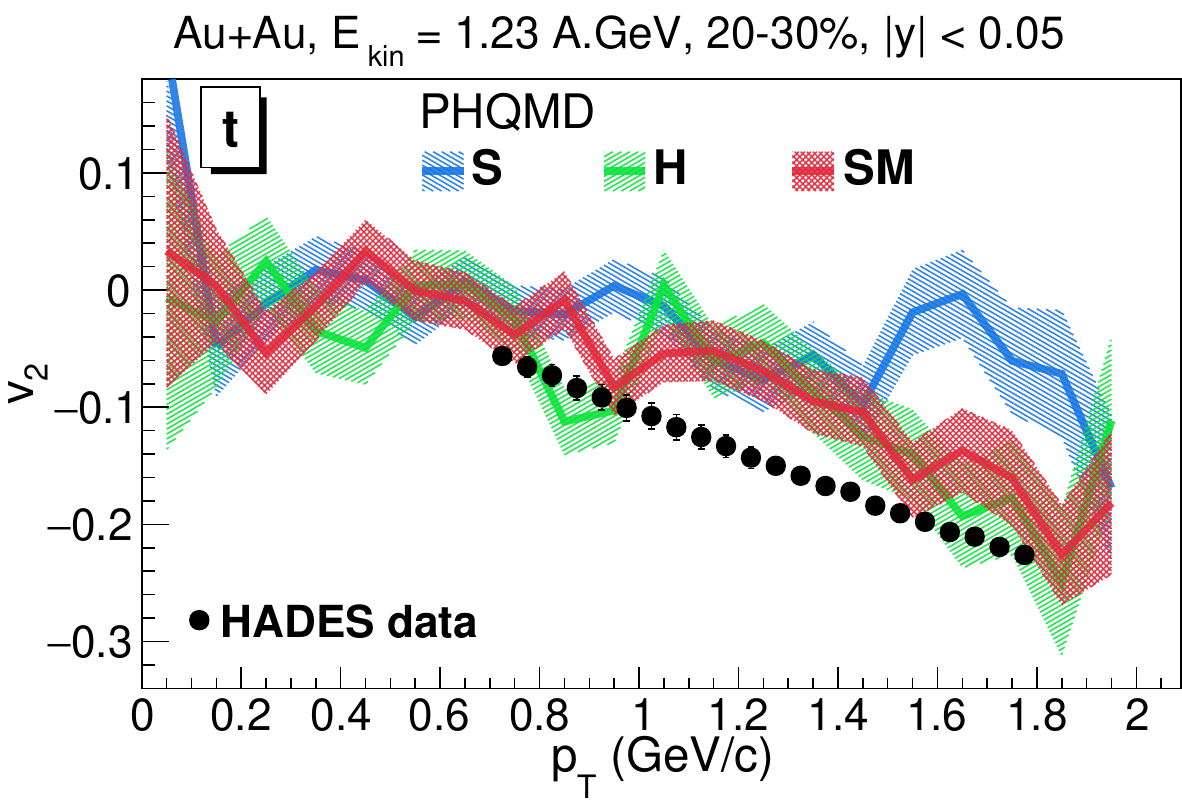}  
\caption{ $v_2$ of protons (left), deuterons (middle) and triton (right) as a function of $p_T$ for rapidity intervals $|y|<0.05$  for 20-30\% central Au+Au collisions at $E_{kin}$=1.23 A GeV. The colour code is the same as in Fig. \ref{fig:hadesv1y}. The HADES experimental data are taken from Ref. \cite{HADES:2020lob,HADES:2022osk}.
 The figure is adopted from Ref.\cite {Kireyeu:2024hjo}.}
\label{fig:v2pt_HADES} 
\end{figure*}

Figure \ref{fig:hadesv1y} presents PHQMD results for the directed flow $v_1(y)$ of protons, deuterons, and tritons from Au+Au collisions at $E_{\text{kin}}=1.23$ A GeV, compared to HADES data \cite{HADES:2020lob}. Calculations for hard, soft, and momentum dependent (MD) equations-of-state are shown for the $p_T$ interval $1.0<p_T<1.5$ GeV/c.
The experimental data confirm the predicted increase in the $v_1(y)$ slope with cluster mass, suggesting deuteron formation near the overlap region border where the nucleonic $v_1$ is strongest. This aligns with earlier findings \cite{Aichelin:1987ti} and recent calculations \cite{Hillmann:2019wlt,Mohs:2020awg}.
For clusters, a soft momentum dependent EoS yields a steeper slope of $v_1(y)$ than a hard EoS, bringing the results closer to the data. The triton slope remains higher than that of deuterons, and even the soft MD interaction underpredicts it. The calculations reproduce the slight non-linearity of $v_1(y)$ at large rapidities.

Figure \ref{pdtv1pt} presents the transverse momentum ($p_T$) dependence of the directed flow $v_1$ for protons (left), deuterons (middle), and tritons (right) in 20-30\% central Au+Au collisions at $E_{\text{kin}}=1.2$ A GeV within the rapidity interval $-0.25<y<-0.15$, calculated for the three equations of state.
The soft momentum dependent (SM) EoS provides the best overall description of the HADES data for both protons and light clusters. For deuterons and tritons, the $v_1(p_T)$ values for the hard and SM EoS are closer to each other than in the proton case, yet both remain significantly above the results from the soft EoS, which substantially underestimates the data.

Figure \ref{fig:HADESv2yclast} presents the elliptic flow $v_2(y)$ for protons, deu\-terons, and tritons from 20-30\% central Au+Au collisions at $E_{\text{kin}}=1.23$ A GeV and $1.0 < p_T < 1.5$ GeV/c, compared to HADES data \cite{HADES:2020lob}. The soft (S) EoS substantially underestimates $v_2$ for all particles. While the hard (H) EoS also underestimates the proton $v_2$, the soft momentum dependent (SM) EoS provides a good description of the data, agreeing within 10\% for protons, deuterons, and tritons. The $v_2(y)$ values for deuterons and tritons are similar for the SM and hard EoS.
The bottom right panel shows that for the SM EoS, the $v_2(y)$ of deuterons produced kinetically or via the MST mechanism are nearly identical, consistent with the behavior observed for $v_1(y)$ in Fig. \ref{fig:hadesv1y}.

\begin{figure*}[h!]
    \centering
        \includegraphics[scale=0.3]{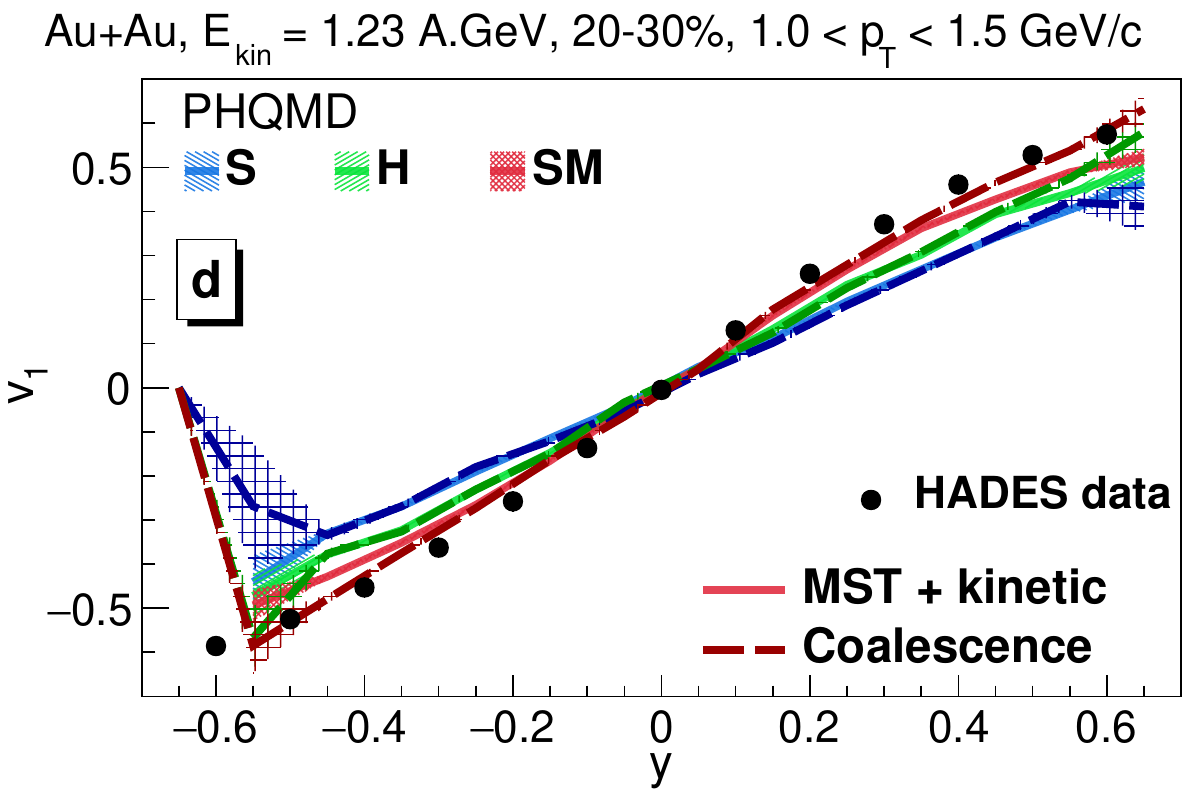} \hspace*{5mm}
         \includegraphics[scale=0.3]{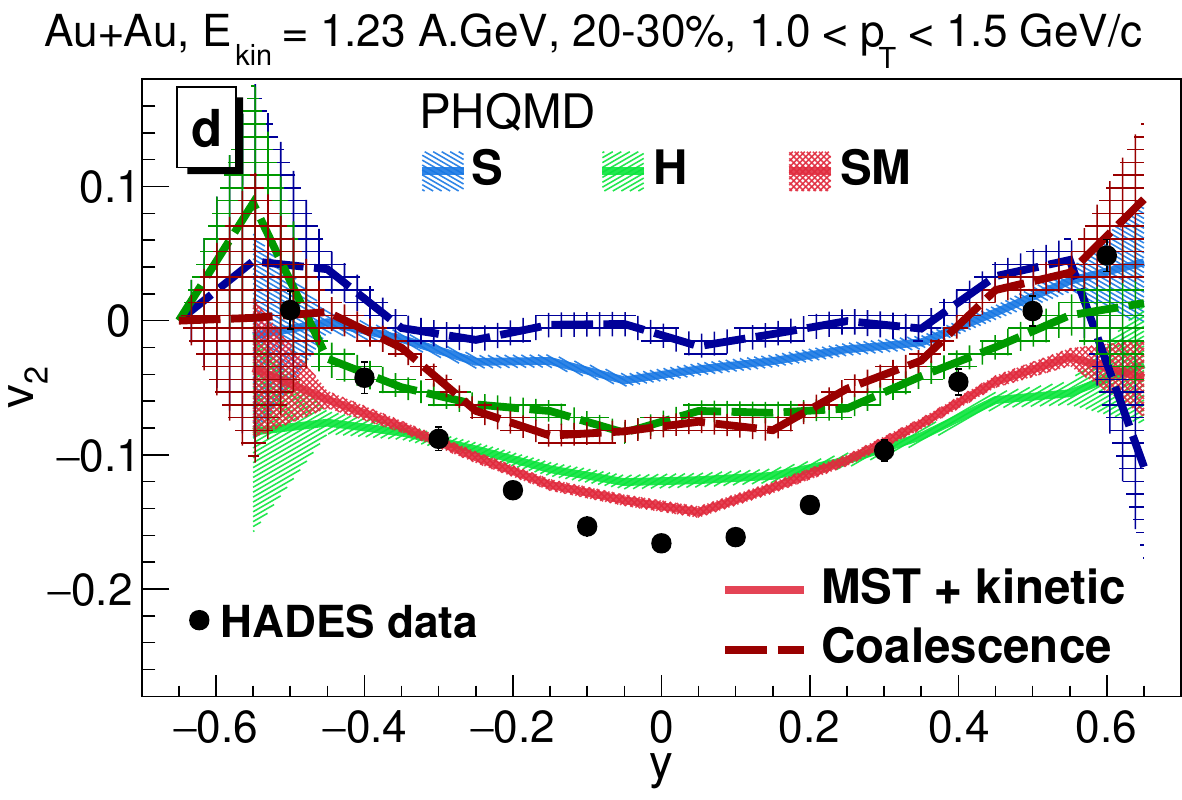} \\
        \includegraphics[scale=0.3]{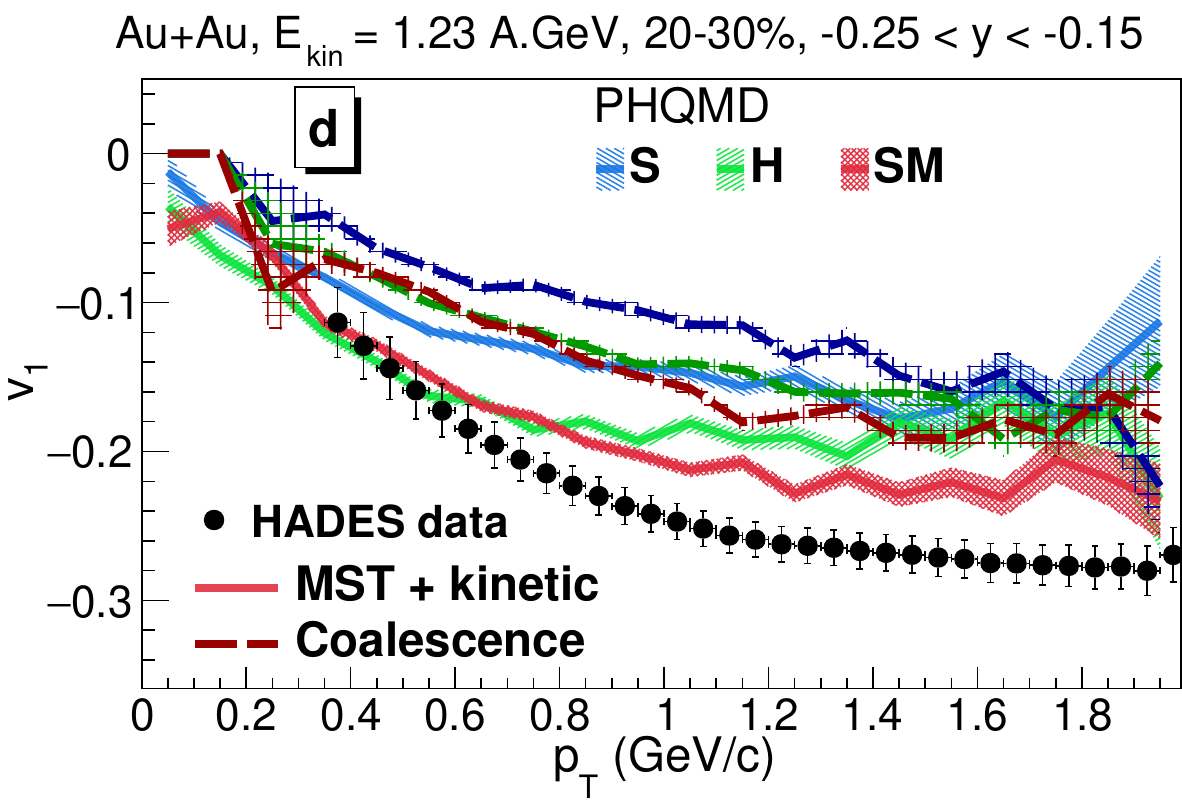}  \hspace*{5mm}
        \includegraphics[scale=0.3]{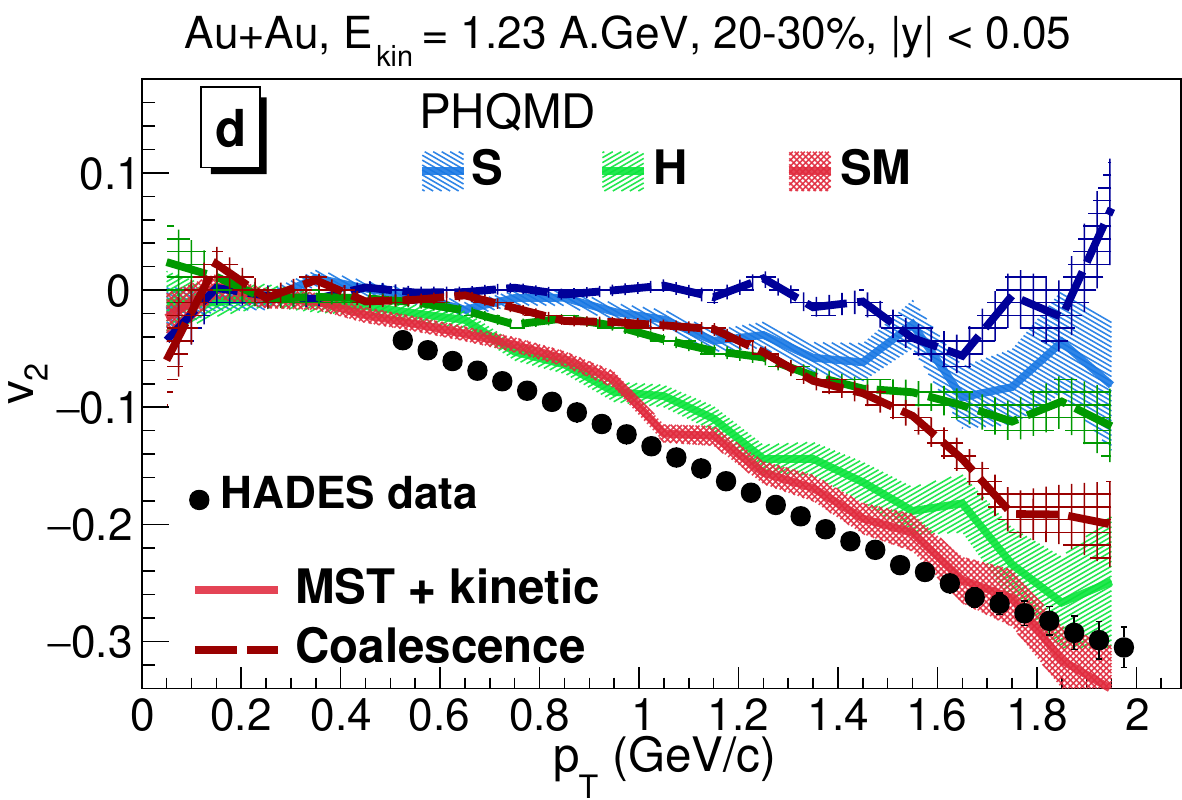}         
    \caption{
    Upper plot: Comparison of the $v_1(y)$ (left) and  $v_2(y) (right)$  of deuterons produced by kinetic + MST mechanisms (solid lines) with the coalescence mechanism (dashed lines)  for 20-30\% central Au+Au collisions at $E_{kin}=1.23$ A GeV for  $1.0 < p_T < 1.5$ GeV/c. 
    Lower plot: The comparison of $v_1(p_T)$  for $-0.25 < y < -0.15$ (left) and  $v_2(p_T)$ for $|y| < 0.05$ (right) of deuterons produced by kinetic + MST mechanisms (solid lines) with the  coalescence mechanism (dashed lines) for 20-30\% central Au+Au collisions.
   The blue lines "S" correspond to the PHQMD calculations with the "soft" EoS, the green lines "H" show the "hard" EoS, the red lines "SM" represent the momentum dependent "soft" EoS. 
   The color coding for the coalescence results is the same but in dark colors.
    The HADES experimental data are taken from Ref. \cite{HADES:2020lob}.
     The figure is adopted from Ref.\cite {Kireyeu:2024hjo}.    }
    \label{fig:coalescence_v1ypt}
\end{figure*}

We note that the PHQMD  results agree with UrQMD calculations using a static hard EoS \cite{Hillmann:2019wlt}, which leads to similar flow observabels as using a momentum dependent soft EoS \cite{Aichelin:1987ti}. However, they are in tension with SMASH results \cite{Mohs:2020awg,Mohs:2024gyc,Tarasovicova:2024isp}, where a static hard EoS underestimate a negative $v_2$. 
This discrepancy may arise from differences in the implementation of the momentum dependent potential in SMASH.
Moreover, we stress that  SMASH and UrQMD cluster production is based coalescence mechanism for deuteron production while the PHQMD clusters are produced by the MST and kinetic mechanisms.

Figure \ref{fig:v2pt_HADES} presents the transverse momentum distribution of elliptic flow $v_2$ for protons (left), deuterons (middle), and tritons (right) in 20–30\% central Au+Au collisions at $E_{\text{kin}}=1.23$ A GeV, for the midrapidity interval $|y|<0.05$. 
A strong $p_T$ dependence and significant sensitivity to the equation-of-state  are observed for all particle species. The soft EoS substantially underestimates the HADES $v_2(p_T)$ data for protons, deuterons, and tritons, with the discrepancy increasing with $p_T$. The hard EoS fails to reproduce the proton $v_2$ at high $p_T$, consistent with SMASH results \cite{Mohs:2020awg, Mohs:2024gyc}. In contrast, the soft momentum dependent (SM) EoS brings $v_2$ into agreement with experimental data.
While $v_2(p_T)$ for protons differs between hard and soft momentum dependent EoS, deuterons and tritons show nearly identical flow patterns for both EoS. The hard EoS yields higher $v_2$ for deuterons compared to protons, improving agreement with data—a phenomenon also reported in UrQMD calculations \cite{Hillmann:2019wlt}.

We note that similar conclusions follow from the comparison of the PHQMD results with the FOPI data of $v_1$, $v_2$ as presented in Ref.\cite {Kireyeu:2024hjo}.


In order to investigate the sensitivity of the flow observables $v_1$ and $v_2$ on the deuteron production mechanisms, we compare the PHQMD results for the "default" scenario, where deuterons are produced by the kinetic + MST mechanisms, with that produced by the coalescence mechanism. 
We note that the rapidity distribution of deuterons produced by the kinetic mechanism  is very similar to that from the MST mechanism.
We stress that the PHQMD is a unique laboratory for such a comparison since all scenarios are integrated in the same code. 

In the upper plot of  Fig. \ref{fig:coalescence_v1ypt} we present the comparison of the $v_1(y)$ (left) and  $v_2(y) (right)$  of deuterons produced by the kinetic + MST mechanism (solid lines) with that produced by the coalescence mechanism (dashed lines) for 20-30\% central Au+Au collisions at $E_{kin}=1.23$ A GeV for  $1.0 < p_T < 1.5$ GeV/c.  The lower plot shows the comparison of $v_1(p_T)$  for $-0.25 < y < -0.15$ (left) and  $v_2(p_T)$ for $|y| < 0.05$ (right) of deuterons produced by the kinetic + MST mechanism (solid lines) with that produced by the  coalescence mechanism (dashed lines) for 20-30\% central Au+Au collisions.
For all EoS, the $v_1(y)$ and $v_1(p_T)$ of kinetic + MST deuterons at midrapidity are found to be slightly larger than those of coalescence deuterons. This observation is consistent with our previous results in Ref. \cite{Kireyeu:2024woo}, which demonstrated that only a small subset of nucleons identified as deuterons by coalescence and by MST coincide.

The lower plot of Fig. \ref{fig:coalescence_v1ypt} shows the same comparison for $v_2(p_T)$ for deuterons measured near midrapidity. The both deu\-teron production scenarios are confronted with the HADES data  \cite{HADES:2020lob}.  For $v_2$ one can see the same tendency as for $v_1$ - the absolute value of $v_2$ is larger if one applies the kinetic + MST mechanism than if one applies the coalescence mechanism.
Thus, the rapidity distribution of $v_2$ as well as the $p_T$ dependence of $v_2(p_T)$ at midrapidity opens the perspective  to identify  the production mechanism. 

Summarizing,  the PHQMD calculations validate prior results, demonstrating that the equation-of-state significantly affects the rapidity and transverse momentum dependence of the directed flow ($v_1$) and elliptic flow ($v_2$) for protons and light clusters produced in heavy-ion collisions at SIS energies. 

\subsubsection{ PHQMD results for $v_1$, $v_2$ at FAIR/ BES RHIC  energies}

\begin{figure*}[t!]             
    \centering 
   \includegraphics[width=0.35\linewidth]{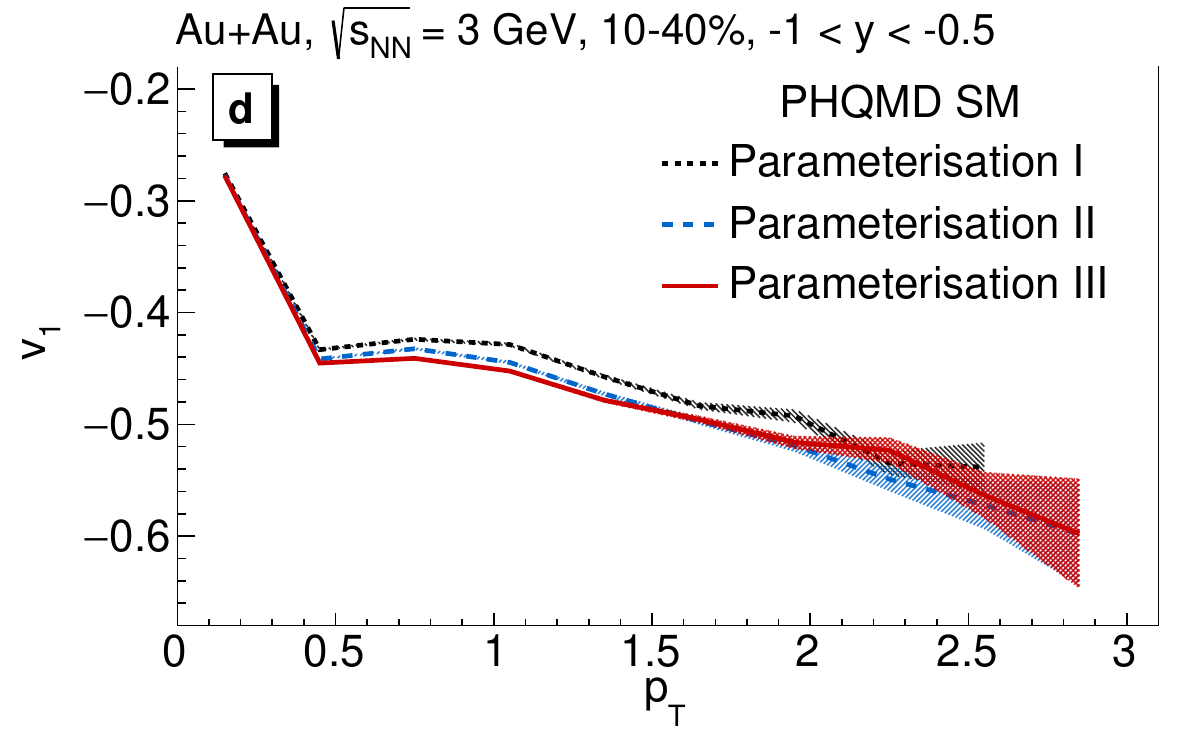}   
   \includegraphics[width=0.35\linewidth]{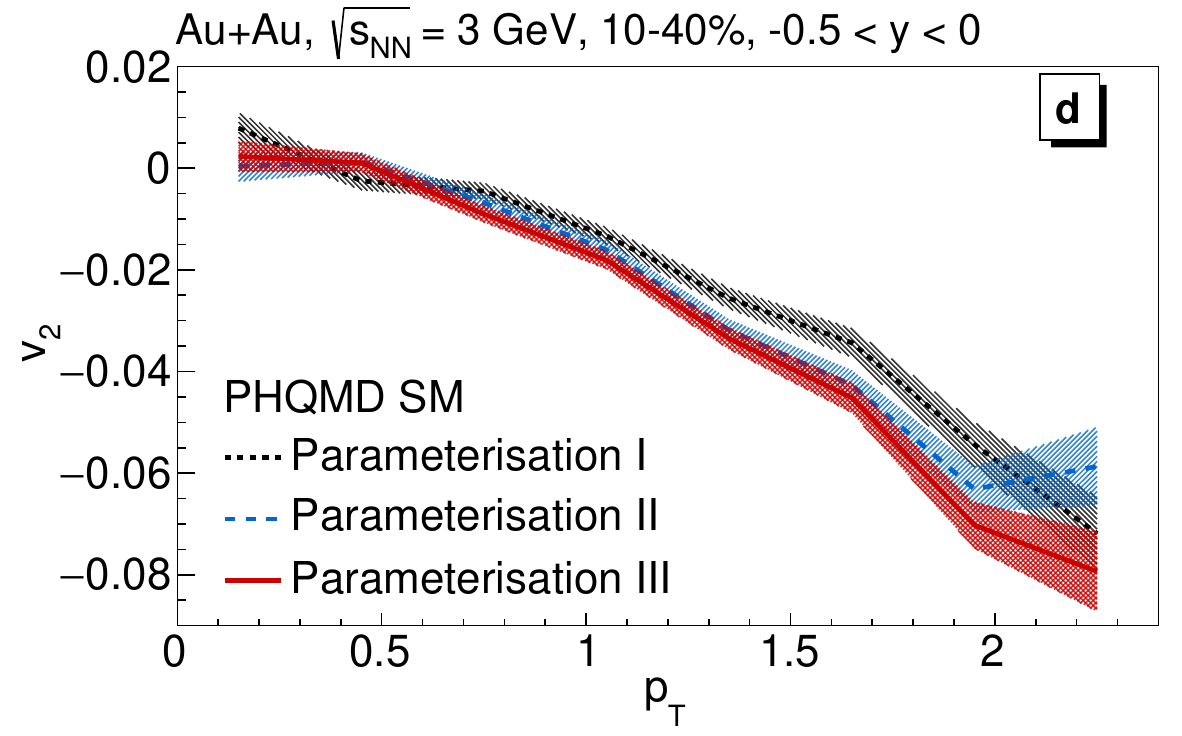} \\
   \includegraphics[width=0.35\linewidth]{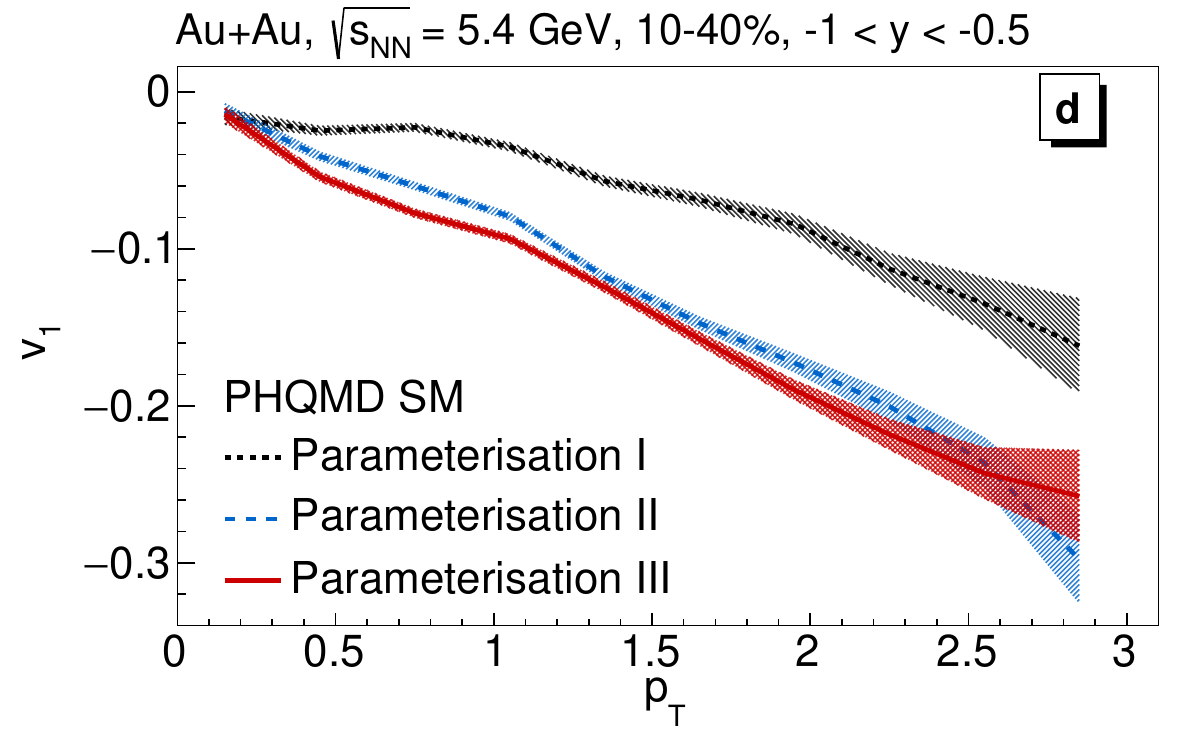}
   \includegraphics[width=0.35\linewidth]{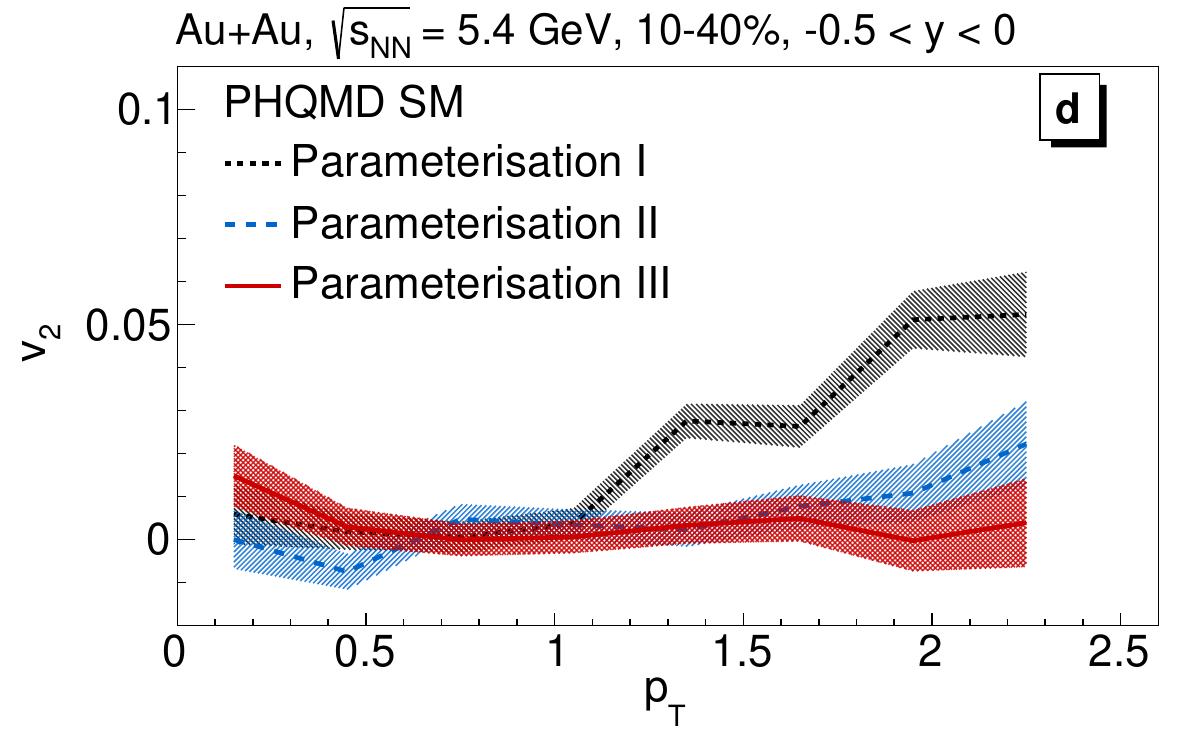}   
    \caption{The in-plane flow $v_1(p_T)$  near target rapidity ($-1<y<-0.5$)  (left) and elliptic flow $v_2(p_T)$ at mid-rapidity ($-0.5<y<0$) (right) of deuterons from PHQMD simulations of Au+Au collisions at $\sqrt{s_{\rm{NN}}}=3$ GeV (left column) and 5.4 GeV (right column)  in the 10–40\% centrality class. Results are shown for three different parameterizations of the nuclear optical potential (Parameterization I, II, and III).
     The figure is adopted from Ref. \cite{Zhou:2025zgn}.    }
    \label{fig:v1pt_optPotential}
\end{figure*}
%
\begin{figure*}[h!]             
    \centering    
    \includegraphics[width=0.33\linewidth]{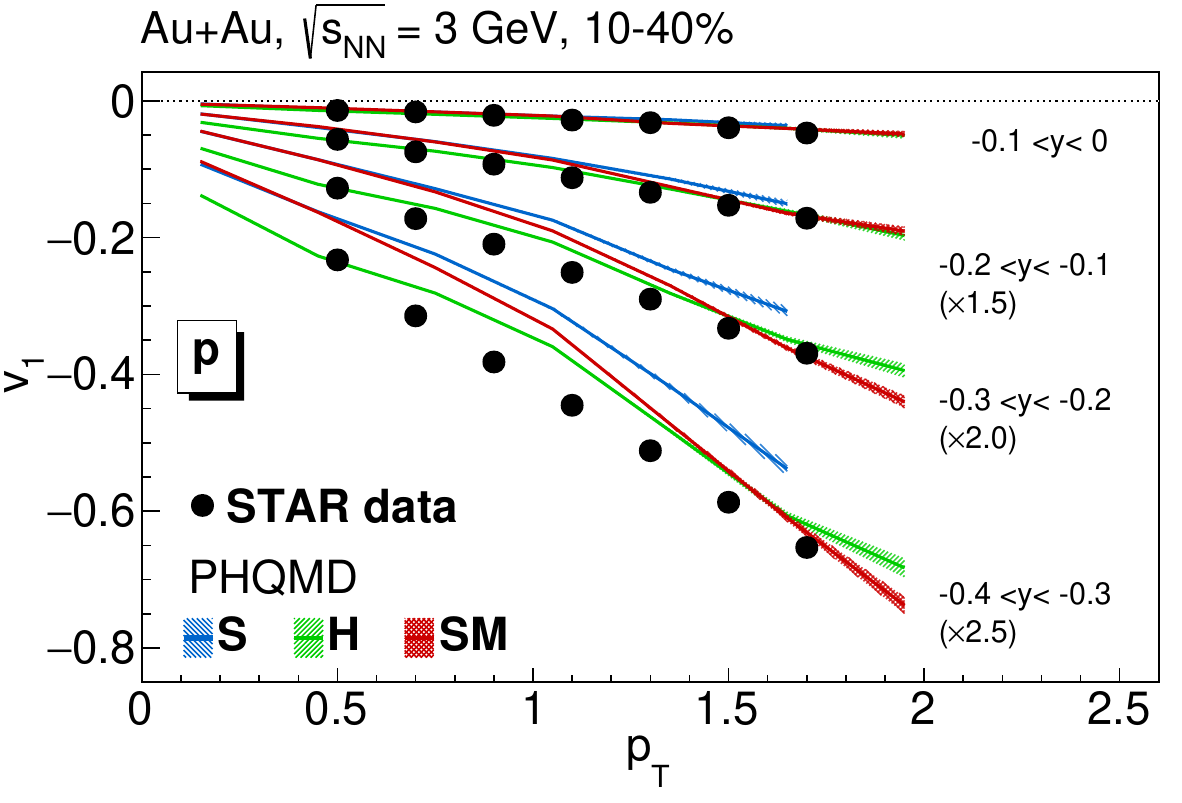}
    \includegraphics[width=0.33\linewidth]{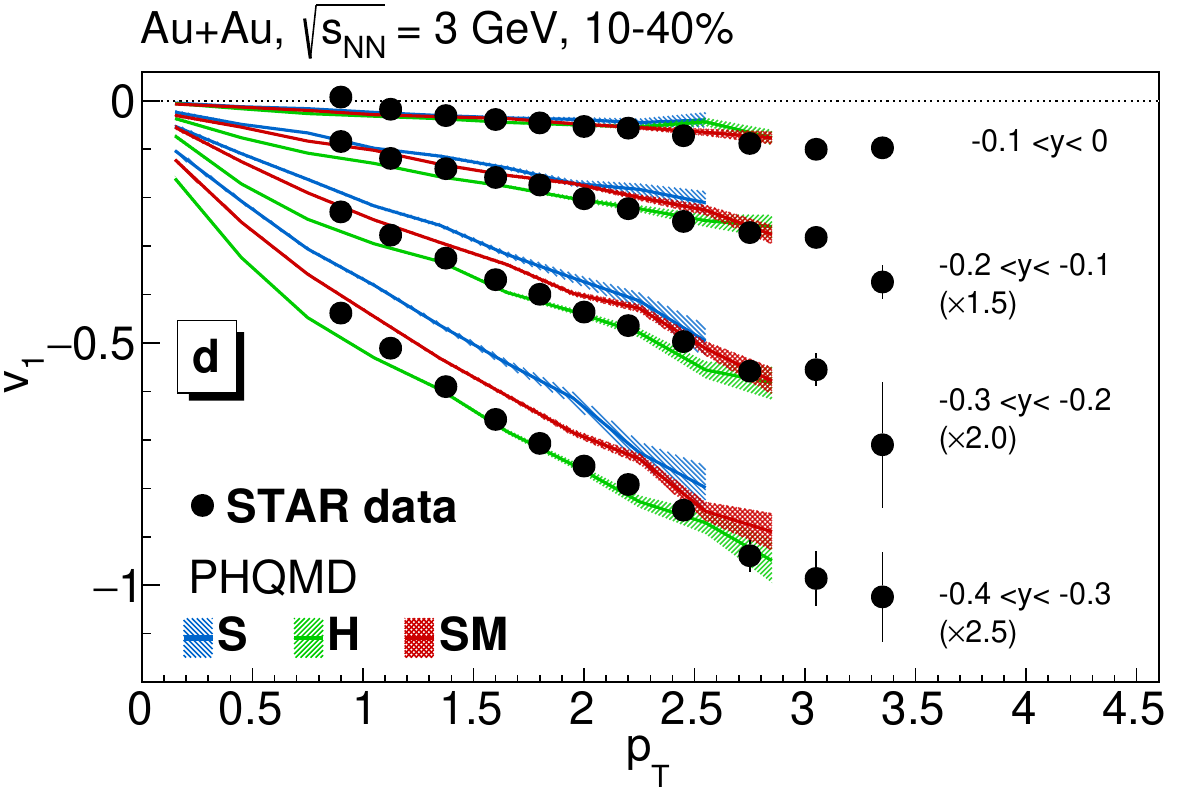}
    \includegraphics[width=0.33\linewidth]{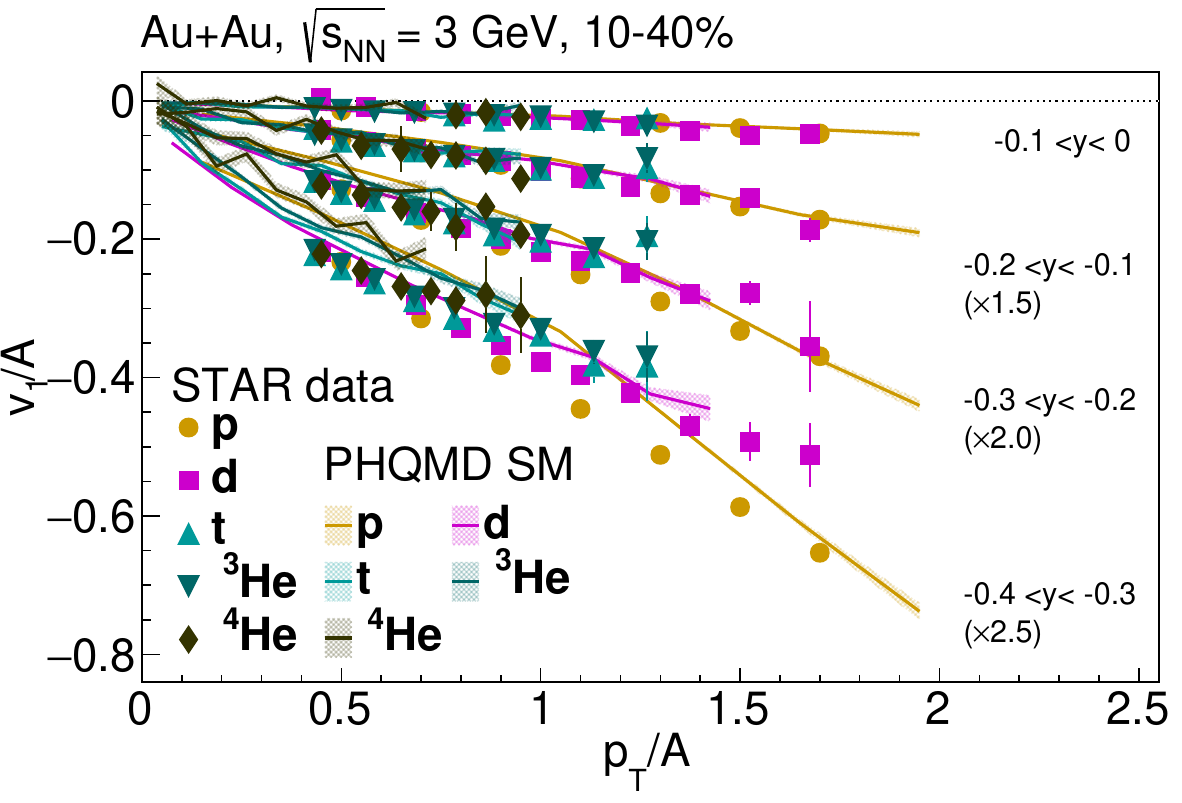}
    \caption{The PHQMD results for the directed flow $v_1$ of protons (left) and deuterons (right)  calculated with S (blue lines), H (green lines), SM (red lines)  EoS as a function of $p_T$ for 4 rapidity intervals in $10-40\%$ mid-central Au$+$Au collisions at $\sqrt{s_{\rm{NN}}}=3$ GeV.
    The right plot  shows the scaled $v_1/A$ for protons, deuterons, triton, $^{3}\rm{He}$, and $^{4}\rm{He}$ versus $p_T$ for 4 rapidity intervals. 
    The STAR data are taken from Ref. ~\cite{STAR:2021yiu}
   The figure is adopted from Ref. \cite{Zhou:2025zgn}. }
    \label{fig:v1pt_pdth34}
\end{figure*}
\begin{figure*}[h!]             
    \centering 
    \includegraphics[width=0.33\linewidth]{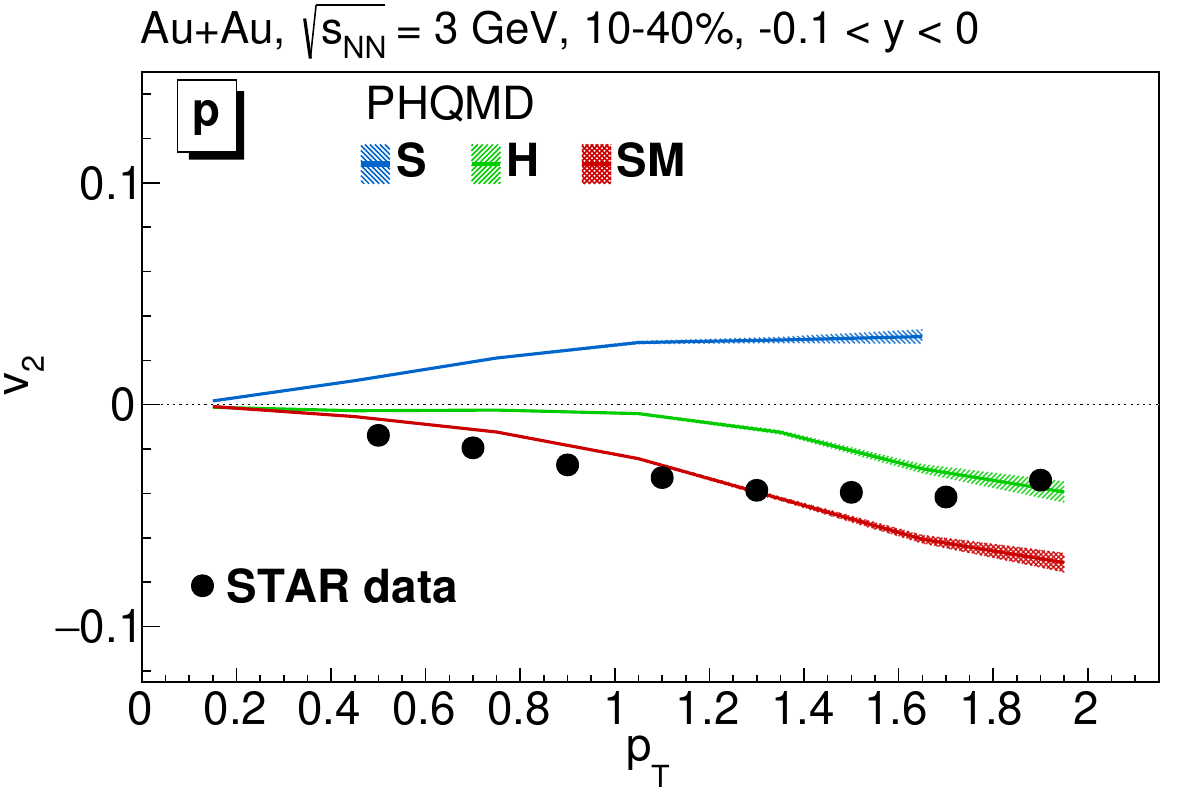}
    \includegraphics[width=0.33\linewidth]{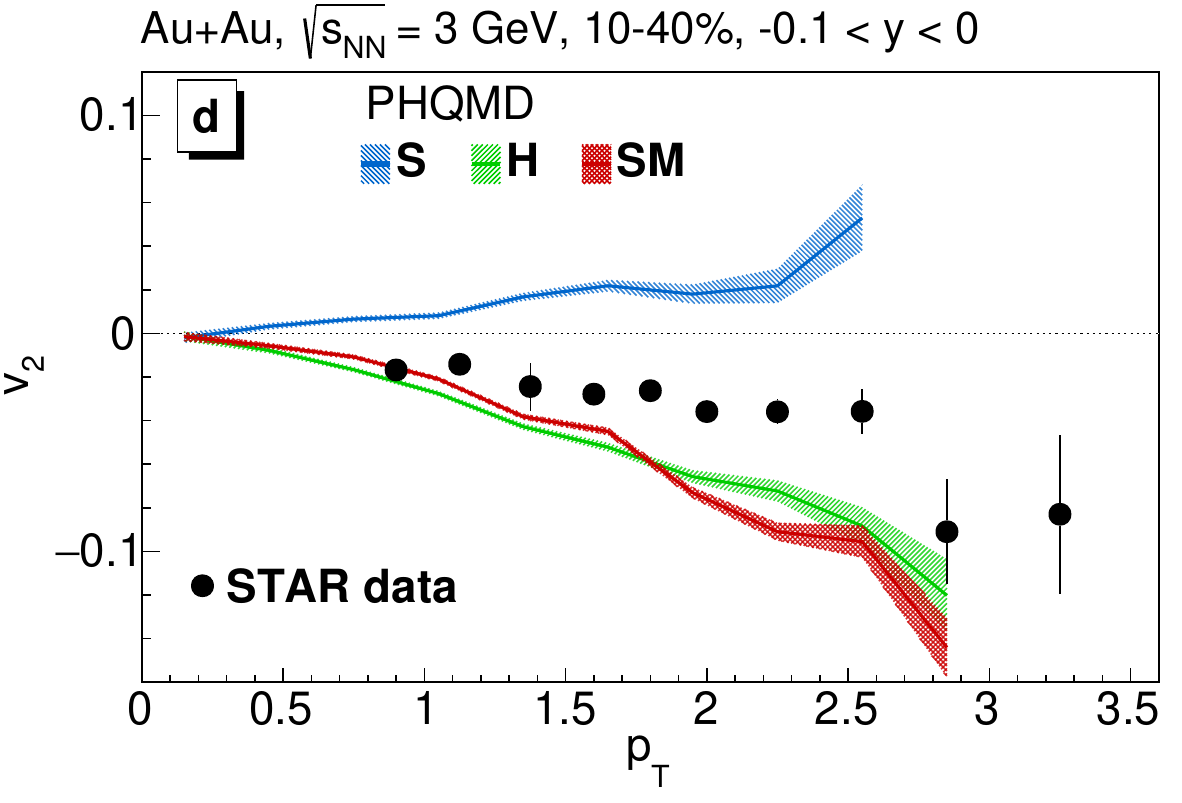}
   \includegraphics[width=0.33\linewidth]{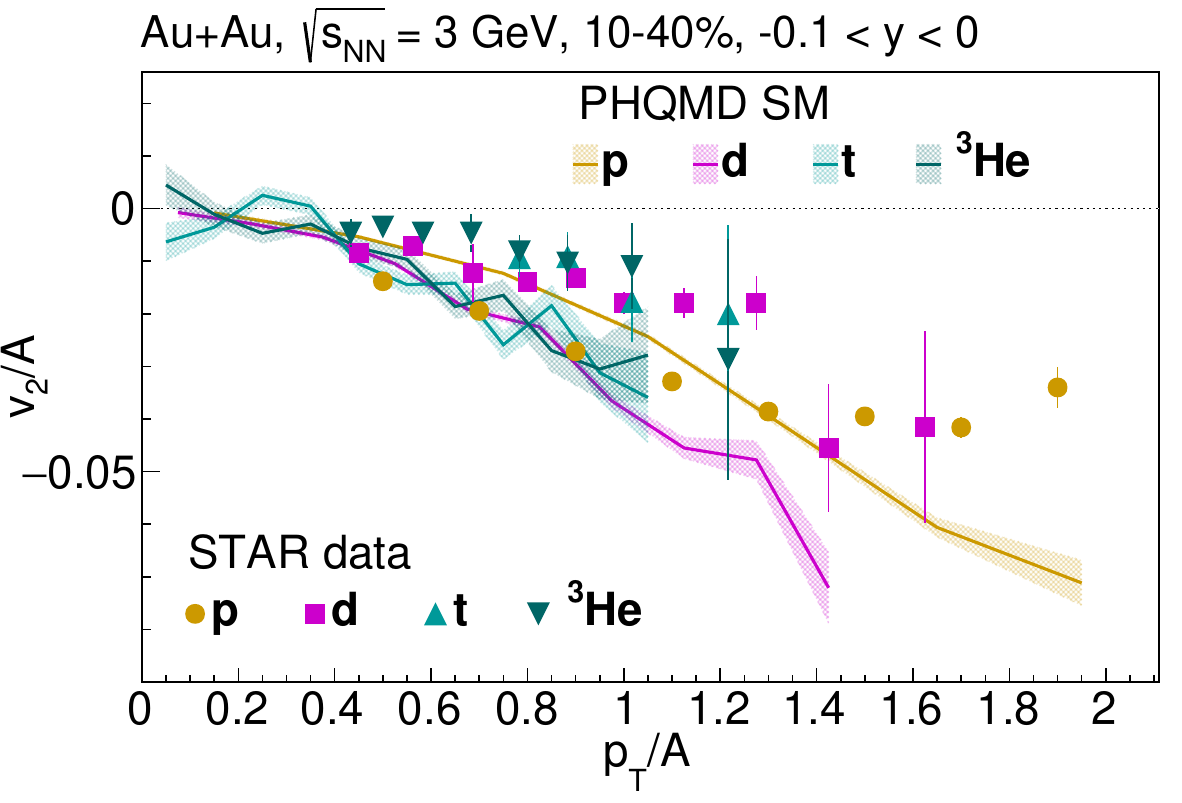} 
    \caption{The PHQMD results for the elliptic flow $v_2$ of protons (left) and deuterons (right)  calculated with S (blue lines), H (green lines), SM (red lines)  EoS as a function of $p_T$ for rapidity interval $-0.1<y<0$ in $10-40\%$ mid-central Au$+$Au collisions at $\sqrt{s_{\rm{NN}}}=3$ GeV.
    The right plot  shows the scaled $v_2/A$ for protons, deuterons, triton, $^{3}\rm{He}$ versus $p_T$ for $-0.1<y<0$. 
    The STAR data are taken from Ref. ~\cite{STAR:2021yiu}
   The figure is adopted from Ref. \cite{Zhou:2025zgn}. }    
    \label{fig:v2pt_pdth3}
\end{figure*}


As demonstrated in the previous section, PHQMD results at SIS energies exhibit significant sensitivity to the equation-of-state, implemented through either static or momentum dependent potentials. At higher collision energies, however, this sensitivity becomes intertwined with uncertainties in the optical potential $U_{\text{opt}}(p)$ at large momenta, as illustrated in Fig. \ref{fig:uopt}.
In Ref. \cite{Zhou:2025zgn} the model study has been performed to investigate three different parametrizations of $U_{\text{opt}}(p)$ and their impact on observables. The study revealed that rapidity distributions of $v_1$ and $v_2$ remain practically unaffected by these uncertainties. In contrast, the transverse momentum dependence of both directed ($v_1$) and elliptic ($v_2$) flow for deuterons shows a visible sensitivity to the high momentum behavior of $U_{\text{opt}}(p)$ when the collision energy increases.

This energy-dependent effect is illustrated in Fig. \ref{fig:v1pt_optPotential}, which compares $v_1(p_T)$ near target rapidity ($-1 < y < -0.5$, left) and $v_2(p_T)$ at midrapidity ($-0.5 < y < 0$, right) for deuterons produced in Au+Au collisions at $\sqrt{s_{\text{NN}}} = 3$ GeV and $5.4$ GeV (10–40\% centrality). The results from parametrizations I, II, and III begin to diverge with increasing $p_T$, with the splitting becoming more pronounced at the higher energy of $\sqrt{s_{\text{NN}}} = 5.4$ GeV. This demonstrates that precision measurements of flow observables at high $p_T$ and higher collision energies can provide valuable constraints on the momentum dependence of the nuclear potential.

Recently the STAR collaboration provided a high precession data on multiple observables including the flow harmonics for light clusters \cite{STAR:2021yiu}. Here we highlight several results from Ref. \cite{Zhou:2025zgn} where we compared the PHQMD calculations (using Parametrization I for SM potential) to STAR data.

In Figure~\ref{fig:v1pt_pdth34} we present the PHQMD results for protons (left) and deuterons (middle) as a function of $p_T$ for 4 rapidity intervals for in $10-40\%$ central $\sqrt{s_{\rm{NN}}}=3$ GeV Au+Au collisions, in comparison to the STAR data \cite{STAR:2021yiu}.
The right plot shows the scaled $v_1/A$ for protons, deuterons, tritons, $^{3}\rm{He}$, and $^{4}\rm{He}$ versus $p_T$ for 4 rapidity intervals and for the SM EoS. One can see that the PHQMD results follow approximately  an $A$ scaling near mid-rapidity, $-0.3<y<0$, whereas for larger negative rapidities there are clear deviations. 
One can also see that for all cluster sizes the theoretical $v_1$  calculations with a H EoS provide the best description of the data. SM results are very close to those with H EoS, however, slightly underestimate the data, while a S EOS substantially deviates from the experimental data.

In Fig. \ref{fig:v2pt_pdth3} we show results for $v_2(p_T)$ for the rapidity interval $-0.1<y<0$. Here a similar trend is observed - the soft EoS is excluded by the data while the H and SM EoS stay close, especially for the light clusters. We note, however, that $v_2$ of protons is better explained by the SM EoS. Similar conclusions holds for the elliptic flow of $\Lambda$ which favor SM EoS \cite{Zhou:2025zgn}.
 
\subsection{pBUU of Danielewicz}

At lower beam energies, the determination of the equation-of-state has been addressed in the pioneering studies within the pBUU model by Danielewicz et al. \cite{Pan:1992ef,Danielewicz:1992mi, Danielewicz:1999zn, Shi:2001fm, Danielewicz:2002pu}.  
The pBUU is a mean-field transport model which incorporates the momentum dependent potential for baryons which includes scalar and vector parts. 
In Refs. \cite{Danielewicz:1999zn} and \cite{Danielewicz:2002pu} the flow and stopping observables of protons have been examined to constrain the nuclear equation-of-state. 
It has been found  that  the elliptic flow at midrapidity exhibits a particularly strong sensitivity to the mean-field momentum dependence in midperipheral to peripheral collisions.

Here we emphasize a key difference in the treatment of potential interactions in QMD and BUU dynamics, and in the subtraction of $U_{opt}(p)$. In QMD, the two-body potential depends on the relative momentum of the interacting particles and can be  connected to the experimentally reconstructed $U_{opt}(p)$ at normal nuclear density $\rho_0$ within a non-relativistic Schrödinger equation (cf. Section 2.2). In contrast, BUU is a relativistic mean-field theory with covariant scalar and vector potentials (related to the complex self-energy) that depend on the particle’s momentum relative to the medium. In a fully covariant formulation (see detailed explanations in \cite{Cassing:2021fkc}), the scalar part modifies the effective mass, while the vector part alters the hadron momentum. Thus, the momentum dependent potential in BUU is not the same as in QMD in which vector and scalar part are combined to a Schrödinger equivalent potential (Eq.\ref{uschr}). This leads to additional uncertainties in the determination of the EoS and its interpretation.

\begin{figure}
    \centering
    \includegraphics[width=0.7\linewidth]{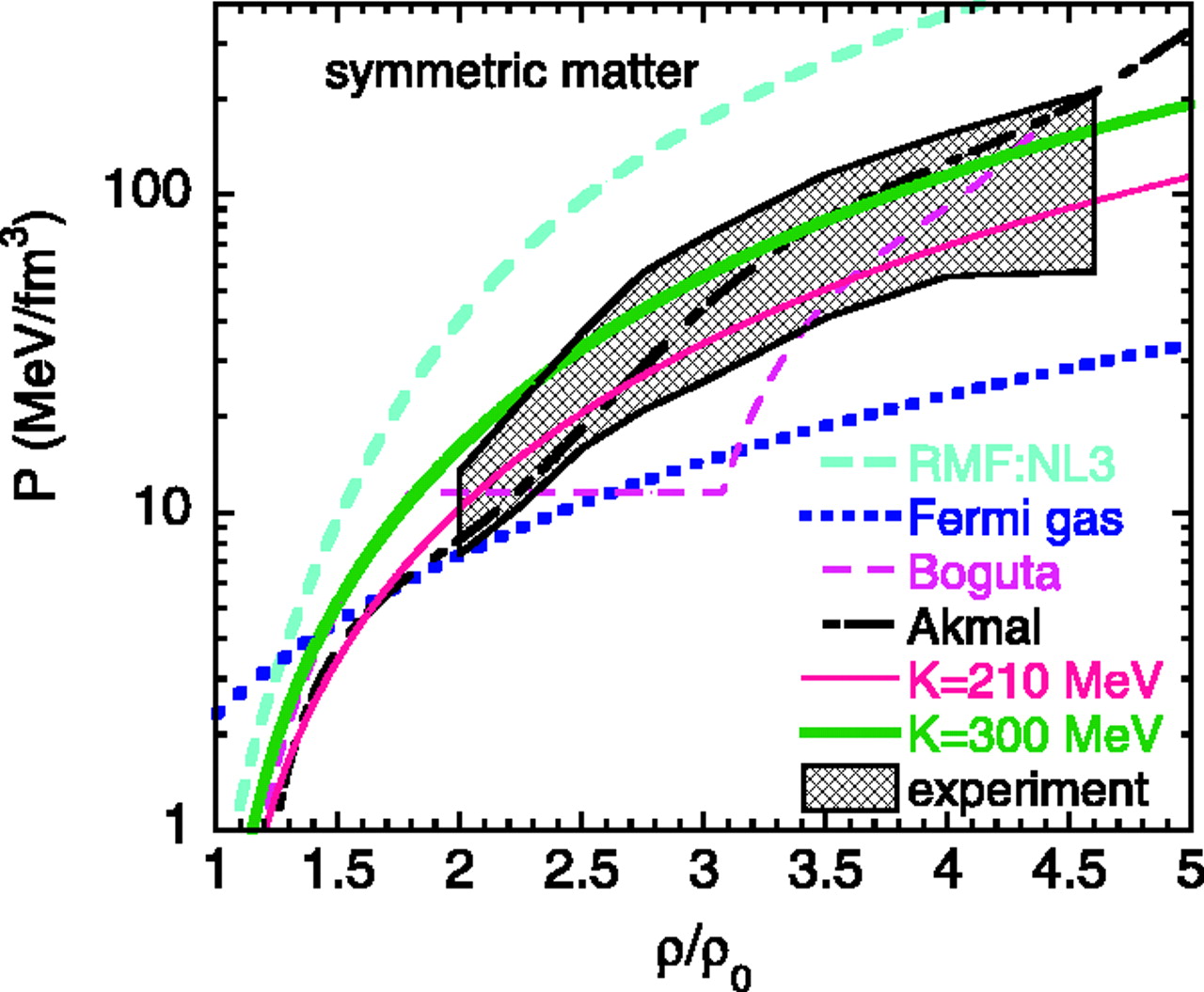}
    \caption{ EOS for symmetric nuclear matter at zero temperature. The shaded region corresponds to the region of pressures consistent with the experimental flow data. The various curves and lines show predictions for different symmetric matter EOS. 
    The figure is adopted from Ref. \cite{Danielewicz:2002pu}.}
    \label{fig:EoS_Pawel}
\end{figure}

Figure  \ref{fig:EoS_Pawel} present the EoS - the pressure $P= \rho ^2 \frac {\partial E/A}{\partial \rho}$  versus scaled baryon density $\rho/\rho_0$ - for zero temperature symmetric matter. 
The sha\-ded region corresponds to the region of pressures consistent with the experimental flow data of protons, in particular for the directed transverse flow 
$F=d\left<{p_x \over A}\right>/d({y\over y_{cm}}),$ 
where  $A$ is an atomic number, $y_{cm}$ is the center-of-mass rapidity of particle) and the elliptic flow ($\left<\cos2\phi\right>$, $\tan\phi=p_x/p_y$ is the angle ) (cf. \cite{Danielewicz:2002pu} in references therein).
The  line shows the pBUU calculations using different EOS with compression modules $\kappa = 210$ (green line) and 300 MeV (magenta line). The EOS with $\kappa =$ 300 MeV generates about 60\% more pressure than the one with $\kappa$ = 210 MeV at densities of 2$\rho_0$ to 5$\rho_0$.

As concluded in Ref. \cite{Danielewicz:2002pu} the pBUU study provided constraints on the EOS of symmetric nuclear matter that rule out very repulsive EoS from relativistic mean-field theory and very soft EoS with a strong phase transition at densities $\rho \le 3\rho_0$, but not a softening of the EOS due to a transformation to quark matter at higher densities. 

The pBUU study has been extended to the production of light fragments  in Ref. \cite{Danielewicz:1992mi}, which explores the blast of light fragments from central heavy-ion collisions. This work employed a density-dependent potential for both protons and deu\-terons, focusing on the emission patterns.  A strong sensitivity of the flow of deuterons on the  equation-of-state has been obtained.
Moreover, in Ref. \cite{Shi:2001fm} the spectator response to the participant blast has been studied  using 3 different potential parametrizations — hard (H), hard momentum dependent (HM), soft (S), and soft momentum dependent (SM) which impact on the emission of light clusters, including deuterons.  The work highlights how the collective dynamics and the formation of light nuclear clusters are influenced by the stiffness and momentum dependence of the nuclear equation-of-state, providing insights into the interplay between the collision dynamics and cluster production mechanisms.

\subsection{RBUU (Giessen)}
    
The equation-of-state  has been also investigated within the Relativistic BUU (RBUU) model \cite{Weber:1990qd,Cassing:1992gf}, which represented a pioneering development of a relativistic transport equation for the baryon phase-space distribution in line with  Dirac– Brueckner theory in semiclassical limit \cite{Cassing:1992gf}.
This approach accounts for the self-consistent mean-field dynamics, including momentum dependent forces, as well as the residual nucleon–nucleon collision history.
For the description of relativistic systems in RBUU the covariant mean-field (RMF) model is used with an attractive scalar self energy $\Sigma_S(\vec r,t)$ and a repulsive 4-vector self energy $\Sigma_\mu(\vec r,t)$ which is proportional to the baryon 4-current $J_\mu$. 

Constraints on the nuclear EoS have been extracted from a systematic analysis of directed and elliptic hadron flow observables in heavy-ion collisions over a broad energy range, from SIS to AGS \cite{Sahu:1998vz,Sahu:1999mq,Cassing:2000bj,Sahu:2002ku}. 
There the self-energies have been evaluated using Lagrangian densities which includes the interactions of nucleon, $\sigma$ (scalar) and $\omega$ (vector) fields as well as nonlinear self-interactions of the scalar field  based on the relativistic mean-field approaches NL2 and NL3  \cite{Lang:1991qa}.  In this framework, baryons interact via attractive scalar mesons and repulsive vector mesons. The scalar field exhibits self-interaction. These mesons give rise to an attractive scalar self-energy $U_s(\vec{r},t)$ and a repulsive four-vector self-energy $U_\mu(\vec{r},t)$, the latter being proportional to the baryon four-current $J_\mu$. Both approaches reproduce nuclear matter properties, but they differ in compression modulus $\kappa$ due to variations in meson masses and coupling constants.  

In order to obtain an agreement of the corresponding Schrö\-dinger - equivalent potential $U_{\text{sep}}(p)$ (Eq.~\ref{uschr}) with  the experimentally observed momentum dependence of the optical potential $U_{opt}(p)$ (Fig \ref{fig:uopt}) the momentum dependent  couplings in terms of vertex form factors have been introduced  in the RBUU model \cite{Sahu:1998vz, Sahu:1999mq, Sahu:2002ku}.
The latter effectively suppresses  $U_{\text{sep}}(p)$ at high momenta ($p \ge 1\ \text{GeV}/c$), being still  compatible with the information from the  experimentally defined $U_{opt}(p)$ and Dirac– Brueckner calculations, and on the other hand, provides a more realistic description of in-medium dynamics at intermediate and relativistic energies.

\begin{figure}[t!]
    \centering   
    \phantom{a}\hspace*{5mm}
    \includegraphics[width=0.7\linewidth]{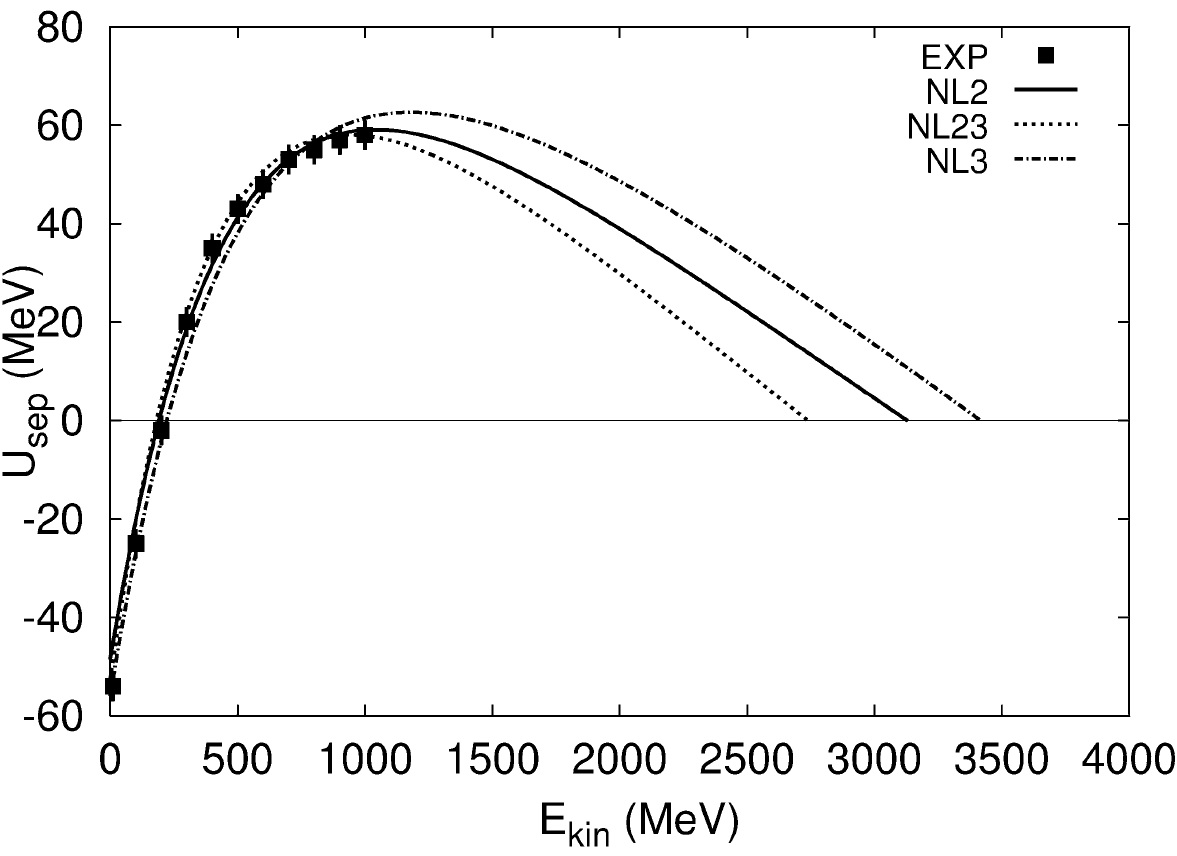}
    \includegraphics[width=0.7\linewidth]{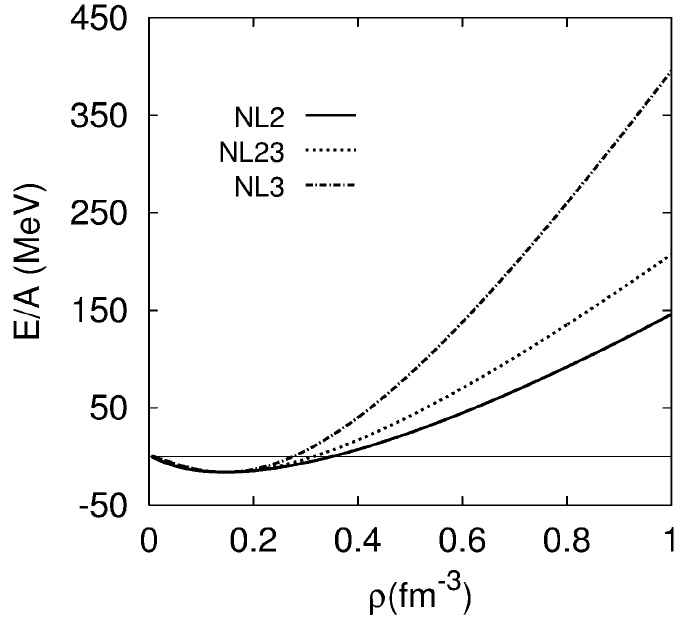}    
    \caption{ Top: The Schrödinger equivalent potential $U_{sep}$ at density $\rho_0$ as a function of kinetic energy in comparison to the experimental data \cite{Hama:1990vr} (full squares). The solid line results from the parameter set NL2 (soft, $\kappa = 210$ MeV) while the dotted and dot-dashed lines correspond to NL23 (medium, $\kappa = 300$ MeV) and NL3 (stiff (hard), $\kappa = 380$ MeV), respectively.
    Bottom: The energy per nucleon $E/A$ as a function of density $\rho$ for the parameter
    sets NL2 (solid line), NL23 (dotted line) and NL3 (dot-dashed line).
    The figure is adopted from Ref. \cite{Sahu:2002ku}.}
    \label{fig:RBUU_EOS}
\end{figure}

The corresponding  Schrödinger equivalent potential $U_{sep}$ (top) and the corresponding EoS at zero temperature as a function of $\rho$  (bottom) are shown in Fig. \ref{fig:RBUU_EOS} for
3 different realizations of relativistic mean-field  potentials NL2 - soft, NL23 - medium, NL3 - hard EoS  including momentum dependent vertex form  factors. 

\begin{figure}[h!]
    \centering 
    \phantom{a}\hspace*{-5mm}
    \includegraphics[width=0.8\linewidth]{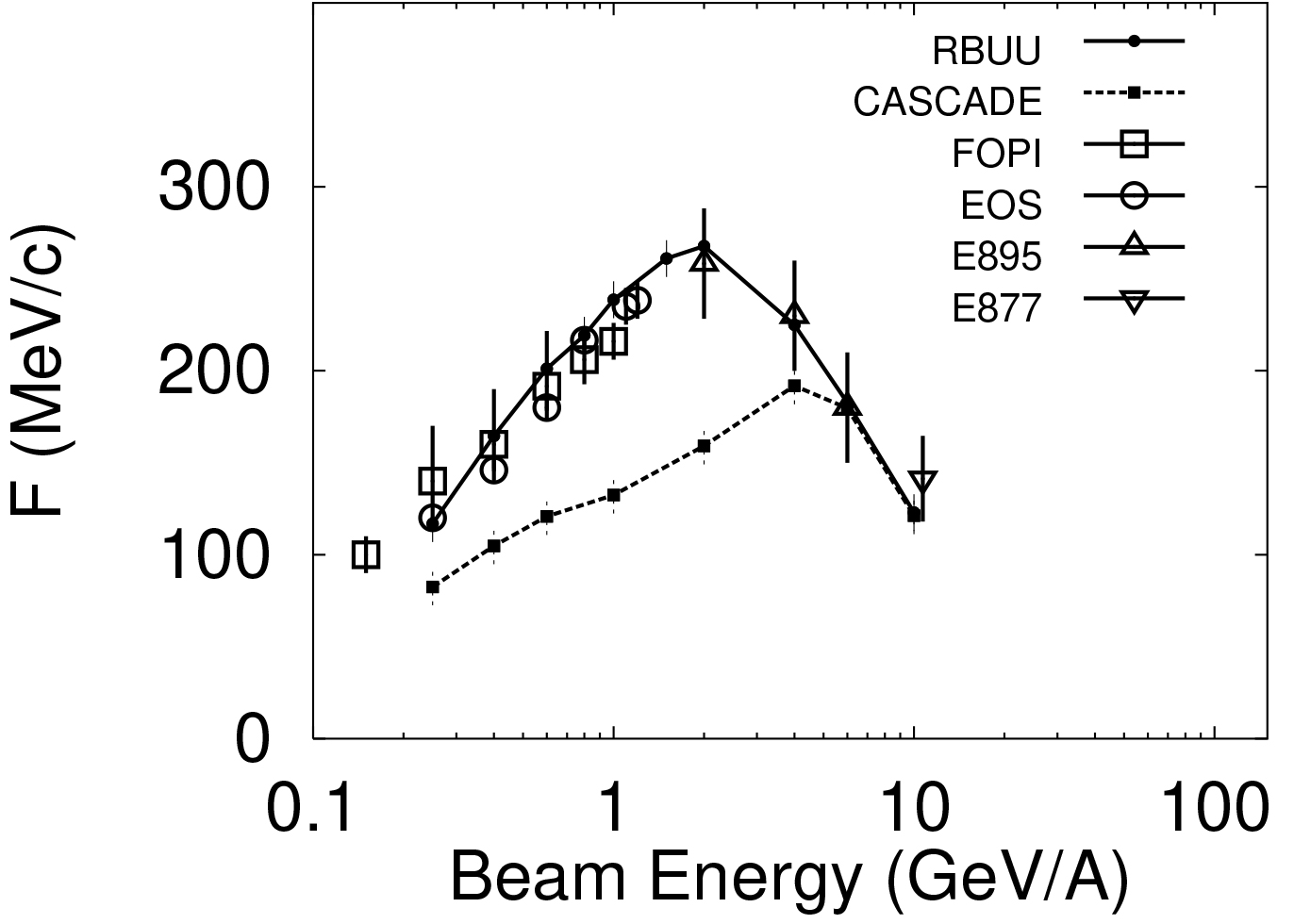}
    \includegraphics[width=0.75\linewidth]{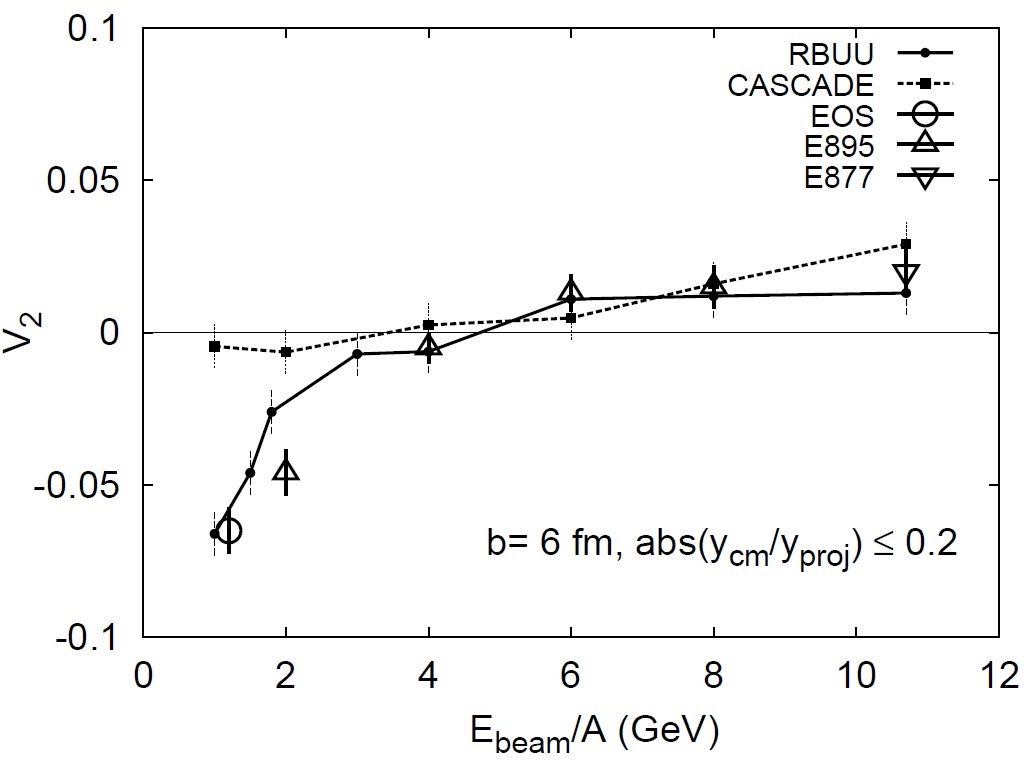}    
    \caption{ The sideward flow $F$ (top) and the elliptic flow $v_2$ (bottom) of protons as a function of the beam energy per nucleon for Au + Au collisions at $b=6$ fm from the RBUU calculations. The solid line results for the RBUU results, the dotted line for a cascade calculation.
The data points are from the FOPI, EOS, E895 and E877 collaborations.
    The figure is adopted from Ref. \cite{Sahu:1999mq}.}
    \label{fig:RBUU_v1v2}
\end{figure}
 
 Figure \ref{fig:RBUU_v1v2} shows the RBUU results for the sideward flow 
 $F=d\left<{p_x \over A}\right>/d({y\over y_{cm}})$ (top)
 and the elliptic flow $v_2$ (bottom) of protons as a function of the beam energy per nucleon for Au + Au collisions at $b=6$ fm (solid line)  with the NL3 set \cite{Sahu:1999mq}  including momentum dependent scalar and vector vertex form factors, which provides  the best description of multiple flow observables such as rapidity and transverse spectra of hadrons, while  the dotted line stands for a cascade calculation which substantially deviates from the data.
 
As follows from the RBUU study \cite{Sahu:1998vz,Sahu:1999mq,Sahu:2002ku} the measured sideward flow is less sensitive to the EoS  compression modules $\kappa$ but strongly dependent on the momentum dependence of the mean-field potential.
They found that at low energies nuclear matter undergoes squeeze-out, producing negative elliptic flow as projectile and target spectators block the in-plane expansion of the fireball. With increasing energy  the spectators move too fast to hinder this expansion, resulting in positive elliptic flow. Thus, at AGS energies, the balance between squeeze-out and in-plane flow is sensitive to the nuclear force, as noted by Danielewicz et al. \cite{Pan:1992ef,Danielewicz:1992mi, Danielewicz:1999zn, Shi:2001fm, Danielewicz:2002pu}. The RBUU momentum-dependent potential (NL3) successfully reproduces both sideward and elliptic flow data.

It is important to emphasize that the momentum dependence of the potential does not manifest itself in the EoS of infinite cold nuclear matter. As already noted in Sec. 2.3, the soft (hard) EoS and the corresponding soft (hard) EoS including momentum dependence exhibit the same density dependence of $E/A(\rho)$. The momentum dependence becomes important, however, for colliding  systems, as seen in flow observables in heavy-ion collisions. 

Moreover, the role of the relevant degrees of freedom must be taken into account: with increasing beam energy, nuclear matter undergoes transitions into resonance matter, string matter, and eventually a quark–gluon plasma (QGP), which must be incorporated when interpreting the EoS within a potential-based framework
(cf.  PHSD/PHQMD framefork).

\subsection{IQMD}

The IQMD (Isospin Quantum Molecular Dynamics) model \cite{Hartnack:1997ez} has been employed to constrain the nuclear  equation-of-state through flow studies of protons and light clusters at SIS energies, corresponding to densities of up to about twice saturation \cite{LeFevre:2015paj,LeFevre:2016vpp}. 
Since the PHQMD model adopts from IQMD model the QMD dynamics with momentum-dependent potential and the MST mechanism for cluster formation, we refer the reader to Section \ref{PHQMD} for details. Here, we briefly highlight several key results obtained with the IQMD approach.

\begin{figure}[h!]
\centering 
\includegraphics[width=0.65\columnwidth]{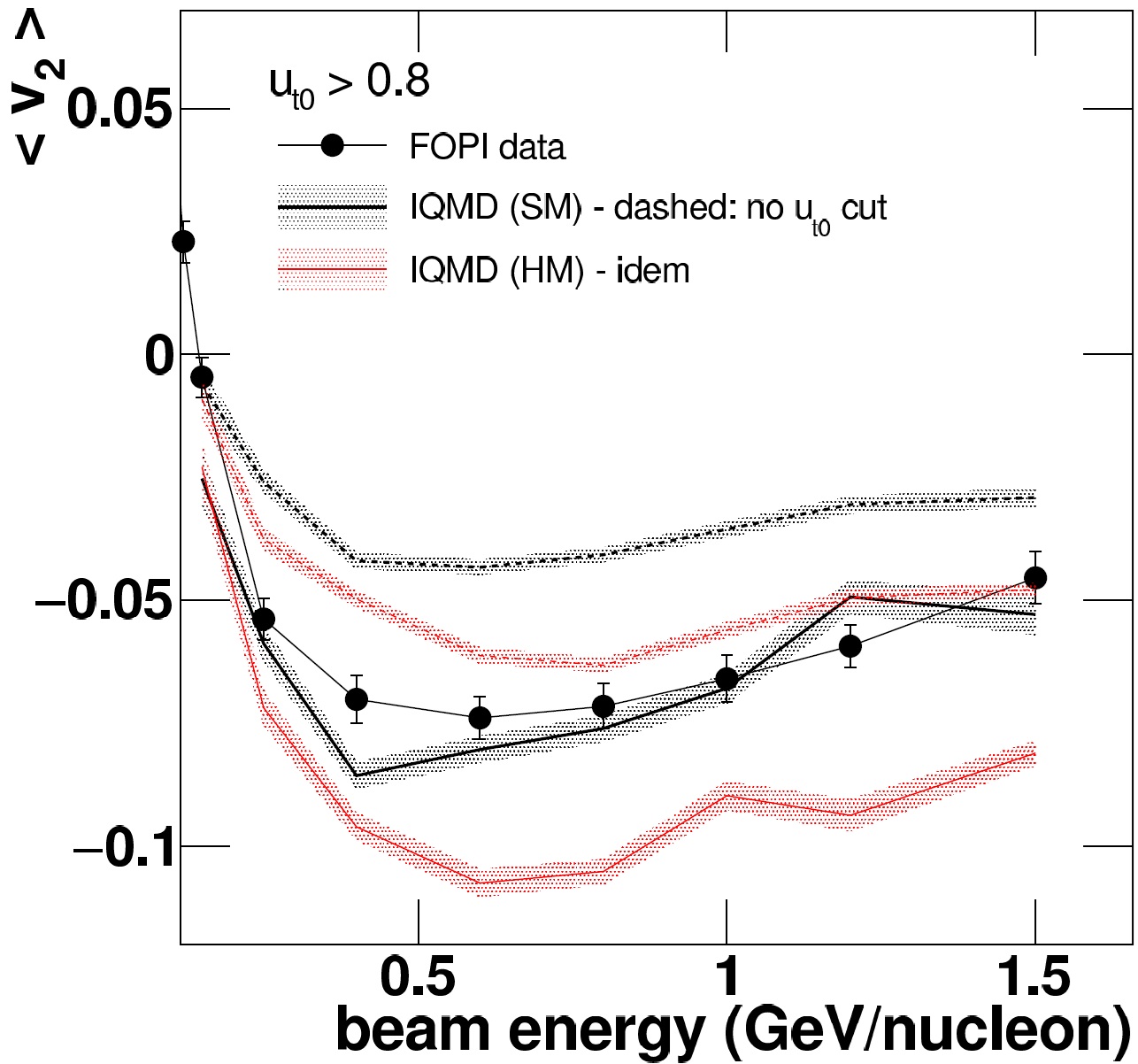}
\caption{Excitation function of the  elliptic flow $v_2$ of protons at mid-rapidity. The experimental data (black circles) are from the FOPI Collaboration \cite{FOPI:2011aa}. The data  is measured in the impact parameter range $3.1 <b<5.6$ fm and a cut on $u_{t0}>0.8$ is applied. IQMD model results  are presented for two different nuclear EOS (HM with red lines and SM  with black lines)  for $b = 4$ fm and with an additional cut on $u_{t0}>0.8$ (full lines) and without any cut (dashed lines).
The figure is adopted from Ref. \cite{LeFevre:2016vpp}. }
\label{fig:IQMD_v2p} 
\end{figure}

\begin{figure}[h!]
\centering 
\includegraphics[width=\columnwidth]{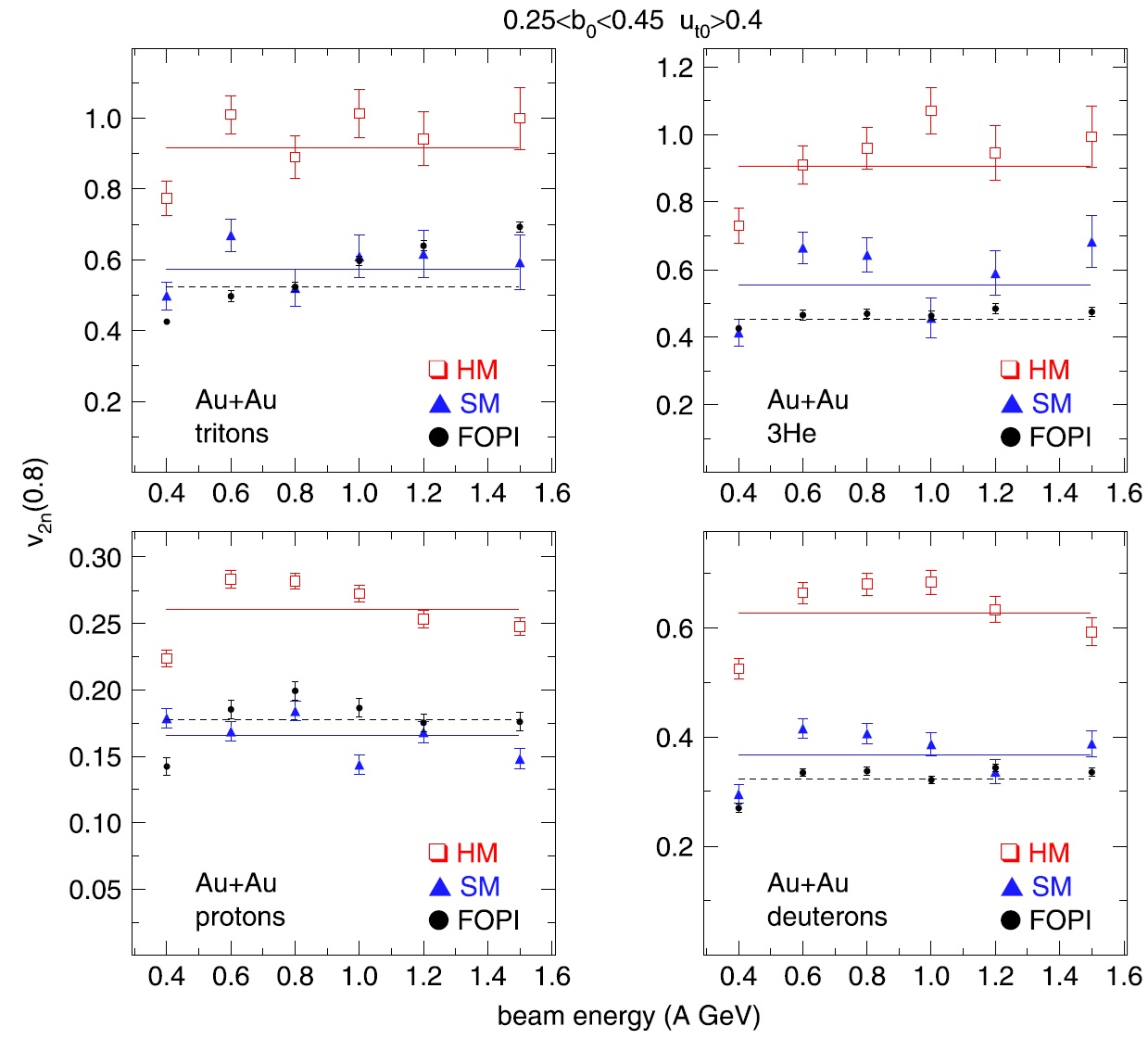}
\caption{Elliptic flow $v_2$ for protons, deuterons, tritons, $^3He$ as function of incident beam energy from IQMD for different EoS: SM - soft momentum dependent (blue), HM - hard momentum dependent (red) in comparison to the FOPI data \cite{FOPI:2011aa}.
The figure is adopted from Ref. \cite{LeFevre:2016vpp}. }
\label{fig:IQMD_v2A} 
\end{figure}

In Fig. \ref{fig:IQMD_v2p} the IQMD results for the excitation function of the elliptic flow $v_2$ of  protons at mid-rapidity are shown in comparison to the FOPI data \cite{FOPI:2011aa}. IQMD results  are presented for two different nuclear EOS, a hard momentum dependent (HM) EoS (red lines) and a soft momentum dependent (SM) EoS  (black lines)  for $b = 4$ fm and with an additional cut on $u_{t0}>0.8$ (full lines) and without any cut (dashed lines). One can see that the FOPI  data favor the SM EoS.

In Fig. \ref{fig:IQMD_v2A}  the elliptic flow $v_2$ is shown for protons, deuterons, tritons, $^3He$ as function of incident beam energy from IQMD for SM (blue) and HM (red) in comparison to the FOPI data \cite{FOPI:2011aa}. Here one can see a similar trend as for protons - the SM EoS provides a better description of the FOPI data in the energy range from 0.4 to 1.5 A GeV.


\begin{figure*}[h!]   
\centering 
\includegraphics[width=0.7\textwidth]{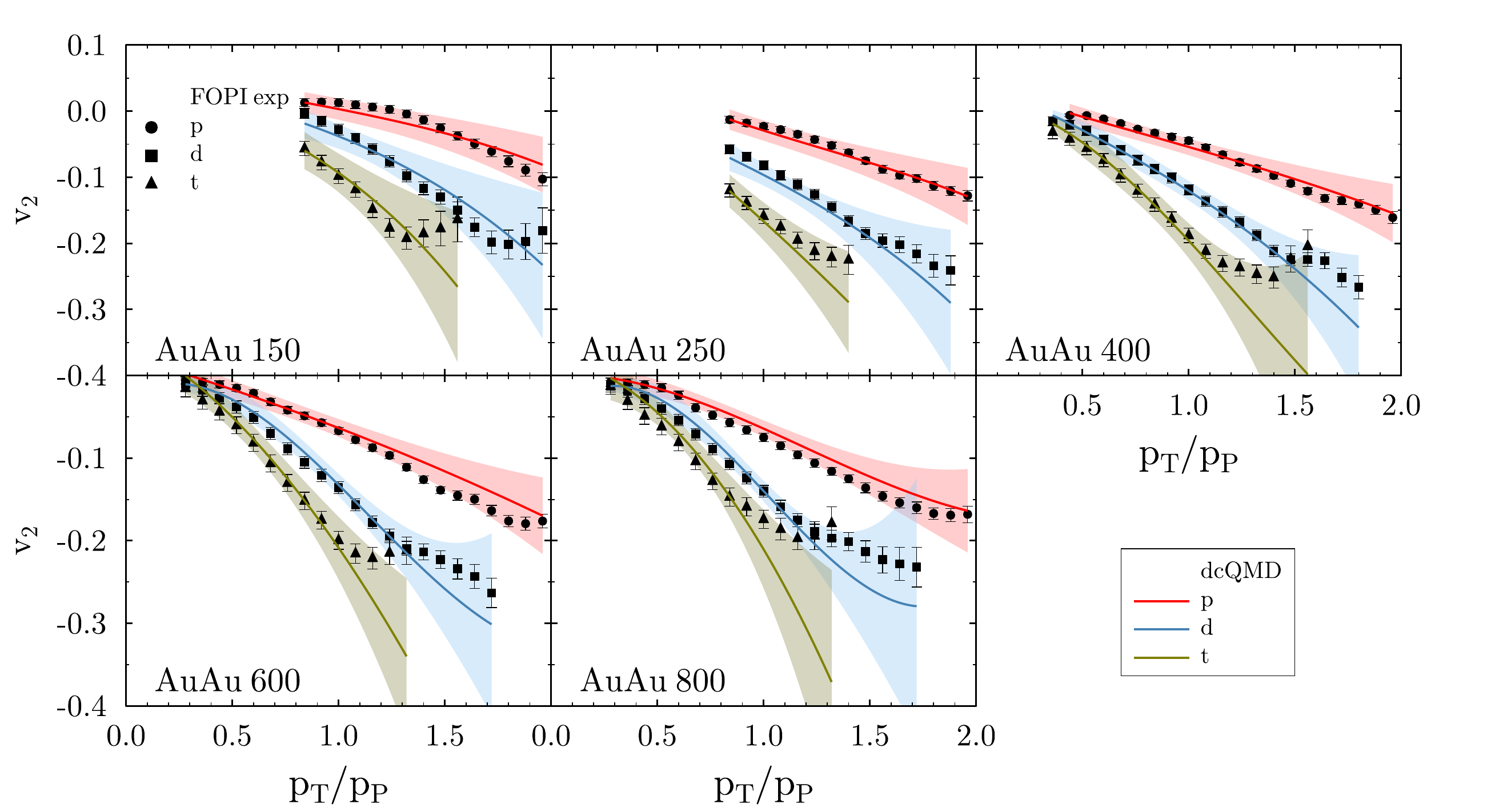}
\caption{  The dcQMD predictions for transverse momentum dependent $v_2$ of protons, deuterons and tritons are compared to the FOPI experimental data ~\cite{FOPI:2011aa}.}
\label{dcFOPI_fopi2_v2pt}
\end{figure*}
\begin{figure*}[h!]
\centering 
\includegraphics[width=0.7\columnwidth]{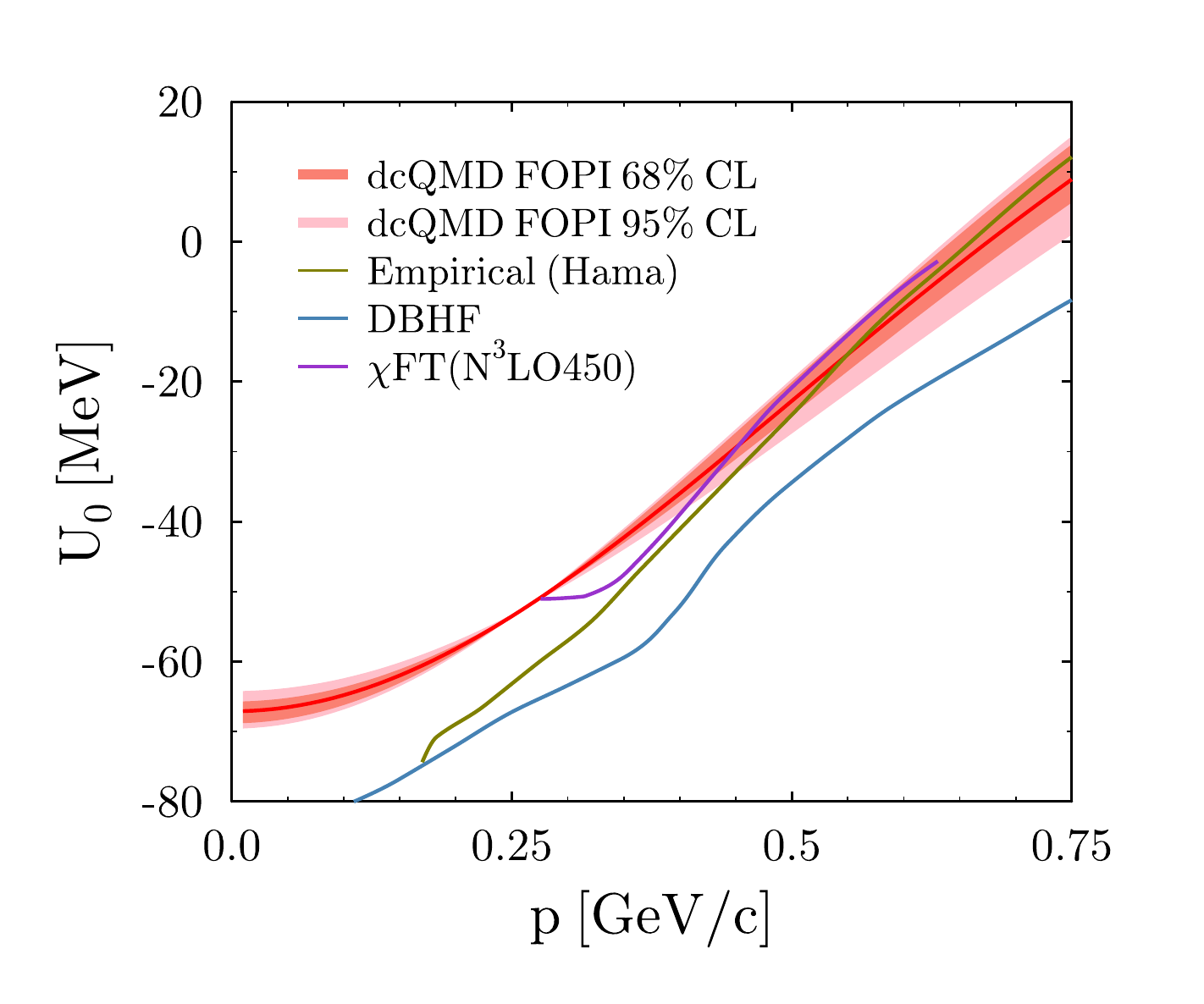} 
\includegraphics[width=0.7\columnwidth]{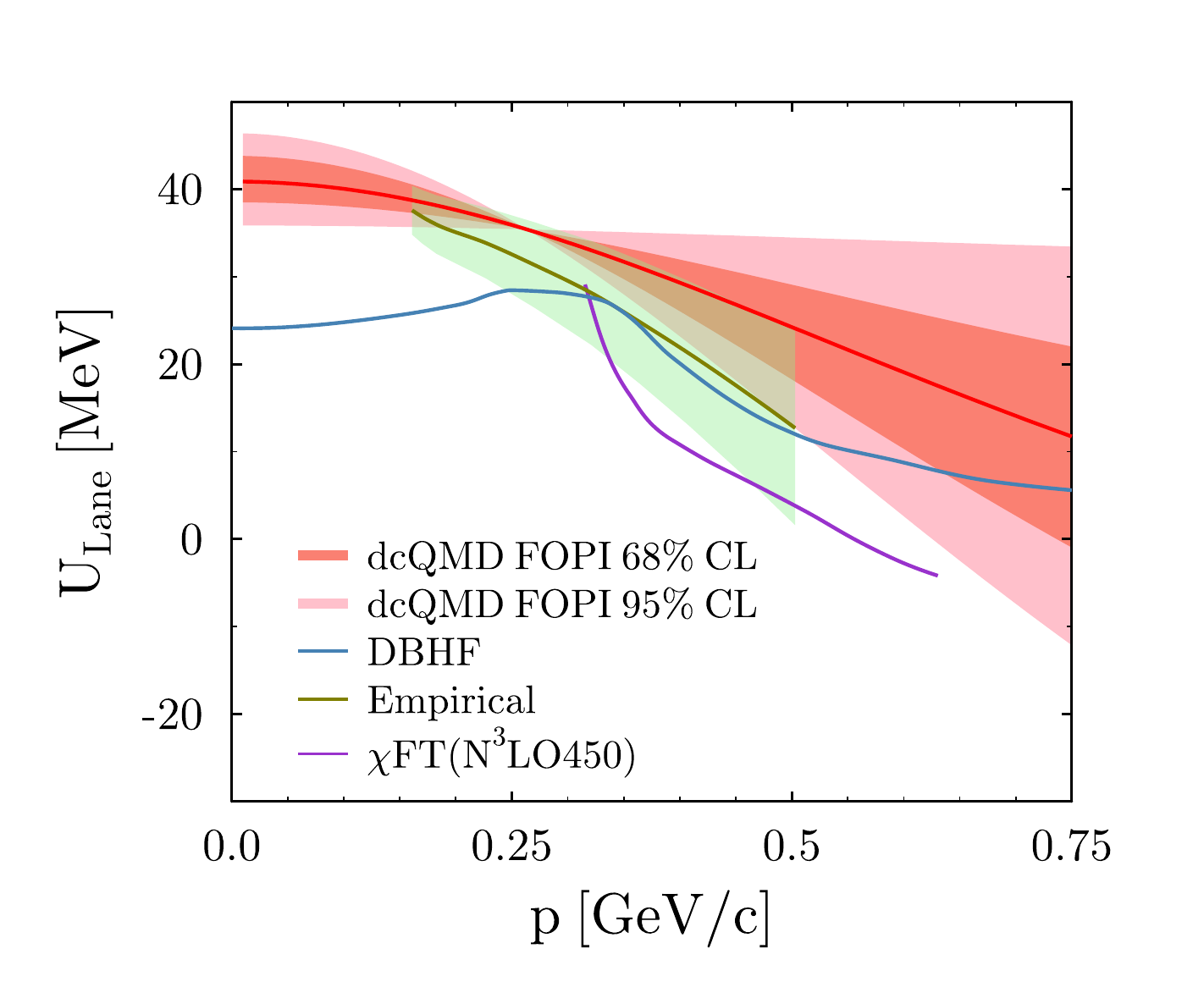}
\caption{Left: Isoscalar optical potential at saturation density as a function of nucleon momentum from four sources: HIC (pcQMD), empirical Hama potential~\cite{Hama:1990vr}, microscopical DBHF ~\cite{vanDalen:2005sk} and $\chi$FT~\cite{Holt:2015dfa} calculations. 
Right: The same as in the left panel but for the symmetry (Lane) potential. The shown empirical Lane potential depicts a parametrization~\cite{Li:2004zi} of analyses of nucleon-nucleus scattering experiments at beam energies below 100 MeV.
The figure is adopted from Ref. \cite{Cozma:2024cwc}. }
\label{dcQMD_optpot} 
\end{figure*}

\subsection{dcQMD}

At lower beam energies (150–800 AMeV), the determination of the nuclear equation-of-state has been addressed within the dcQMD model by Cozma in Ref. \cite{Cozma:2024cwc}. The dcQMD transport model is based on a Gogny-inspired momentum-dependent interaction (MDI2). This form allows for flexible adjustment of key bulk parameters of nuclear matter: the incompressibility $\kappa$, the symmetry energy slope $L$ at saturation density, the isoscalar effective mass $m^*/m$, and the neutron-proton effective mass splitting $\Delta m_{np}^*$. This enables systematic studies of how each independently affects observables if the data are sufficiently precise.
In-medium effects on two body collisions are included by applying a suppression factor $f(\rho,\delta)$ that depends on density $\rho$ and isospin asymmetry $\delta$, so that
\begin{equation}
\sigma_{\text{med}}(\rho,\delta) = f(\rho,\delta)\,\sigma_{\text{vac}}^{\text{mod}}.
\end{equation}

Several classes of observables are used to constrain the EOS with dcQMD. The $\pi^-/\pi^+$ ratio probes the symmetry energy, while neutron-to-proton double ratios $(n/p)$ further constrain isovector dynamics. Collective flow observables, such as directed flow $v_1$ and elliptic flow $v_2$, probe the isoscalar EOS through $\kappa$ and $m^*$. Taken together, these observables may constrain both the density dependence of the symmetry energy and the overall stiffness of nuclear matter, which are, however, not independent. 
For cluster production a MST coalescence algorithm applied at the local freeze-out time (rather than during the entire simulation, as in PHQMD); and threshold effects for elastic scattering.

In Ref. \cite{Cozma:2024cwc} the detailed comparison of the dcQMD model to the FOPI data \cite{FOPI:2011aa} has been presented. The model parameters have been adjusted by comparison to the FOPI data for the rapidity dependent directed flow $v_1(y), v_2(y), v_2(p_T)$ of $Z = 1, Z = 2$, proton, deuteron, $A = 3$ and $\alpha$ particles using low energy data (to stay below  the resonance excitations) for mid-central Ni+Ni, Xe+CsI and Au+Au collisions and for five beam energy $E_{kin}$=0.15, 0.25, 0.40, 0.60 and 0.80 AGeV. 
The quality of the obtained model parameters is demonstrated in Fig. \ref{dcFOPI_fopi2_v2pt}. It shows 
the dcQMD predictions (i.e. the experimental data were not included in the original fit) for the transverse momentum dependent $v_2$ of protons, deuterons and tritons compared to the FOPI experimental data ~\cite{FOPI:2011aa}. One can see a good agreement of the dcQMD model with the FOPI data.

The obtained parameters lead to the optical potential of Fig. \ref{dcQMD_optpot}.
As follows from Fig. \ref{dcQMD_optpot}, comparisons with microscopic calculations \cite{vanDalen:2005sk,Holt:2015dfa} and empirical information \cite{Hama:1990vr,Cooper:1993nx,Li:2004zi,Xu:2014cwa} show good agreement for the isoscalar optical potential (left plot) at higher momenta, although a moderate discrepancy exists near and below the Fermi momentum. 
The HIC isoscalar optical potential agrees well with the empirical Hama potential at higher but not too high momenta, while moderate discrepancies appear near and below the Fermi momentum. Agreement improves significantly when the effective isoscalar mass is set to $m^* = 0.55 m_N$ which is lower than  effective mass $m^*\approx 0.7 m_N $ deduced from model study.

For the isovector symmetry potential, called Lane potential (right plot), the dcQMD result agrees well with empirical data, showing a decrease in repulsion with increasing momentum due to the positive neutron-proton effective mass difference $\Delta m^*_{np} \simeq 0.17 \delta$ GeV. A negative value, as suggested by some ImQMD analyses \cite{Morfouace:2019jky,SpiRIT:2023htl}, would contradict the empirical trends. The figure shows as well DBHF calculations 
and results for $\chi$FT.  

\subsection{UrQMD}

The influence of the EoS on flow observables has been studied within the Ultra-relativistic Quantum Molecular Dynamics (UrQMD) transport model \cite{Bass:1998ca,Bleicher:1999xi} within the static hard and soft EoS \cite{Hillmann:2019wlt} as well as within the momentum dependent (MD) potentials from a parity doubling chiral mean-field (CMF) model \cite{Papazoglou:1998vr}  presented in Ref. \cite{Steinheimer:2024eha}.   From a comparison with HADES data it has been concluded  that the present parametrization of the CMF model leads to a slightly too weak momentum dependence, however, the MD potential reproduces the data better than a static (only density dependent) potential.

\begin{figure}[t!]
    \centering
    \includegraphics[width=0.95\columnwidth]{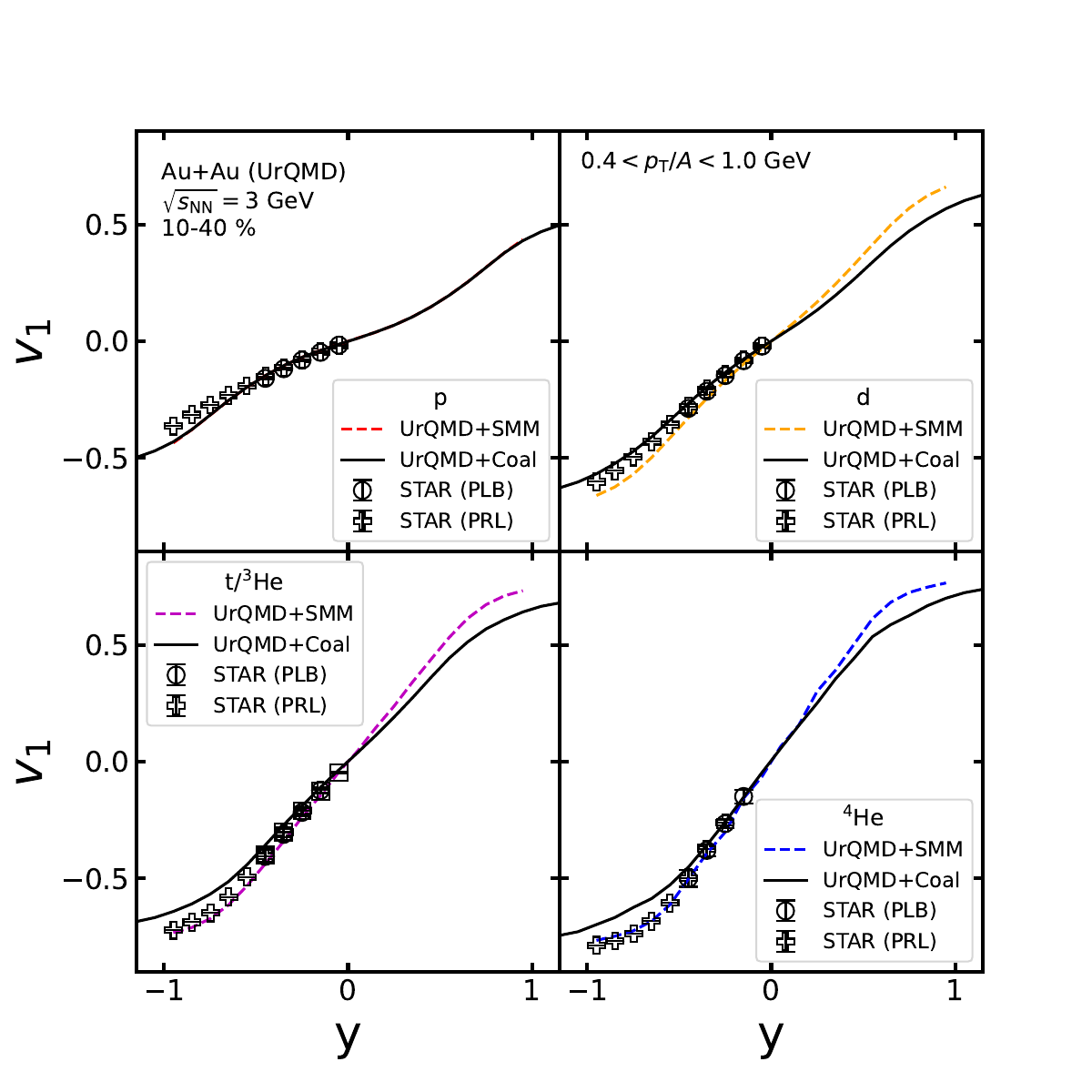}
    \vspace*{-2mm}
    \caption{The directed flow $v_1$ as a function of rapidity of protons (upper left), deuterons (upper right), tritons and $^3$He (lower left) and $^4$He (lower right) from 10-40\% central Au+Au collisions at $\sqrt{s_\mathrm{NN}}=3$ GeV from UrQMD with coalescence (solid lines) and from UrQMD combined with the statistical multi-fragmentation model (dashed lines). Experimental data points are taken from STAR \cite{STAR:2021ozh,STAR:2021yiu,STAR:2022fnj}.
    The figure is adopted from Ref. \cite{Reichert:2025rnw}. }
    \label{UrQMD_v1}
\end{figure}
\begin{figure}[h!]
    \centering
    \includegraphics[width=0.95\columnwidth]{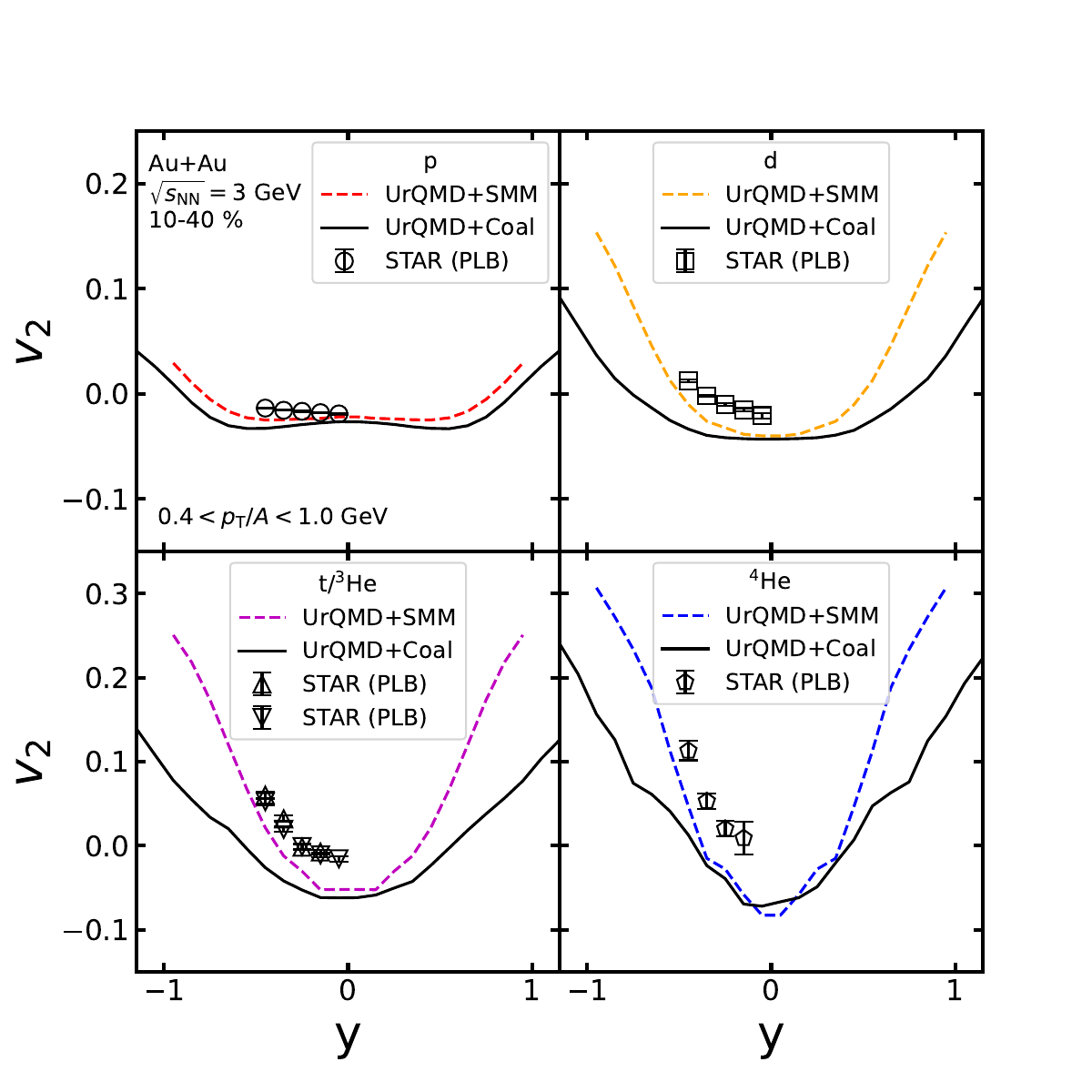}
    \vspace*{-2mm}    
    \caption{The elliptic flow $v_1$ as a function of rapidity of protons (upper left), deuterons (upper right), tritons and $^3$He (lower left) and $^4$He (lower right) from 10-40\% central Au+Au collisions at $\sqrt{s_\mathrm{NN}}=3$ GeV from UrQMD with coalescence (solid lines) and from UrQMD combined with the statistical multi-fragmentation model (dashed lines). Experimental data points are taken from STAR \cite{STAR:2021ozh,STAR:2021yiu}.
    The figure is adopted from Ref. \cite{Reichert:2025rnw}.}
    \label{UrQMD_v2}
\end{figure}

In the recent UrQMD study \cite{Reichert:2025rnw} the flow of light nuclei and hypernuclei has been investigated in Au+Au collisions at $\sqrt{s_\mathrm{NN}}=3$ GeV. There the $v_1$ and $v_2$ of  light clusters and hypernuclei have been compared for two different production mechanisms employed in the UrQMD simulations: the coalescence mechanism  \cite{Sombun:2018yqh} versus  a statistical multi-fragmentation (SMM) approach \cite{Bondorf:1995ua}. There the  momentum dependent potentials from a CMF model has been used. 
The calculations show a good agreement with the STAR data - as demonstrated in Figs. \ref{UrQMD_v1},\ref{UrQMD_v2}, which present the directed flow $v_1$ and $v_2$, respectively,
as a function of the rapidity of protons (upper left), deuterons (upper right), tritons and $^3$He (lower left) and $^4$He (lower right) from 10-40\% central Au+Au collisions at $\sqrt{s_\mathrm{NN}}=3$ GeV from UrQMD with coalescence (solid black lines) and from UrQMD combined with the statistical multi-fragmentation model (dashed lines).

Moreover, it has been found that the directed flow $v_1$ of p, d, t, $^3$He, $^4$He as well as of $\Lambda$, $^3_\Lambda$H and $^4_\Lambda$H  approximately scales with the mass number $A$ of the light cluster in both calculations - coalescence and SMS. This is in line with the  directed flow measured by STAR as well as the PHQMD calculations.


\begin{figure*}[h!]
    \centering
    \includegraphics[width=0.35\textwidth]{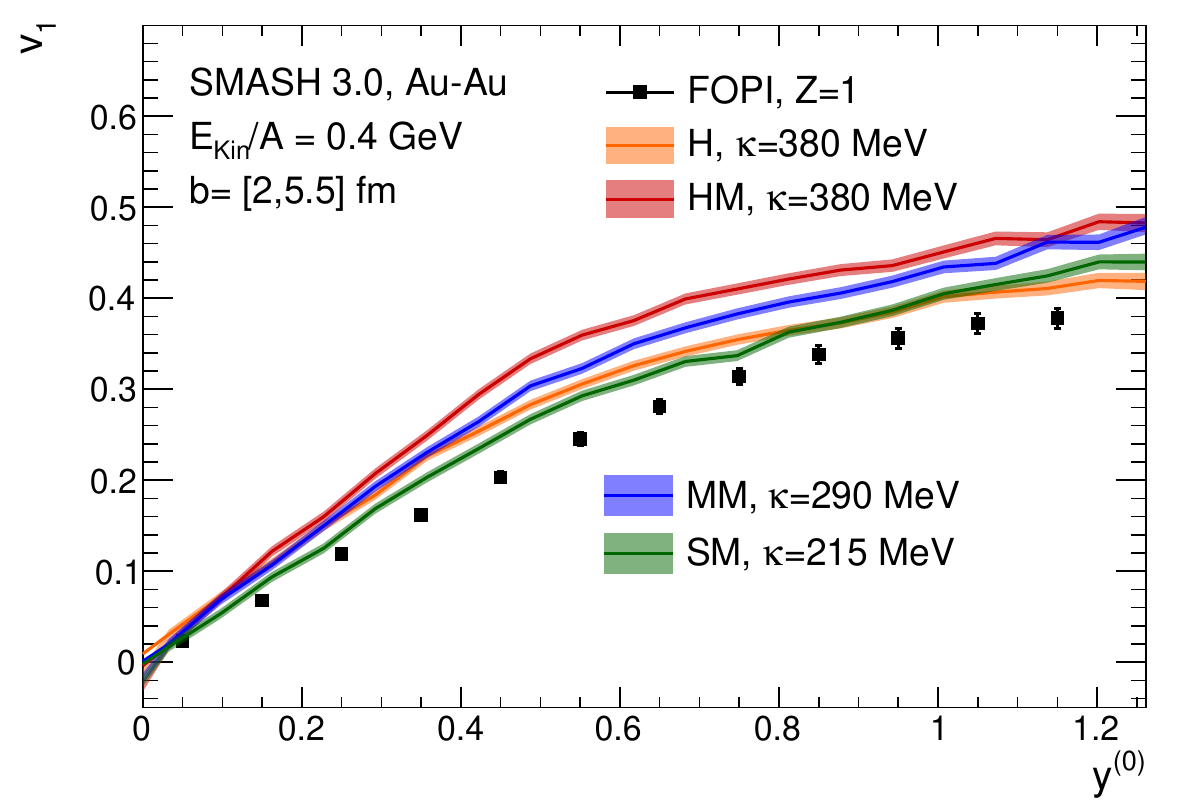} \hspace*{3mm}
     \includegraphics[width=0.35\textwidth]{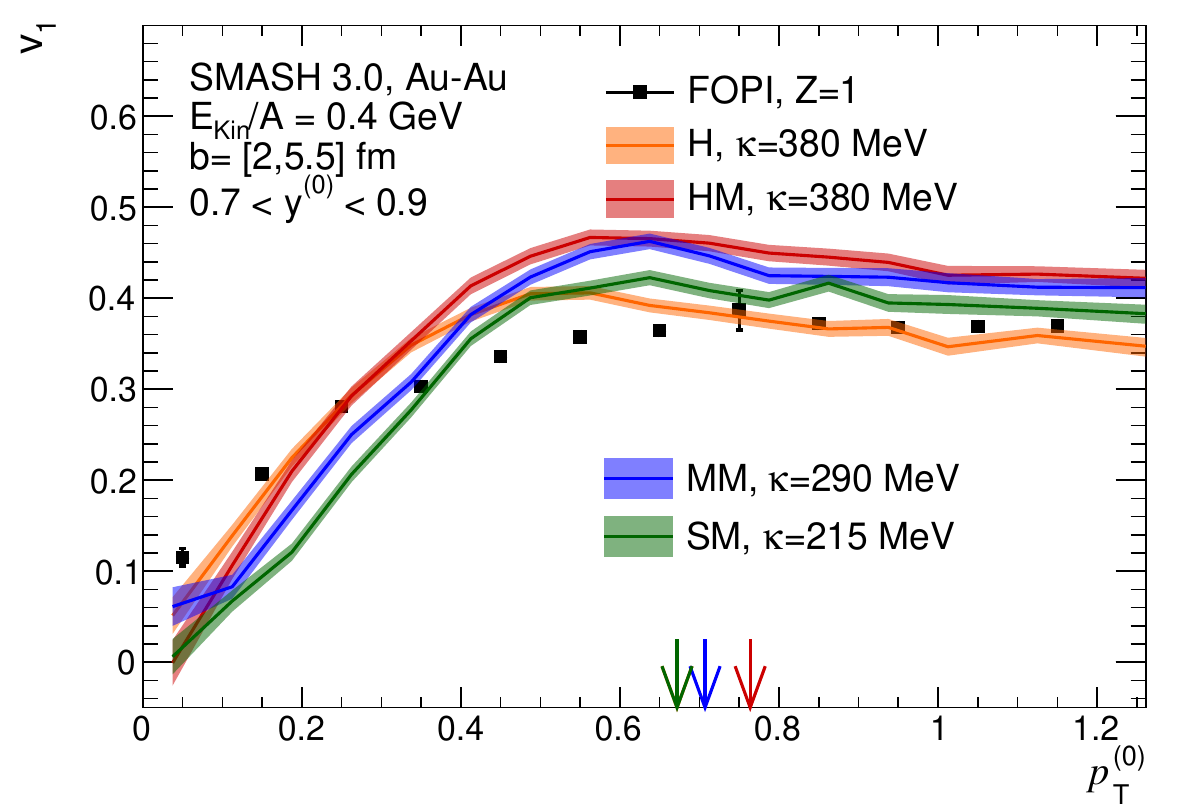}
    \caption{SMASH results for the directed flow $v_1$ for $Z=1$ for different EoS as a function of normalized rapidity $y^{0}$ (left) and transverse momentum $p_T$ (right) compared to the FOPI data~\cite{FOPI:2003fyz} for Au+Au collisions at $E_{kin}=0.4$ AGeV.
    The figure is adopted from Ref. \cite{Tarasovicova:2024isp}.    }
    \label{fig:SMASH_v1_AuAu}
\end{figure*}
\begin{figure*}[h!]
    \centering
    \includegraphics[width=0.34\textwidth]{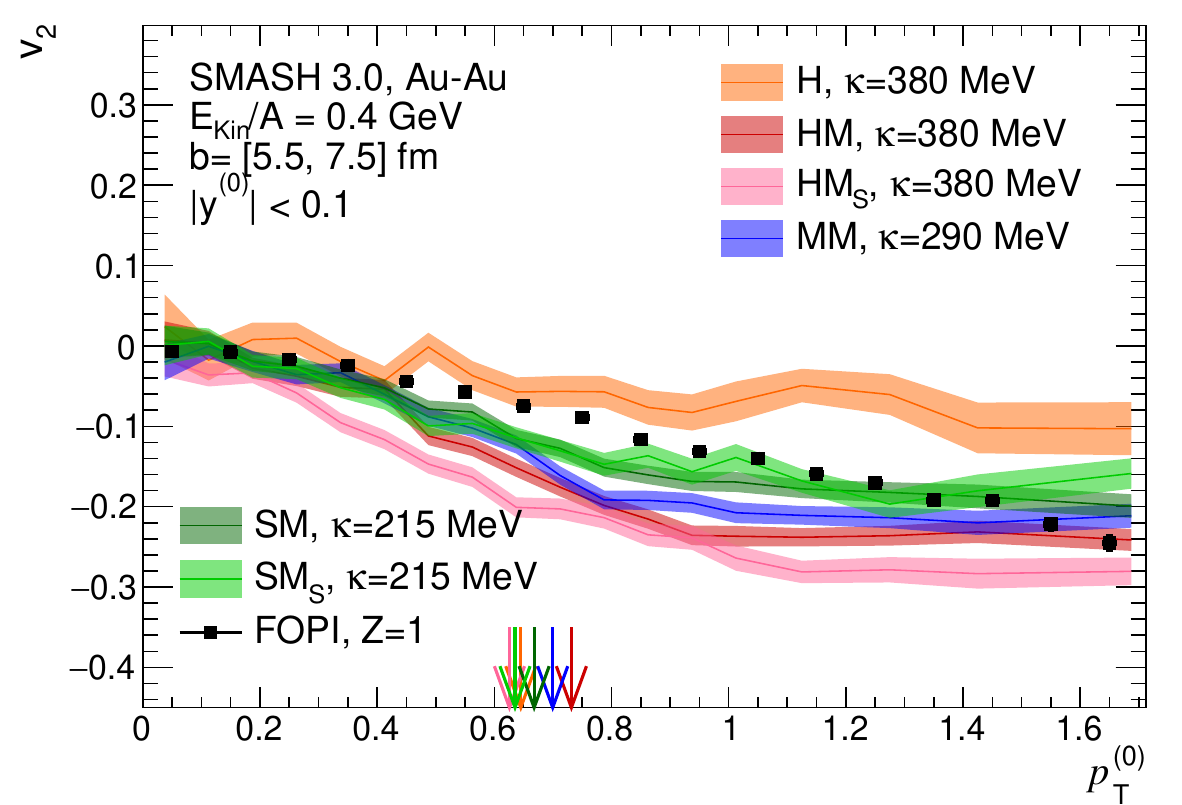}\hspace*{-2mm}
     \includegraphics[width=0.34\textwidth]{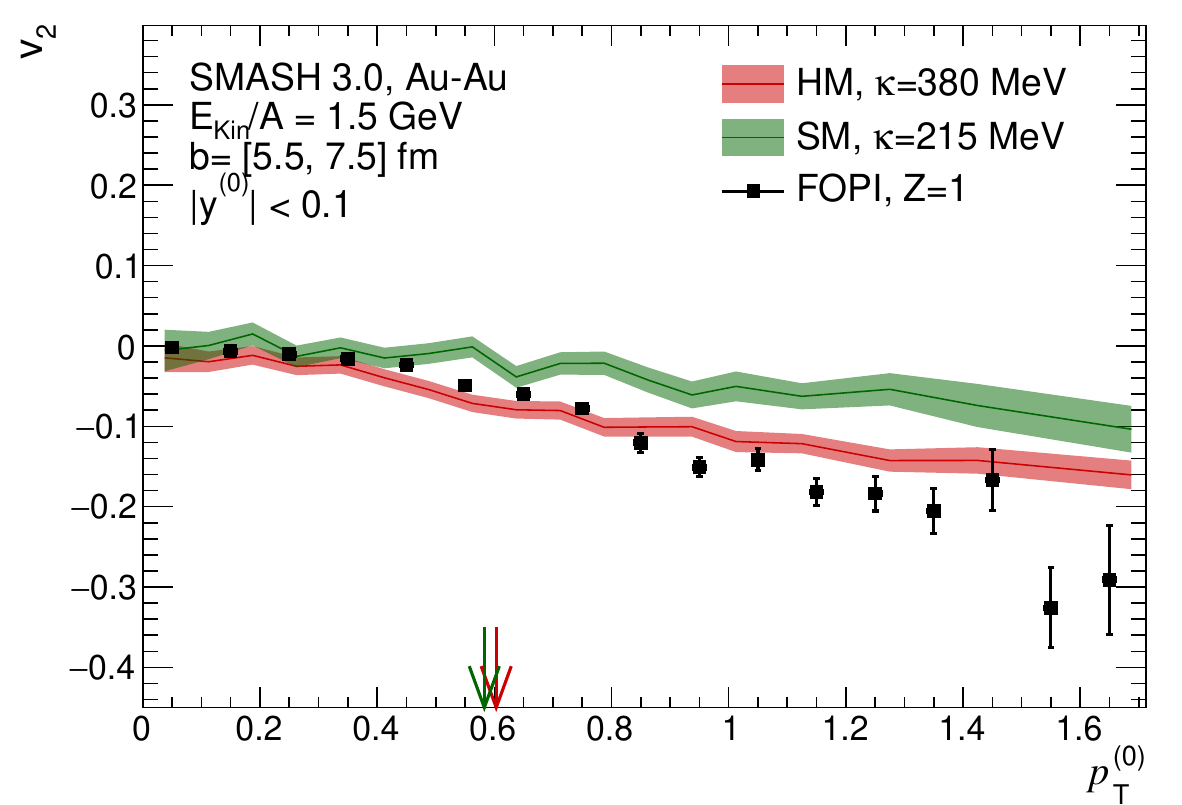}\hspace*{-2mm}
     \includegraphics[width=0.34\textwidth]{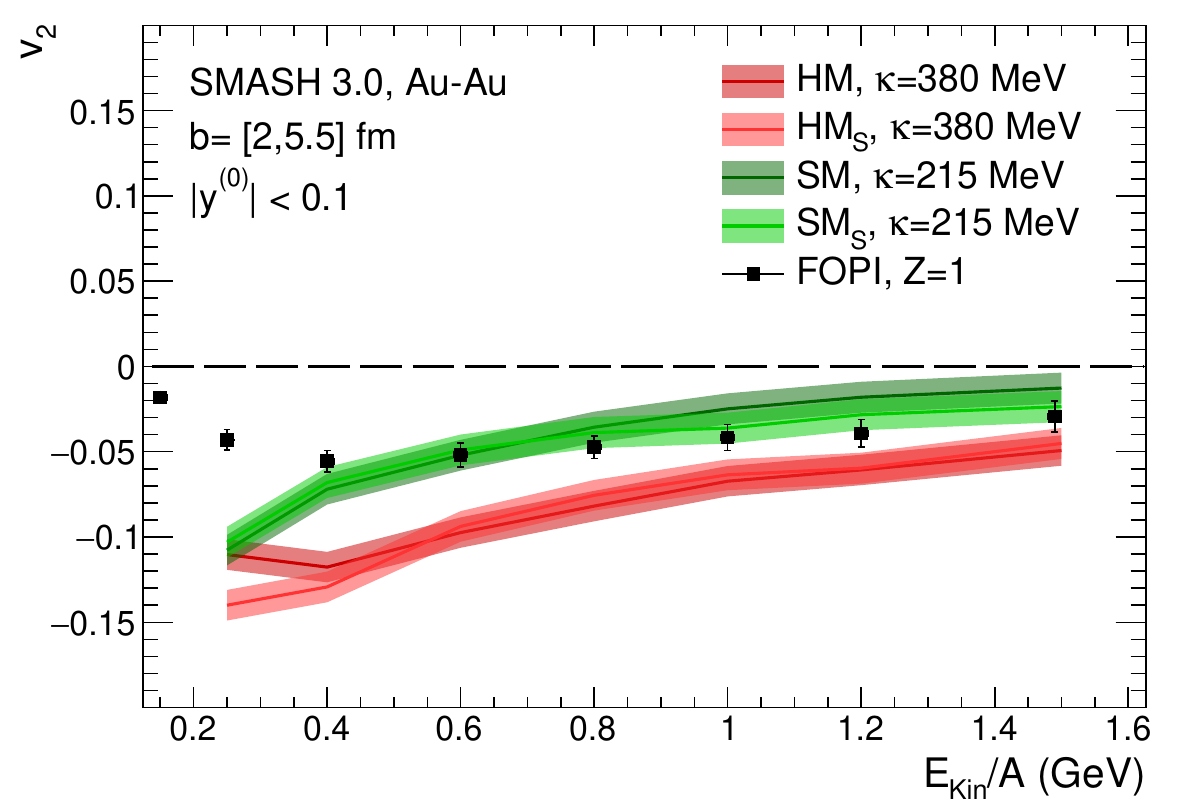} 
    \caption{SMASH results for the elliptic flow $v_2$ for different EoS  as a function of transverse momentum $p_T$ at $E_{kin}=0.4$ AGeV (left) and $E_{kin}=1.0$ AGeV (right) compared to FOPI data~\cite{FOPI:2004bfz}. 
    Here "HMs" stands for Hard EoS with momentum dependent potentials and the stochastic collision criterion. 
   Right plot: the elliptic flow coefficient for $Z=1$ particles as a function of the beam kinetic energy for mid-central Au+Au collisions compared with FOPI data~\cite{FOPI:2004bfz}.  
    The figure is adopted from Ref. \cite{Tarasovicova:2024isp}.}
    \label{fig:SMASH_v2_AuAu}
\end{figure*}

\subsection{SMASH}

The EoS  has been studied within  SMASH (Simulating Many Accelerated Strongly-interacting Hadrons) model  \cite{SMASH:2016zqf}, which is a BUU-type hadronic transport approach based on vacuum degrees of freedom. SMASH has been  applied for the study of flow coefficients of protons and light clusters at SIS energies in Refs. \cite{Mohs:2020awg,Mohs:2024gyc,Tarasovicova:2024isp}.  
The light cluster production in SMASH occurs via a coalescence mechanism \cite{Mohs:2020awg}, or by the kinetic fusion reactions realized either via a fictitious $d'$ resonance as intermediate state \cite{Oliinychenko:2020znl} or by multiparticle  $3 \leftrightarrow 2$ reactions for deuterons, tritons, Helium-3 and hypertriton production \cite{Staudenmaier:2021lrg}.

Recently, in addition to the static Skyrme (and symmetry) potential) with compression moduli $\kappa$ from a soft ($\kappa=$215 MeV) to a hard ($\kappa=$380 MeV) EoS, a momentum dependent potential for nucleons and baryonic resonances has been implemented in Refs. \cite{Mohs:2020awg,Mohs:2024gyc,Tarasovicova:2024isp} in the form:
\begin{equation}
U_\mathrm{MD}(\rho,\mathbf{p})=\frac{2C}{\rho_0}g\int\frac{d^3p'}{(2\pi\hbar)^3}\frac{f(\mathbf{r}, \mathbf{p}')}{1+\left(\frac{\mathbf{p}-\mathbf{p}'}{\hbar\Lambda}\right)^2} \,,
\label{eq:momentum_dependence}
\end{equation}
where $\mathbf{p}$ is the 3-momentum of the considered hadron, $g=4$ is the degeneracy factor and $C$ and $\Lambda$ are parameters  (see Table 1 in Ref. \cite{Tarasovicova:2024isp}), which are adjusted to reproduce the nuclear ground state properties and to describe the Schrödinger equivalent optical potential $U_{opt}(p)$  extracted from pA collisions \cite{Hama:1990vr, Clark:2006rj,Cooper:1993nx}. 
In order to simplify the numerical calculation, an important assumption has been adopted in SMASH: following the implementation in GiBUU~\cite{Buss:2011mx} (used for pA reactions),  a cold nuclear matter approximation for the distribution function $f(\mathbf{r},\mathbf{p}') = \Theta(p_F(\rho_B(\mathbf{r}))-|\mathbf{p}'|)$ has been applied, which allows to solve the integral over $d^3 p'$ analytically. 
We stress that this assumption, certainly justified for pA reactions, is questionable for heavy-ion collisions and might be the origin for the  different results for similar momentum dependent interactions  in SMASH and PHQMD, which leads to different  conclusions - as pointed out in Section 3.1. Moreover, since the potential (\ref{eq:momentum_dependence}) is not covariant,  the numerical realization is sensitive to the calculation of local baryon density (here an additional Gaussian smearing is used in SMASH which is beyond the semi-classical BUU equations), This requires an averaging over many test particles. The latter, however, makes  the cluster description in the BUU framework challenging.

Here we highlight the most recent SMASH results from Ref. \cite{Tarasovicova:2024isp} where detailed comparison with the FOPI data for Au+Au, Ni+Ni, Xe+CsI for bombarding energies $E_{kin}=0.4 - 1.5$ A GeV have been presented.

Figure \ref{fig:SMASH_v1_AuAu} shows the SMASH results for the directed flow $v_1$ for $Z=1$ particles, calculated for different EoS parametrizations and potentials as a function of the normalized rapidity $y^{0}=y/y_{proj}$ (left) and the transverse momentum $p_T$ (right) in comparison to the FOPI data~\cite{FOPI:2003fyz} for Au+Au collisions at $E_{kin}=0.4$ AGeV. The color lines correspond to the static hard "H" EoS with $\kappa=$380 MeV (orange) and the momentum dependent EoS with different $\kappa=$: 
hard "HM" with $\kappa=$380 MeV (red), medium  "MM" with $\kappa=$290 MeV (blue), soft "SM" with $\kappa=$215 MeV (green).
As follows from the Fig. \ref{fig:SMASH_v1_AuAu}, the FOPI data  for $v_1(y^{0})$  and $v_1(p_T)$ at 0.4 AGeV favor soft momentum dependent potential, while the hard EoS is also in line with data. This conclusion is consistent with the PHQMD results and  other models.

Figure \ref{fig:SMASH_v2_AuAu} present the SMASH results for the elliptic flow $v_2$ for different EoS  as a function of transverse momentum $p_T$ at $E_{kin}=0.4$ AGeV (left) and $E_{kin}=1.5$ AGeV (middle) compared to FOPI data~\cite{FOPI:2004bfz}. 
The right plot shows $v_2$ for $Z=1$ particles as a function of the beam kinetic energy for mid-central Au+Au collisions compared with FOPI data~\cite{FOPI:2004bfz}. Here the conclusions are not straightforward - the FOPI data favor the SM EoS at low energy while with increasing beam energy the  hard momentum dependent EoS becomes more consistent with the data. This
is explicitly seen in the $p_T$ distribution at $E_{kin}=1.5$ AGeV.

\begin{figure}
    \centering
    \includegraphics[width=0.7\linewidth]{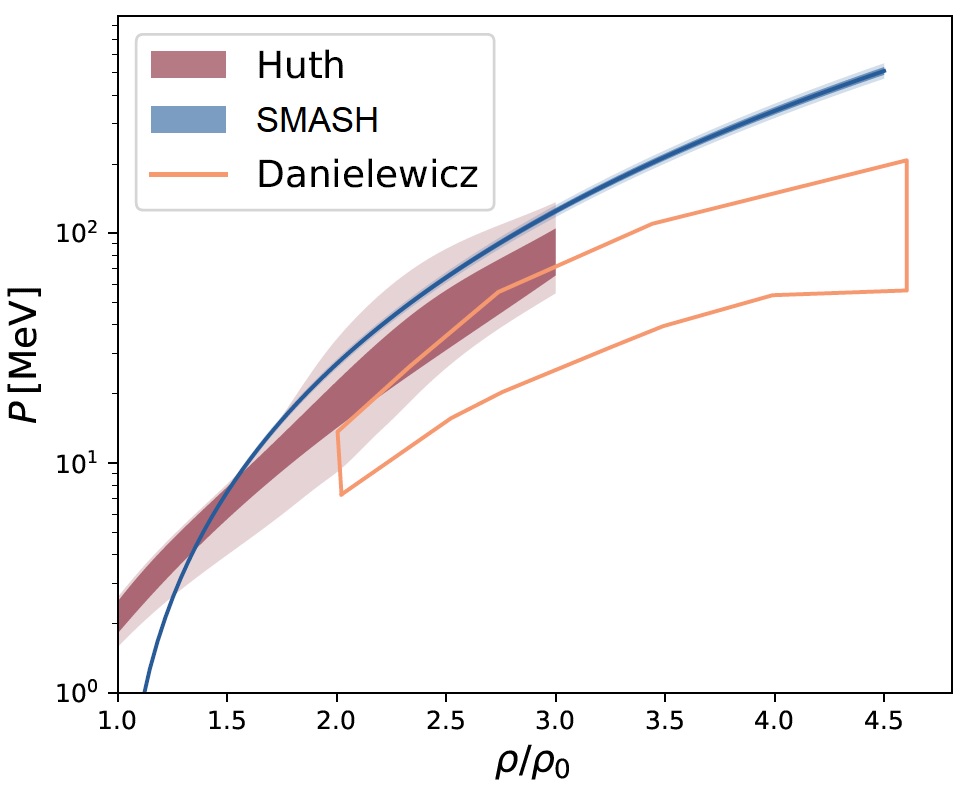}
    \caption{Comparison of EoS - a pressure  as a function of the baryon density - for symmetric nuclear matter at vanishing temperature  reconstructed via Bayesian analysis by SMASH (blue line) to the estimate from the BUU model by Danielewicz et al. \cite{Danielewicz:2002pu} (aria bounded by orange line) and the IQMD model by Huth et al. \cite{Huth:2021bsp} (brown area).
    The figure is adopted from Ref. \cite{Mohs:2024gyc}.}
    \label{fig:SMASH_eos}
\end{figure}

A similar trend has been found by the SMASH model in Ref. \cite{Mohs:2024gyc} based on a Bayesian analysis of the HADES data from Ref. \cite{HADES:2020lob}. The reconstructed EoS  for symmetric nuclear matter at vanishing temperature is shown in Figure \ref{fig:SMASH_eos} by the blue line.  For comparison, it is confronted with the constraints from the BUU model by Danielewicz et al. \cite{Danielewicz:2002pu} (area bounded by the orange line) and the IQMD model by Huth et al. \cite{Huth:2021bsp} (brown-shaded area).

Thus, the SMASH results, obtained by a Bayesian analysis using the HADES data \cite{Mohs:2024gyc}, as well as by the comparison to the FOPI data \cite{Tarasovicova:2024isp}, indicate a transition from a soft (for $E_{kin} \leq 1$ A GeV) to a hard EoS with increasing energy.
 A similar trend was predicted by the BUU transport model of Danielewicz et al. \cite{Danielewicz:1998vz,Danielewicz:2002pu,FOPI:2004bfz}, while the IQMD model \cite{LeFevre:2015paj} favored instead a soft EoS with momentum dependent potential across this range, similar to the PHQMD model.  As pointed out in \cite{Mohs:2024gyc}
the  conclusions about the EoS stiffness also depend on the amount of resonances employed in the transport approach \cite{Hombach:1998wr}.


\section{Summary}
\label{conclusions}

In this review, we have summarized the status of microscopic transport approaches employed to constrain the nuclear equation-of-state from collective flow observables of nucleons and light clusters in the few-GeV energy domain. Particular attention has been devoted to the concepts and recent developments of the PHQMD model, which is currently being advanced by the authors. In addition, results obtained with other transport frameworks have been discussed, including QMD-based approaches (IQMD, dcQMD, UrQMD) as well as BUU-type mean-field models (pBUU, RBUU, SMASH).

Within transport descriptions, the EoS is implemented via effective potentials, ranging from static soft and hard parametri\-zations to their momentum-dependent extensions, the latter being linked to the nuclear optical potential. The empirical knowledge of the momentum-dependent optical potential is restricted to elastic 
pA scattering data up to projectile energies of about 1 GeV. At higher momenta, its functional form necessarily relies on extrapolations constrained by comparisons with heavy-ion observables.
This situation underlines the need for new high-precision elastic pA scattering measurements at higher energies. Such data are indispensable for refining the momentum-dependent potential, which represents a key ingredient in constraining and interpreting the nuclear EoS.

The flow harmonics of protons and light clusters obtained from transport calculations display a pronounced sensitivity to the underlying nuclear EoS. Most studies indicate that a purely soft static potential is incompatible with the experimental data. In contrast, the soft momentum-dependent potential provides a consistent description of both $v_1$ and $v_2$ at lower beam energies. In many cases, a hard static EoS also yields a satisfactory reproduction of flow observables, particularly for light clusters. Certain models, however, point to a gradual hardening of the EoS with increasing incident energy.

Historically static hard and soft EoS have been introduced to reproduce the compressibility deduced from Plastic Ball data and giant monopole vibrations, which test quite different regimes of $\rho/\rho_0$. We see that the introduction of the momentum dependent potential reconciles the compressibility of both data sets, which have, however, considerable error bars. That data at low and high beam energy can be described by the same EoS opens the possibility to apply a Bayesian analysis with the available data sets to fix the compressibility in a specific transport approach.

Despite of qualitative agreements of the transport model results, drawing unique and definitive conclusions about the nuclear EoS remains challenging for several reasons:  
\begin{itemize}
    \item \textbf{flow observables are sensitive to other ingredients.}
    The in-plane and elliptic flow are small signals, which  not only depend
    on the EoS but as well on the ingredients of the transport approaches. They include the parametrization of elastic nucleon-nucleon collisions and the initial
    phase space distribution of the nucleons, quantities which are differently parametrized in different transport approaches. It is a very good sign,
    which shows the maturity of the transport approaches, that in most of the models a SM EoS gives the best agreement with data.
    
    \item \textbf{Definition of the EoS.}  
    The EoS is rigorously defined only for infinite, cold, and equilibrated nuclear matter, whereas heavy-ion collisions probe a rapidly expanding, off- equilibrium medium that is both hot and dense. In equilibrium matter, the momentum dependence of the potential is not manifest because the Fermi momentum is a function of the density and therefore a momentum dependent potential can be expressed as a density dependent potential. Consequently,
    soft (hard) static and soft (hard) momentum dependent parametrizations lead to the same density dependence of $E/A(\rho)$. In contrast, in heavy-ion reactions the situation is reversed: a soft momentum dependent potential can reproduce flow patterns similar to those generated by a hard static potential.  

    \item \textbf{Model-specific interpretations of the optical potential.}  
    Transport approaches differ in their treatment of the momentum dependence of the optical potential, the only quantity directly constrained by experiment. In QMD-type models, $U_{\mathrm{opt}}(p)$ arises from a two-body semiclassical potential depending on the relative momentum of the colliding particles. BUU-type models, in contrast, employ invariant scalar and vector mean fields that depend on the particle momentum with respect to the medium.  

    \item \textbf{Energy dependence and degrees-of-freedom.}  
    The properties of strongly interacting matter created in heavy-ion collisions depend critically on the relevant degrees-of- freedom and their in-medium modifications, which evolve dynamically with beam energy. At higher energies, resonances and string excitations contribute increasingly, while deconfinement phenomena and the possible onset of a quark-gluon plasma  become relevant. Accordingly, the role of ha\-dronic mean-field potentials diminishes with increasing energy.  
\end{itemize}

Another interesting perspective offers the study of hypernuclei. They are rarely produced but carry very valuable information because they contain a produced hyperon, the $\Lambda$, whose dynamical properties are different from that of protons \cite{Zhou:2025zgn}. The study of the dynamical variables of hypernuclei will provide therefore further inside into the reaction mechanism and the expansion of the mid-rapidity interaction zone.

In order to achieve further progress and obtain precise quantitative constraints on the nuclear EoS, two key steps are required:  
\begin{enumerate}
    \item High-precision measurements of $v_1(y,p_T)$ and $v_2(y,p_T)$ in the relevant beam-energy range of a few~AGeV must be performed. Such data are expected from the upcoming FAIR facility.  
    
    \item A systematic comparison of transport approaches with high precession  data will allow to shrink substantially the range of uncertainties in model parameters that  affect the calculated flow observables. 
\end{enumerate}

\phantom{a}\\[1mm]
The authors thank to our colleagues whose model results have been used in the review. We are grateful to W. Cassing for a careful reading of the manuscript and valuable comments. 


\bibliographystyle{epj}
\bibliography{references} 
\end{document}